\documentclass[a4paper,10pt]{article}

\usepackage[utf8]{inputenc}

\usepackage{authblk}

 \usepackage{ifthen,ifpdf}
 
 \ifpdf
   \usepackage{hyperref}
   \hypersetup{colorlinks=false,allcolors=blue}
   \usepackage{hypcap}
   \pdfpagewidth=\paperwidth
   \pdfpageheight=\paperheight
 \fi

\usepackage[utf8]{inputenc}
\usepackage[T1]{fontenc}

\usepackage{amsmath, amsthm, amssymb}
\usepackage{mathtools}
\usepackage{enumerate}
\usepackage{afterpage}
\usepackage{tabularx}
\usepackage{dsfont}
\usepackage{graphicx}

\usepackage{algorithm}
\usepackage{algpseudocode}
\usepackage{caption}
\usepackage[labelformat=simple]{subcaption}

\usepackage{color}
\usepackage{units}
\usepackage{array, multirow}
\usepackage{booktabs}

\newcolumntype{M}[1]{>{\vspace{3pt}\raggedleft\arraybackslash}m{#1}}

\usepackage{siunitx}
\usepackage{tikz}
\usetikzlibrary{matrix}
\usepackage{adjustbox}
\usepackage{pgfplots}
\usepackage{pgfplotstable}
\usepgfplotslibrary{units}
\pgfplotsset{compat=1.9}
\pgfplotstableset{col sep=comma}

\usepackage{cite}

\usepackage{anysize} 
\marginsize{2.5cm}{2.5cm}{2cm}{2cm}

\usepackage{wrapfig}

\theoremstyle{plain}

\theoremstyle{remark}

\numberwithin{equation}{section}

\newcommand{\R}{\mathds{R}}

\newcommand{\C}{\mathds{C}}

\newcommand{\argmin}{\text{argmin}}
\newcommand{\symgrad}{\nabla^s}

\newcommand{\lb}{\left(}
\newcommand{\rb}{\right)}
\newcommand{\sv}{\lb\begin{array}{cccccccccc}}
\newcommand{\ev}{\end{array}\rb}

\DeclareMathOperator{\Id}{\textrm{Id}}

\renewcommand{\div}{\textrm{div }}
\newcommand{\divm}{\textrm{div}^-\,}

\newcommand{\mean}[1]{\left\langle{#1}\right\rangle_{\Y}}

\newcommand{\Y}{Y}
\newcommand{\CCMF}{CCMF}
\newcommand{\EffCrack}{{Effective crack energy}}
\newcommand{\effCrack}{{effective crack energy}}
\newcommand{\geff}{\gamma_\text{eff}}
\newcommand{\crackres}{{crack resistance}}
\newcommand{\damping}{\delta}

\DeclareMathOperator{\eps}{\varepsilon}

\usepackage{comment}

\title{An FFT-based method for computing the {effective}\\ crack energy of a heterogeneous material\\ on a combinatorially consistent grid}

\author[1]{Felix Ernesti}
\author[1,*]{Matti Schneider}

\affil[1]{Karlsruhe Institute of Technology (KIT), Institute of Engineering Mechanics}
\affil[*]{correspondence to: \texttt{matti.schneider@kit.edu}}

\date{\today}

\begin{document}

\maketitle 

\begin{abstract}
\noindent We introduce an FFT-based solver for the combinatorial continuous maximum flow discretization applied to computing the minimum cut through heterogeneous microstructures.\\
Recently, computational methods were introduced for computing the \effCrack{} of periodic and random media. These were based on the continuous minimum cut-maximum flow duality of G. Strang, and made use of discretizations based on trigonometric polynomials and finite elements. For maximum flow problems on graphs, node-based discretization methods avoid metrication artifacts associated to edge-based discretizations.\\
We discretize the minimum cut problem on heterogeneous microstructures by the combinatorial continuous maximum flow discretization introduced by Couprie et al. {Furthermore, we introduce an associated FFT-based ADMM solver and provide several adaptive strategies for choosing numerical parameters. We demonstrate the salient features of the proposed approach on problems of industrial scale.}\\ \quad \\
{\noindent\textbf{Keywords:} FFT-based computational homogenization; \EffCrack{}; Combinatorial continuous maximum flow; Alternating direction method of multipliers}
\end{abstract}

\newpage

\section{Introduction}
\label{sec:intro}

\subsection{State of the art}

\begin{figure}[t]
	\begin{minipage}{.3\textwidth}
		\begin{tikzpicture}[      
        	every node/.style={anchor=south west,inner sep=0pt},
        	x=1mm, y=1mm,
      	]   
     	\node (fig1) at (0,0) {\includegraphics[width=\textwidth, trim = 550 60 650 220, clip, draft = false]{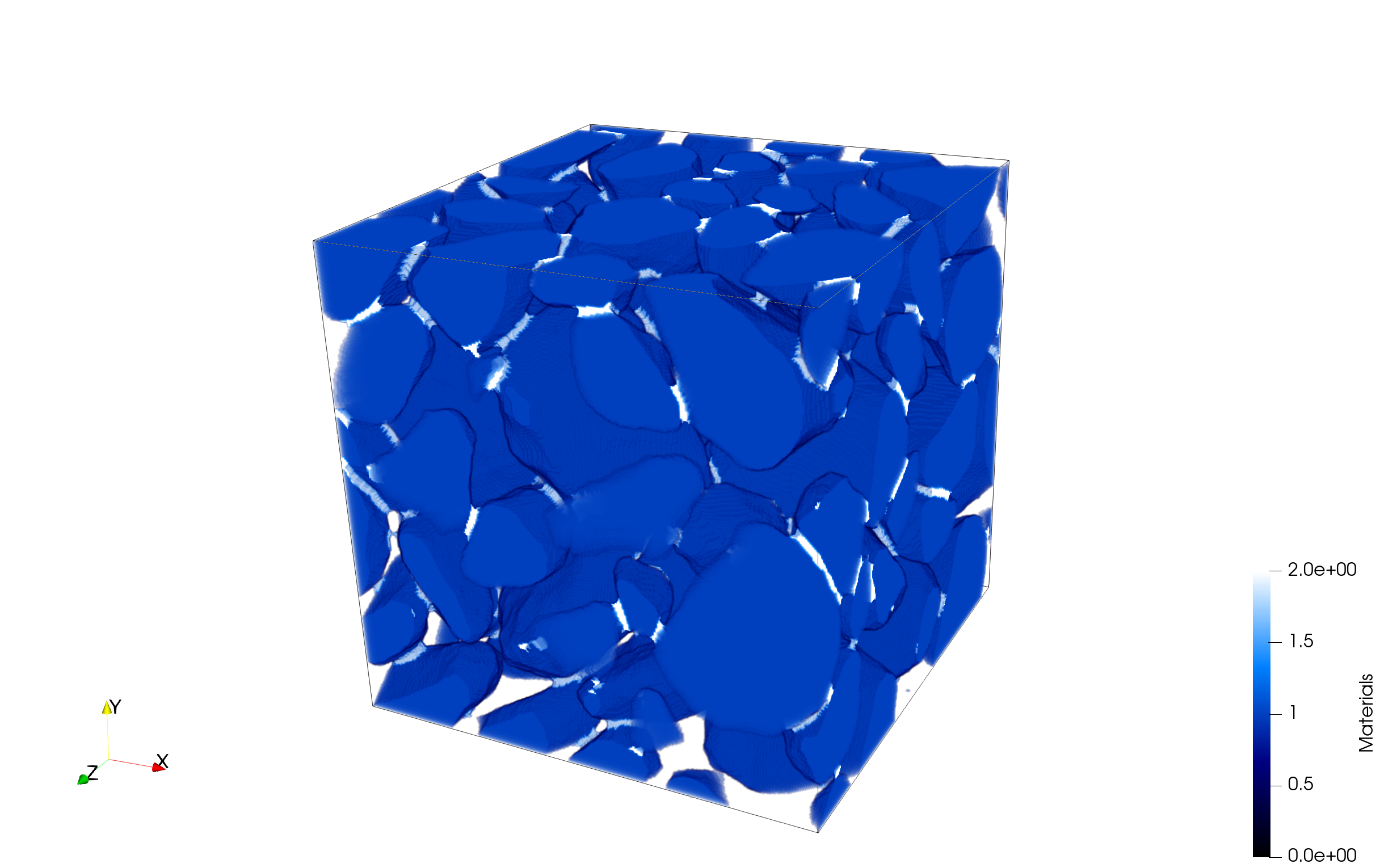}};
     	\node (fig2) at (0,0) {\includegraphics[width=0.15\textwidth, draft = false]{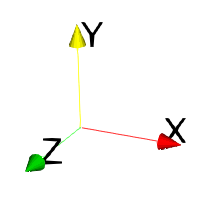}};  
		\end{tikzpicture}
    	\caption{Complex microstructure of bound sand~\cite{Sand,SandMultiPhysics}}
    	\label{fig:SandStructure}
	\end{minipage}
	\hspace{0.015\textwidth}
	\begin{minipage}{.64\textwidth}
		\includegraphics[width=\textwidth, draft = false]{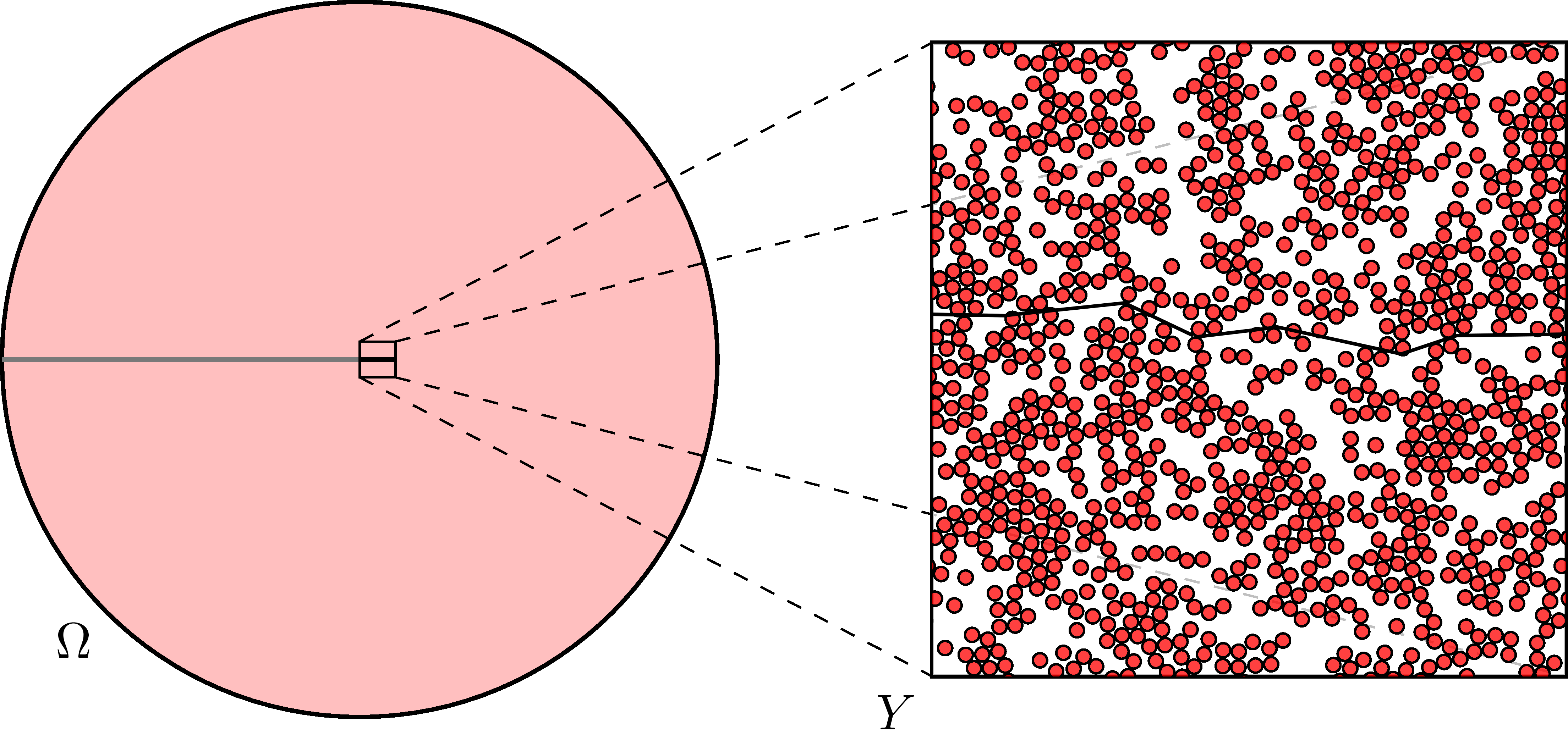}
		\caption{Schematic of a crack increment in a microstructured \\material}			
		\label{fig:homogenizationSchema}
	\end{minipage}
\end{figure}

Modern fracture mechanics~\cite{GrossSeelig} originated from the pioneering work of Griffith~\cite{Griffith}, who postulated a criterion for the quasi-static growth of a pre-existing crack in a brittle, isotropic and elastic solid. More precisely, his postulate is based on an energetic reasoning. For a finitely-sized increment in loading, the increase in elastic stored-energy and the increase in crack length need to be balanced in terms of an appropriate proportionality constant, the critical energy-release rate (also called fracture toughness or crack resistance). More precisely, Griffith's criterion postulates that the crack grows whenever it is energetically more favorable to increase the surface energy of the crack than to increase the elastic energy stored in the body.\\
Irwin~\cite{Irwin} extended Griffith's reasoning to three-dimensional isotropic elasticity and a semi-infinite planar pre-existing crack in an infinite medium. Irwin realized that, for the so-called crack modes, analytic solutions for the strain and stress fields are accessible, and, upon loading under either mode I, mode II or mode III, the semi-infinite planar crack grows by homogeneously advancing the crack front, essentially preserving the semi-infinite planar shape of the crack. Irwin also introduced stress-intensity factors associated to each mode, which correspond to factors of proportionality in front of the analytic solutions. Expressing the energy-release rate in terms of these stress-intensity factors, Irwin suggested a reformulation of Griffith's criterion solely based on the stress-intensity factors~\cite{Irwin62}.\\
Independently, Cherepanov~\cite{Cherepanov1967} and Rice~\cite{Rice68} proposed a clever way to compute the local energy-release rate at a crack trip in terms of a contour integral around the crack tip, the so-called J-integral, which is actually independent of the chosen path, and may be used to compute the fracture toughness of a material.\\
Griffith's criterion presupposes the crack path to be known in advance, i.e., it may only be used for assessing \emph{when} a crack propagates, and not for predicting \emph{how} it grows. To predict the latter, the most common approaches exploit the principle of local symmetry~\cite{PLS} or follow the postulate of maximum energy-release~\cite{Hussain74}.\\
Linear elastic fracture mechanics was also extended to account for elastoplastic effects, see Dugdale~\cite{Dugdale} and Barenblatt~\cite{Barenblatt} for early contributions. Classical finite-element methods may be used in computational approaches to fracture mechanics, for instance by computing the stress-intensity factors numerically~\cite{RiceTracey}. However, it turns out to be difficult to resolve the singularity at the crack tip. For this purpose, enriched~\cite{EnrichtedFEM} or extended~\cite{GXFEM} finite-element discretizations were developed, which account for the crack-tip by adding special ansatz functions to cracked elements.\\
To alleviate the burden of characterizing the mechanical behavior of anisotropic materials, multiscale methods, in particular homogenization approaches, proved to be very effective. We refer to Matou\v{s} et al.~\cite{MatousSummary} for a recent overview. As a general note, upscaling the mechanical behavior of solids is well understood in the absence of localization, for instance for elastoplasticity with hardening. In case of localization, additional difficulties arise, see Gitman et al.~\cite{Gitman}.\\
One way to avoid these problems consists of trying to understand crack propagation in a homogeneous elastic medium with an anisotropic stiffness tensor. However, Irwin's approach~\cite{Irwin} based on  a modal decomposition and associated stress-intensity factors does not work for this general scenario. Indeed, Sih, Paris \& Rice~\cite{AnisotropicSIF} showed that the associated stress-intensity factors may attain complex, i.e., unphysical, values. Physically speaking, this is rooted in the incompatibility of a semi-infinite planar crack and anisotropic elasticity. Thus, despite its sweeping success, the strong dependence of classical linear elastic fracture mechanics on specific, analytical solutions appears to preclude handling multiscale materials in a natural way.\\
Francfort and Marigo~\cite{FrancfortMarigo} revisited Griffth's original proposition from the perspective of "modern" mathematical analysis. More precisely, for a given body $\Omega$ and after a discretization in time, they seek the displacement $u$ and the crack surface $S$ as minimizers of the functional
\begin{equation}\label{eq:FM}
	FM(u,S) = \frac{1}{2} \int_\Omega \symgrad u:\C(x):\symgrad u \, dx + \int_{S} \gamma(x) \,d A
\end{equation}
under the constraint of crack irreversibility, i.e., that the crack set $S$ must contain the crack set of the previous time step. Some care has to be taken with the formulation \eqref{eq:FM}, as Griffith's original proposal concerns only critical points of the functional \eqref{eq:FM}, whereas a rigorous mathematical treatment~\cite{Chambolle2018b} of the Francfort-Marigo model \eqref{eq:FM} appears to be limited to \emph{global minimizers}. Please note that the formulation \eqref{eq:FM} accounts for heterogeneities in a natural way.\\
To improve upon classical linear elastic fracture mechanics, the formulation \eqref{eq:FM} must also go beyond analytical solutions. However, this is not that simple, as the formulation \eqref{eq:FM} is based on an evolving singular surface $S$. A powerful strategy, closely resembling the Ambrosio-Tortorelli approximation of the Mumford-Shah functional, was introduced by Bourdin~\cite{Bourdin}. The proposed approach may be interpreted as a non-local damage model~\cite{DimitrijevicHackl,Bazant}. Due to its similarity to phase-field models, however, this class of numerical approximations is nowadays referred to as phase-field fracture models~\cite{Kuhn2010,Miehe}.\\
Due to their ability to produce complex crack patterns, phase-field fracture models were subject to a flurry of activities, see Ambati et al.~\cite{Ambati2015} for a review. In particular, strategies to account for material anisotropy in the phase-field framework~\cite{AnisotropicPhaseField,AnisotropicPhaseField2} were proposed. Still, these anisotropic models on the macroscale are phenomenological, i.e., they do not emerge from an upscaling or a homogenization procedure. In particular, the involved (anisotropic) material parameters still need to be identified.\\
As already indicated, upscaling softening damage or fracture is non-trivial, in particular for complex materials as shown in Fig.~\ref{fig:SandStructure}.
This becomes apparent from different perspectives. Classically, a sufficient condition for homogenization to be applicable is a scale separation between the heterogeneities on the microscale and the typical change of the fields of interest on the macroscale. For a crack, however, the stress is singular at the crack tip, and this sufficient condition does not hold. From another perspective, let us consider a non-local damage model. The non-locality is necessary to arrive at mesh-independent results for a corresponding finite-element model. Thus, two scales are present in such a multiscale non-local damage model - the typical scale of heterogeneity and the length scale of the nonlocality. In upscaling, the scale of heterogeneity is small, and we wish to pass to the limit of vanishing heterogeneity size. Fixing the non-local length scale upon homogenization means that the nonlocality essentially screens the influence of the heterogeneities. Tying the non-local length scale to the size of the heterogeneities, however, means that the mesh-dependence of the damage model is recovered upon homogenization, rendering the procedure illegitimate. Of course, computations of non-local damage models~\cite{Berthier,BoeffEtAl} or phase-field fracture~\cite{Yvonnet2016,Chen,Ernesti2020} on microstructures may be pursued. Unfortunately, connecting the results to macroscopic material properties appears challenging.\\
More to the point, the actual form of the effective model is at the heart of the problem. Only in the isotropic and linear elastic case, there appears to be a universal agreement on the physics of brittle fracture. In particular, the notions of crack resistance, critical energy-release rate and fracture toughness coincide, and represent a scalar material parameter. Upon generalization, such an agreement of the different notions may be lost. Consider, for instance, a heterogeneous material. For a crack crossing a material boundary, in general, the energy-release rate will be a function of time (under quasi-static loading). Then, it remains to \emph{define} the macroscopic, i.e., \emph{effective}, crack resistance, for instance as the maximum or the mean of the encountered values. We do not intend to evaluate these proposals here, but rather stress that there appears to be some disagreement on the notion of "effective crack resistance" in the literature, and care has to be taken when comparing these notions.\\
In the context of classical linear elastic fracture mechanics, Bower-Ortiz~\cite{BowerOrtiz1991} provided a perturbative solution for a semi-infinite crack passing through a single, tough inclusion in a matrix. Roux et al.~\cite{Roux2003} discussed an emerging effective crack resistance for a material whose elastic properties are isotropic and homogeneous, and only the crack resistance is heterogeneous. In this context, a self-consistent method for estimating the effective fracture toughness of a planar crack propagating through inclusions is established. In a Gaussian medium they identified regions of weak pinning, where the fracture toughness is given by the arithmetic mean of the local toughness, and strong pinning, where a much higher toughness emerges, see also D\'emery et al.~\cite{Demery2014} for a related study. Lebihain~\cite{LebihainDiss} extended the mentioned studies by accounting for cracks which bypass an inclusion, based on a perturbative, co-planar approach~\cite{Rice1985}.\\
To account for a heterogeneity in the elastic properties, Hossain et al.~\cite{Hossain2014EffectiveToughness} performed phase-field fracture computations on heterogenous microstructures with specific, so-called "surfing" boundary conditions. The emerging effective crack resistance equals the maximum in time of the J-integral evaluated along the crack tip, see also Brach et al.~\cite{Brach2019}.\\
As an alternative to these approaches, Braides et al.~\cite{Braides1996} proved a mathematical homogenization result for the Mumford-Shah functional, which corresponds to the Francfort-Marigo~\cite{FrancfortMarigo} model upon antiplane-shear loading. More precisely, after a discretization in time, Braides et al.~\cite{Braides1996} consider a fixed periodic microstructure (with non-degenerate stiffness and crack resistance), and identify the  $\Gamma$-limit for vanishing period as the functional
\begin{equation}\label{eq:FM_eff}
	FM_\text{eff}(u,S) = \frac{1}{2} \int_\Omega \symgrad u:\C_\text{eff}:\symgrad u \, dx + \int_{S} \gamma_\text{eff}(n) \, d A,
\end{equation}
where $n$ denotes the unit normal to the crack surface $S$. Here, the (possibly anisotropic) effective stiffness tensor $\C_\text{eff}$ arises from the usual elastic homogenization formula based on the classical cell problem~\cite{Milton2002}. The integrand $\gamma_\text{eff}$ of the surface term is a function of a unit vector, and may be computed by a corrector problem involving the local crack resistances only. However, in contrast to the elastic contribution, the latter surface term involves an infinite-volume limit, as standard for stochastic homogenization~\cite{KanitRVE,Owhadi2003,BourgeatPiatnitski}, also for periodic materials. For cells of finite size, the surface term $\gamma_\text{eff}(n)$ may be interpreted as finding the $\gamma$-weighted minimal surface with average normal $n$ cutting the microstructure~\cite{HomFrac2019}.\\
In particular, the volumetric and the surface energies decouple upon homogenization, as a result of the different scalings of these terms in the model \eqref{eq:FM}. Subsequently, the homogenization statement was extended to the case of stationary and ergodic random materials~\cite{Cagnetti2019}, i.e., in engineering terms, the existence of representative volume elements~\cite{HillRVE,DruganWillis}, separately for the bulk and the surface part, is ensured. Please note that this volume-surface decoupling is a consequence of the assumed non-degeneracy of the integrands. In case of degeneracy, an interaction of the two terms is not excluded, see Barchiesi et al.~\cite{CIP_homogenisation_soft_inclusions} and Pellet et al.~\cite{CIP_high_contrast_MumfordShah}.\\
Let us also highlight that the effective model \eqref{eq:FM_eff} also emerges when homogenizing the Ambrosio-Tortorelli approximation of the Francfort-Marigo model, i.e., phase-field fracture models, see Bach et al.~\cite{Bach2021}.\\
Let us put the homogenization result \eqref{eq:FM_eff} into context. The heterogeneous fracture problem involves two prominent length scales: the correlation length of the heterogeneities and the typical size of a displacement increment. Classically, owing to the quasi-static framework, the size of the displacement increment is assumed infinitesimal. In this interpretation, a crack propagates through a microstructure, and its progress may be hindered by various factors, like being pinned to an interface. This interpretation is implicit in Hossain et al.~\cite{Hossain2014EffectiveToughness}, for example. In practical applications, however, the displacement increment is typically of the order of the macroscopic scale. In contrast, Braides et al.~\cite{Braides1996} fix the displacement increment once and for all, and pass to the limit of infinitesimally small heterogeneities, see Fig.~\ref{fig:homogenizationSchema} for an illustration.\\
Another difference is the understanding of the emerging \emph{effective} properties. As Hossain et al.~\cite{Hossain2014EffectiveToughness} always consider a time-continuous problem, their crack resistance is defined as the maximum in time of the local J-integral. In contrast, Braides et al.~\cite{Braides1996} practically work with an energy equivalence between the macroscopic fracture energy and the microscopic fracture energy, as a result of their energetic framework.\\
Last but not least, let us remark that $\Gamma$-convergence implies the convergence of \emph{absolute} minimizers, but does not predict what happens to \emph{local} minimizers. Although an energy equivalence between the microscopic fracture energy and the macroscopic fracture energy appears in a natural way, the (absolutely) minimal surface in the cell problem appears to be a byproduct of $\Gamma$-convergence. From a physical point of view, it might be more appropriate to work with crack surfaces that are just local minima of the weighted area. Still, with assessing the safety of microstructured components in mind, the absolutely minimal surface serves as a lower bound for the (real) \effCrack{}, and is furthermore robust w.r.t. stochastic fluctuations in the microstructure. To stress the difference to the crack resistance, we will reserve the terminology "\effCrack{}" for the surface integrand $\gamma_\text{eff}$.\\
A method for computing the \effCrack{} was proposed by Schneider~\cite{HomFrac2019} based on a convex reformulation of the minimum-cut problem~\cite{Strang1983}. More precisely, a primal-dual hybrid gradient method~\cite{EsserZhangChan,PockCremersBischofChambolle} was used, extending previous FFT-based computational homogenization methods.
 
\subsection{Contributions}

In this work, we present an FFT-based solver for a more favorable discretization of the cell problem for computing the \effCrack{} on cells of finite size. More precisely, the latter cell problem is closely related to the minimum-cut problem put forward by Strang~\cite{Strang1983} in his analysis of the continuous maximum-flow problem, see Section \ref{sec:theory_cell_formulae}. Actually, the maximum-flow problem is typically considered for graphs, where powerful algorithms are available~\cite{Ford}. Unfortunately, these graph-based maximum flow algorithms cannot be used for the continuous maximum-flow problem due to metrication artifacts~\cite{Kolmogorov}. As a remedy, Couprie et al.~\cite{CCMF} introduced the \emph{combinatorial continuous maximum-flow} (CCMF) discretization, as discussed in Section \ref{sec:theory_CCMF}. The CCMF discretization naturally avoids metrication errors, and may be implemented into standard convex optimization solvers~\cite{Boyd2004}. However, as the authors remark themselves: "In 3D, our \CCMF{} implementation is suffering from memory limitations in the direct solver we used, limiting its performances."~\cite[Sec. 4.4.3]{CCMF}\\
In this work, we overcome the mentioned limitations by proposing an FFT-based solver exploiting the alternating direction method of multipliers (ADMM) with adaptive penalty selection. Section \ref{sec:numerics} explains how to treat the \CCMF{} discretization by FFT-based computational solvers. This is non-standard, and we propose a strategy based on doubling the dimension of the fields involved. As a byproduct, we arrive at an expression for the minimum-cut problem that is much simpler than in Couprie et al.~\cite{CCMF}, see Section \ref{sec:numerics_primal}. We discuss the alternating direction method of multipliers (ADMM) in Section \ref{sec:numerics_polarization}, enriched by various adaptive parameter-selection strategies, recently studied in Schneider~\cite{ADMM2021}. Finally, we demonstrate the capabilities of our approach in applications of industrial size, see Section \ref{sec:computations}. We find that an adaptive parameter-selection strategy is critical for high performance, improving upon the standard ADMM used by Willot~\cite{Willot2020StronglyNonlinearMedia}, who treats the closely related graph-based maximum-flow problem (not the continuous one).

\section{The \effCrack{} of a heterogeneous material}
\label{sec:theory}

\subsection{Cell formulae for the minimum cut and the maximum flow}
\label{sec:theory_cell_formulae}

\begin{figure}
	\begin{center}
		\includegraphics[width=.5\textwidth]{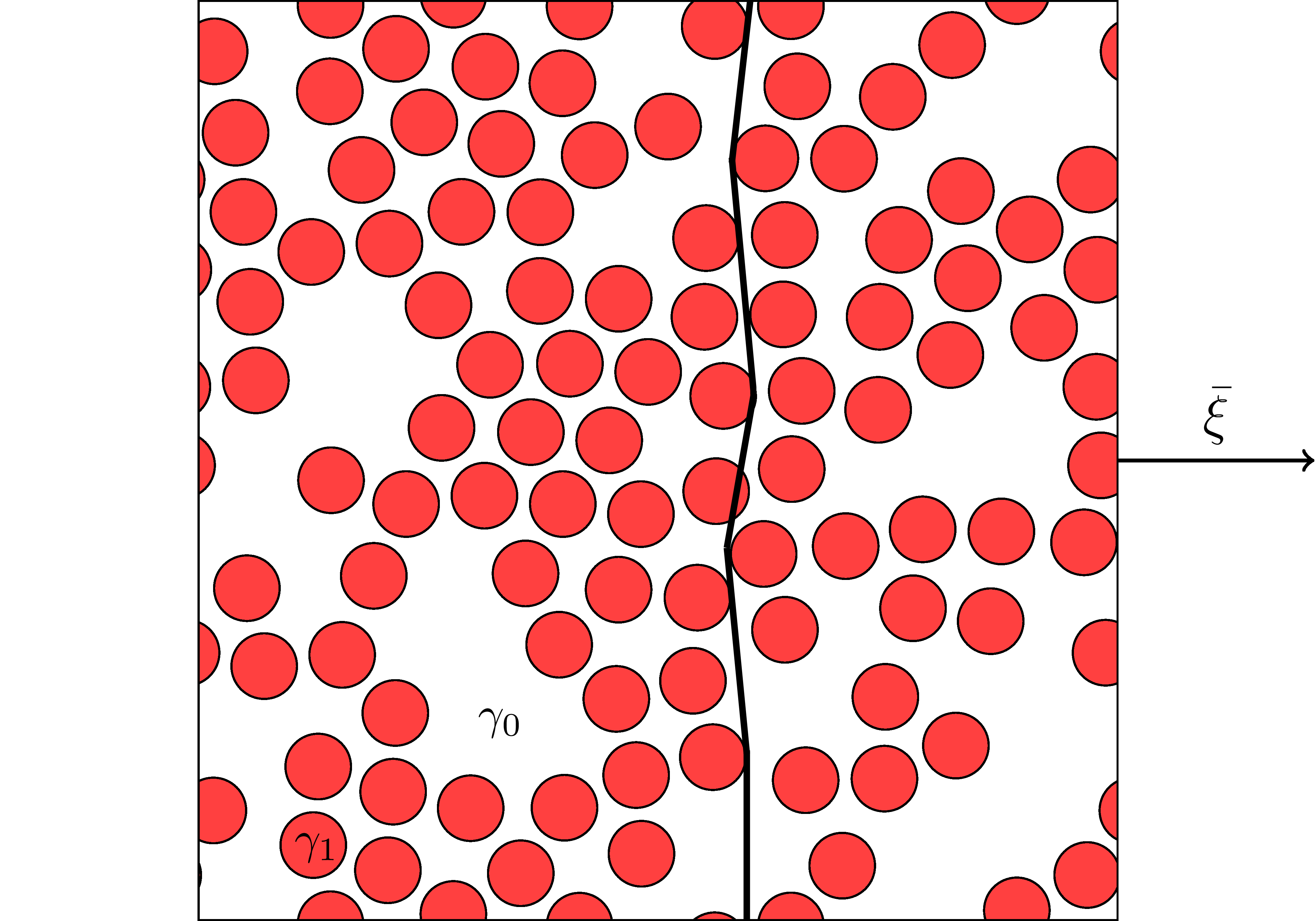}
		\caption{Schematic of a potentially minimal crack traversing a two-phase microstructure for prescribed normal $\bar{\xi}$}
		\label{fig:cellFormula}
	\end{center}
\end{figure}

Let us consider a cuboid cell $\Y=[0,L_1]\times [0,L_2] \times [0,L_3]$, on which a heterogeneous field of \crackres{}s\footnote{{For homogeneous, isotropic materials under certain loading conditions, the term \crackres{}, also known as critical energy-release rate, may be used interchangeably with the notion of fracture toughness, as Griffith's~\cite{Griffith} energetic criterion may be equivalently rewritter in terms of stress-intensity factors~\cite{Irwin}, see Gross \& Seelig~\cite[Ch. 4.6]{GrossSeelig}. In the heterogeneous case, however, this is may not be the case. We therefore restrict to the term \crackres{}, avoiding the term fracture toughness altogether.}} $\gamma:\Y \rightarrow \R$ is given. For heterogeneous materials, as of today, there is no universal agreement upon whether brittle fracture can be homogenized, and how to unambiguously extract the parameters of a suitably up-scaled, \emph{effective} model. For the work at hand, we follow the proposal of Braides et al.~\cite{Braides1996} and consider the \emph{\effCrack{}} of a heterogeneous material. The latter notion is mathematically well-defined, and even gives rise to reasonable large-scale properties for stationary stochastic materials satisfying an ergodic hypothesis, see Cagnetti et al.~\cite{Cagnetti2019} for details. 
The computed \effCrack{} is expected to yield a lower bound for the "effective \crackres{}"~\cite{LebihainDiss,Hossain2014EffectiveToughness}, since the energy released during cracking is proportional to the (weighted) crack area, and is, in particular, bounded from below by the minimum (weighted) crack area.\\
For mathematical reasons, we suppose that there are positive constants $\gamma_\pm$, s.t. the inequalities
\[
	\gamma_- \le \gamma(x) \le \gamma_+ \quad \text{hold for all} \quad x\in \Y.
\]
We define the \effCrack{} $\geff$~\cite{Braides1996,Cagnetti2019,HomFrac2019}, a function on the unit sphere $S^2 \subseteq \R^3$, by
\begin{equation}\label{eq:theory_cell_formulae_cell_formula}
	\geff(\bar{\xi}) = \inf \frac{1}{|\Y|} \int_\Y \gamma\,\left\| \bar{\xi} + \nabla \phi \right\|\, dx, \quad \bar{\xi} \in S^2,
\end{equation}
see Fig.~\ref{fig:cellFormula}, where $|\Y| = L_1L_2L_3$ denotes the volume of the cell and the infimum is evaluated over all smooth scalar fields $\phi:\Y \rightarrow \R$ which are periodic, together with all their derivatives. Please note that the integrand in the right hand side of equation \eqref{eq:theory_cell_formulae_cell_formula} is homogeneous of degree one. This contrasts with thermal conductivity~\cite{Milton2002}, where a homogeneity of degree two leads to a linear Euler-Lagrange equation associated to the variational problem.\\
For the problem at hand \eqref{eq:theory_cell_formulae_cell_formula}, additional complications arise. For a start, due to the one-homogeneity, the functional in the definition \eqref{eq:theory_cell_formulae_cell_formula} is not differentiable. In particular, the first-order necessary conditions are (strongly) non-linear. Furthermore, the one-homogeneity permits \emph{localization} to appear for minimizers of the variational problem \eqref{eq:theory_cell_formulae_cell_formula}. This localization is not unwarranted, as such minimizers actually represent minimum cuts through the microstructure~\cite{Strang1983}, weighted by the \crackres{} , and enable computing the \effCrack{} by the local \crackres{} averaged over the minimum cut. In fact, the minimum cut need not be unique. However, the computed \effCrack{} is unique as a consequence of the convexity of the functional to be minimized.\\ 
To circumvent the inherent lack of differentiability characterizing the functional \eqref{eq:theory_cell_formulae_cell_formula}, dual and primal-dual formulations may be exploited~\cite{Chambolle2016Overview}.  As an example, the (formal) dual to the variational problem is given by the maximum flow problem~\cite{Strang1983}
\begin{equation}\label{eq:theory_cell_formulae_max_flow}
	\frac{1}{|\Y|} \int_\Y v \cdot \bar{\xi}\, dx \longrightarrow \max_{\substack{\div v = 0\\ \|v\|\le \gamma}},
\end{equation}
where the maximum is evaluated over all smooth and solenoidal vector fields $v:\Y \rightarrow \R^3$ which satisfy the point-wise constraint
\begin{equation}\label{eq:theory_cell_formulae_ptw_constraints}
	\|v(x)\| \le \gamma(x) \quad \text{for (almost) all} \quad x \in \Y.
\end{equation}
The dual problem \eqref{eq:theory_cell_formulae_max_flow} maximizes the total flow in direction $\bar{\xi}$ through the microstructure under the point-wise constraints \eqref{eq:theory_cell_formulae_ptw_constraints}. The advantage of the dual formulation \eqref{eq:theory_cell_formulae_max_flow} over the primal formulation \eqref{eq:theory_cell_formulae_cell_formula} is that it represents a smooth (in fact linear) optimization problem with linear and quadratic constraints, for which powerful solution methods are available~\cite{Boyd2004}. However, some caution is advised, as the primal \eqref{eq:theory_cell_formulae_cell_formula} and the dual problem \eqref{eq:theory_cell_formulae_max_flow} are strongly dual in the continuous setting only for a \emph{continuous} \crackres{} $\gamma$~\cite{Strang1983}. As soon as the \crackres{} $\gamma$ is discontinuous, explicit counterexamples~\cite{Nozawa1994} to strong duality are known, i.e., the maximum computed in the dual problem \eqref{eq:theory_cell_formulae_max_flow} is strictly less than the minimum computed for the primal problem \eqref{eq:theory_cell_formulae_cell_formula}.\\
For practical considerations, this delicacy does not play much of a role. Indeed, in finite dimensions, convex optimization problems with convex constraints \emph{always} satisfy strong duality provided Slater's condition is satisfied~\cite[Sec. 5.2]{Boyd2004}. Slater's condition states that there is a strictly feasible point, i.e., a point where all inequality constraints are satisfied as strict inequalities. Due to our prerequisite $\gamma \ge \gamma_- > 0$, the field $v \equiv 0$ is strictly feasible for the dual problem \eqref{eq:theory_cell_formulae_max_flow}, and strong duality holds upon discretization. In particular, we may exploit the maximum flow formulation \eqref{eq:theory_cell_formulae_max_flow}, as long as it arises by formal Lagrangian dualization~\cite[Ch. 5]{Boyd2004} of a discretization of the cell problem \eqref{eq:theory_cell_formulae_cell_formula}.

\subsection{The combinatorial continuous maximum flow discretization}
\label{sec:theory_CCMF}

\begin{figure}
	\begin{center}
		\includegraphics[width = 0.4\textwidth]{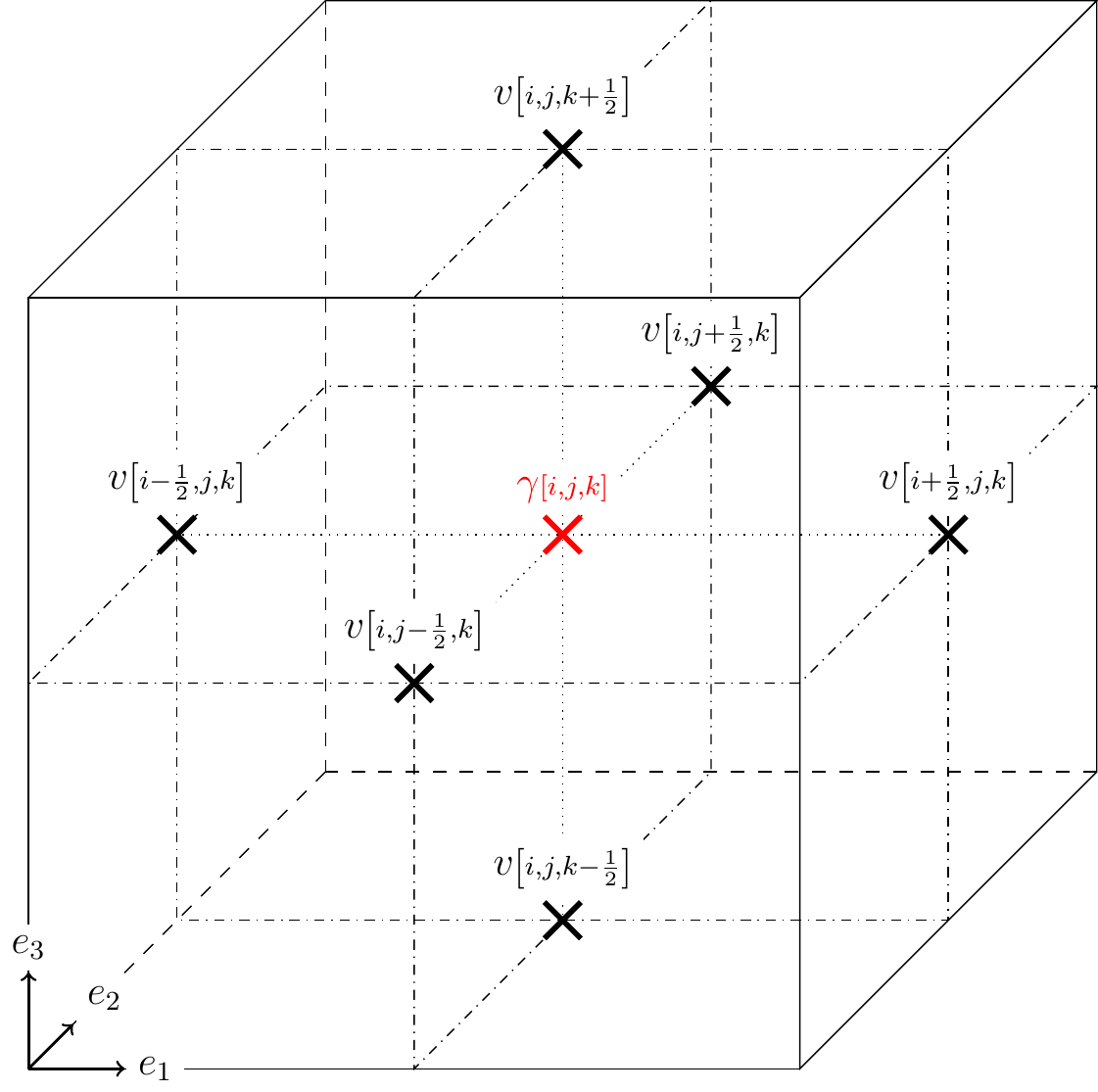}
		\caption{Consistent placement of the flow-field variables on a generic voxel cell}
		\label{fig:discr}
	\end{center}
\end{figure}

In this section, we discuss the combinatorial continuous maximum flow discretization (\CCMF{})~\cite{CCMF} for the special case of regular grids and in the periodic setting. The discretization scheme naturally approximates the maximum flow formulation \eqref{eq:theory_cell_formulae_max_flow}, and we take it as our point of departure.\\
For this purpose, suppose that the unit cell $\Y=[0,L_1] \times [0,L_2] \times [0,L_3]$ is discretized by a regular grid with $N_i$ ($i=1,2,3$) voxels for each coordinate direction. Each voxel is assumed to be cubic with edge length $h$, i.e., the conditions $h=L_i/N_i$ ($i=1,2,3$) are assumed to hold. In a finite volume discretization, where each individual voxel serves as a control volume, the flow between adjacent cells is quantified by a flow field $v$, which is located at the voxel faces{, see Fig.~\ref{fig:discr}}. The conservation of mass is encoded by the balance of in- and outflow
\begin{equation}\label{eq:theory_CCMF_mass_conservation}
	0 = v{\scriptstyle\left[i + \frac{1}{2},j,k\right]} - v{\scriptstyle\left[i - \frac{1}{2},j,k\right]}
	+ v{\scriptstyle\left[i,j + \frac{1}{2},k\right]} - v{\scriptstyle\left[i,j - \frac{1}{2},k\right]}
	+ v{\scriptstyle\left[i,j,k + \frac{1}{2}\right]} - v{\scriptstyle\left[i,j,k - \frac{1}{2}\right]},
\end{equation}
where we tacitly assume the integer indices $i,j,k$ to satisfy
\[
	0\le i < N_1, \quad 0 \le j < N_2 \quad \text{and} \quad 0\le k < N_3,
\]
and the equation \eqref{eq:theory_CCMF_mass_conservation} should be interpreted in a periodic fashion. Let us denote by
\[
	\gamma{\scriptstyle[i,j,k]} = \gamma\left( (i+\tfrac{1}{2})h, (j+\tfrac{1}{2})h, (k+\tfrac{1}{2})h \right)
\]
the evaluations of the \crackres{} $\gamma$ at the voxel centers, which we sample on a discrete grid $\Y_N$. Then, for the \CCMF{}-discretization, the constraint \eqref{eq:theory_cell_formulae_ptw_constraints} is approximated by the $N_1N_2N_3$ constraints
\begin{equation}\label{eq:theory_CCMF_original_bound}
	v{\scriptstyle\left[i + \frac{1}{2},j,k\right]}^2 + v{\scriptstyle\left[i - \frac{1}{2},j,k\right]}^2
	+ v{\scriptstyle\left[i,j + \frac{1}{2},k\right]}^2 + v{\scriptstyle\left[i,j - \frac{1}{2},k\right]}^2
	+ v{\scriptstyle\left[i,j,k + \frac{1}{2}\right]}^2 + v{\scriptstyle\left[i,j,k - \frac{1}{2}\right]}^2 \le 2\,\gamma{\scriptstyle[i,j,k]}^2.
\end{equation}
The factor two on the right hand side appears due to the doubling of terms appearing on the left hand side of the constraint \eqref{eq:theory_CCMF_original_bound} compared to the continuous version \eqref{eq:theory_cell_formulae_ptw_constraints}.\\
Then, for prescribed average crack normal $\bar{\xi}\in S^2$, the \CCMF{} discretization approximates the maximum flow problem \eqref{eq:theory_cell_formulae_max_flow} by the maximization problem
\begin{equation}\label{eq:theory_CCMF_max_flow}
	\frac{1}{N_1N_2N_3}
	\sum_{i,j,k} \bar{\xi}_x \, v{\scriptstyle\left[i + \frac{1}{2},j,k\right]}
	+ \bar{\xi}_y \, v{\scriptstyle\left[i,j + \frac{1}{2},k\right]}
	+ \bar{\xi}_z \, v{\scriptstyle\left[i,j,k + \frac{1}{2}\right]}
	 \longrightarrow \max_{\text{$v$ satisfying \eqref{eq:theory_CCMF_mass_conservation} and \eqref{eq:theory_CCMF_original_bound}}}.
\end{equation}
With FFT-based solution methods, to be discussed in section \eqref{sec:numerics}, in mind, we transform the natural finite volume formulation into a more compact representation that is simpler to manipulate algebraically. For this purpose, we regard the flow field $v$ as a vector field located at the voxel centers, with the identification
\[
	\begin{split}
		v_x {\scriptstyle[i,j,k]} &= v{\scriptstyle\left[i + \frac{1}{2},j,k\right]},\\
		v_y {\scriptstyle[i,j,k]} &= v{\scriptstyle\left[i,j + \frac{1}{2},k\right]},\\
		v_z {\scriptstyle[i,j,k]} &= v{\scriptstyle\left[i,j,k + \frac{1}{2}\right]}.\\
	\end{split}
\]
We also introduce a (backwards) divergence-type operator $\divm{}$ via
\[
	\left(\divm v\right){\scriptstyle[i,j,k]} = v_x {\scriptstyle[i,j,k]} - v_x {\scriptstyle[i-1,j,k]}
	+ v_y {\scriptstyle[i,j,k]} - v_y {\scriptstyle[i,j-1,k]}
	+ v_z {\scriptstyle[i,j,k]} - v_z {\scriptstyle[i,j,k-1]}.
\]
Then, the mass conservation \eqref{eq:theory_CCMF_mass_conservation} is satisfied precisely if $\divm v = 0$ holds. To encode the constraint \eqref{eq:theory_CCMF_original_bound}, we introduce the backwards shift operator $S$, which operates as follows
\begin{equation}\label{eq:theory_CCMF_shift_operator}
	S(v){\scriptstyle[i,j,k]} = \left[
		\begin{array}{c}
			v_x{\scriptstyle[i-1,j,k]}\\
			v_y{\scriptstyle[i,j-1,k]}\\
			v_z{\scriptstyle[i,j,k-1]}\\
		\end{array}
	\right].
\end{equation}
Then, the constraint \eqref{eq:theory_CCMF_original_bound} is equivalent to the condition
\begin{equation}\label{eq:theory_CCMF_bound}
	\left\| v{\scriptstyle[i,j,k]} \right\|^2 + \left\| S(v){\scriptstyle[i,j,k]} \right\|^2 \le 2\,\gamma{\scriptstyle[i,j,k]}^2,
\end{equation}
expressed in terms of the Euclidean norm of the involved vectors. Last but not least, let us introduce the inner product on such vector fields
\begin{equation}\label{eq:theory_CCMF_inner_product}
	\langle v, \tilde{v}\rangle = \frac{1}{N_1N_2N_3}\sum_{i,j,k}  \left(
	v_x{\scriptstyle[i,j,k]}  \tilde{v}_x{\scriptstyle[i,j,k]}
	+ v_y{\scriptstyle[i,j,k]}  \tilde{v}_y{\scriptstyle[i,j,k]}
	+ v_z{\scriptstyle[i,j,k]}  \tilde{v}_z{\scriptstyle[i,j,k]} \right).
\end{equation}
With this notation at hand, we may express the maximization problem \eqref{eq:theory_CCMF_max_flow} in the compact form
\begin{equation}\label{eq:theory_CCMF_max_flow}
	\langle \bar{\xi},v\rangle
	 \longrightarrow \max_{\substack{\divm v = 0\\ \|v\|^2 + \|Sv\|^2\le 2\gamma^2}},
\end{equation}
where we regard $\bar{\xi}$ as a constant vector field. In the latter formulation, the similarities (and differences) to the continuous formulation \eqref{eq:theory_cell_formulae_max_flow} become apparent. Indeed, both the objective function and the divergence constraint are discretized in the natural way. The norm constraint, however, is replaced by a "non-local" constraint which involves neighboring values of the flow field, as well. Please note that this is a feature rather than a bug, as the flow-field variables are naturally located on the voxel faces, whereas the \crackres{} is associated to the voxel center. Instead of \emph{interpolating} the flow-field variables, the \CCMF{} discretization averages the squares of the flow fields. Such an approach has its merits, as will become clear in section \ref{sec:computations}.

\section{An FFT-based solver for the \CCMF{} discretization}
\label{sec:numerics}

\subsection{The primal formulation for the \CCMF{} discretization}
\label{sec:numerics_primal}

On a voxel grid, we consider the maximum flow problem \eqref{eq:theory_CCMF_max_flow}
\begin{equation}\label{eq:CCMF_FFT_problem}
	\langle \bar{\xi},v\rangle \longrightarrow \max_{\substack{\divm v = 0\\ \|v\|^2 + \|Sv\|^2\le 2\gamma^2}}
\end{equation}
in the combinatorial continuous maximum flow (\CCMF{}) discretization. With FFT-based resolution in mind, we compute the corresponding Lagrangian dual, i.e., the associated minimum cut problem.\\
For later reference please notice that the adjoint of the backward shift operator $S$ \eqref{eq:theory_CCMF_shift_operator} w.r.t. the inner product \eqref{eq:theory_CCMF_inner_product}
is given by the (periodized) \emph{forward} shift operator
\begin{equation}\label{eq:CCMF_FFT_SOperator_adjoint}
		S^*(v){\scriptstyle[i,j,k]} = \left[
		\begin{array}{c}
			v_x{\scriptstyle[i+1,j,k]}\\
			v_y{\scriptstyle[i,j+1,k]}\\
			v_z{\scriptstyle[i,j,k+1]}\\
		\end{array}
	\right].
\end{equation}
In particular, as backward and forward shifting are mutual inverses, the equation $S^* S = \Id$ holds.\\
The shift operator is non-local, which makes the inequality constraint in the maximum flow problem \eqref{eq:CCMF_FFT_problem} non-local, as well. With computational resolution in mind, we seek a local formulation that relies upon a doubling of dimension. For this purpose, we introduce the linear extension operator $A$, acting on vector fields $v$ via
\begin{equation}\label{eq:CCMF_FFT_AOperator}
	(Av) = \frac{1}{\sqrt{2}} \, \left[ 
		\begin{array}{c}
			v\\
			Sv
		\end{array}	
	\right],
\end{equation}
and producing a vector field with six scalar components per voxel. Then, the problem \eqref{eq:CCMF_FFT_problem} may be expressed in the equivalent form
\begin{equation}\label{eq:CCMF_FFT_problemA}
	\langle \bar{\xi},v\rangle \longrightarrow \max_{\substack{\divm v = 0\\ \|Av\|\le \gamma}},
\end{equation}
where the factor two in front of the \crackres{} \eqref{eq:CCMF_FFT_problem} was transferred into the $A$-operator \eqref{eq:CCMF_FFT_AOperator} and the norm in the constraint refers to the Euclidean norm of vectors with six components. For later reference, let us remark that the adjoint of the operator $A$ \eqref{eq:CCMF_FFT_AOperator} w.r.t. the six-component version of the inner product \eqref{eq:theory_CCMF_inner_product} is given by
\begin{equation}\label{eq:CCMF_FFT_AOperator_adjoint}
	A^* \left[ 
		\begin{array}{c}
			w_1\\
			w_2
		\end{array}	
	\right] = \frac{1}{\sqrt{2}}( w_1 + S^*w_2)
\end{equation}
in terms of the backward shift operator \eqref{eq:CCMF_FFT_SOperator_adjoint}. In particular, it holds
\[
	A^*A v = \frac{1}{2}( v + S^*Sv) = v,
\]
i.e., $A^*A = \Id$ and $\|A\| = 1$ in operator norm. Thus, the operator $A$ is an isometric embedding, and the operator $A^*$ is a left inverse to the operator $A$. In turn, the operator $AA^*$ is the orthogonal projector onto the image of the operator $A$.\\
With the necessary terminology at hand, we turn our attention to deriving the Lagrangian dual of the maximum flow problem \eqref{eq:CCMF_FFT_problem} in the constrained form 
\begin{equation}\label{eq:CCMF_FFT_problemA_constrained}
	\langle \bar{\xi},v\rangle - \iota_{\{ \divm v = 0 \}}(v) - \iota_{\mathcal{C}_\gamma}(w) \longrightarrow \max_{ w + Av = 0},
\end{equation}
where the indicator function $\iota_S$ of a non-empty, closed and convex set $S$ takes the value $\iota_S(u) = 0$ for $u \in S$ and $+\infty$ otherwise, and we denote by $\mathcal{C}_\gamma$ the set
\begin{equation}\label{eq:CCMF_FFT_Cgamma}
	\mathcal{C}_\gamma = \left\{ w: \Y_N \rightarrow \R^6 \, \middle| \, \|w{\scriptstyle[i,j,k]}\| \le \gamma{\scriptstyle[i,j,k]} \quad \text{for all} \quad i,j,k\right\}.
\end{equation}
The associated Lagrangian function reads
\begin{equation}\label{eq:CCMF_FFT_problemA_Lagrangian}
	L(v,w, \xi) = \langle \bar{\xi},v\rangle - \iota_{\{ \divm v = 0 \}}(v) - \iota_{\mathcal{C}_\gamma}(w) - \langle \xi,Av + w\rangle
\end{equation}
in terms of the Lagrangian multiplier field $\xi:\Y_N \rightarrow \R^6$. To evaluate the dual function
\[
	\varphi(\xi) = \sup_{v,w} L(v,w,\xi),
\]
we rearrange the expression of the Lagrangian \eqref{eq:CCMF_FFT_problemA_Lagrangian}
\[
	\begin{split}
	\varphi(\xi) &= \sup_{v} \langle \bar{\xi},v\rangle - \iota_{\{ \divm v = 0 \}}(v) - \underbrace{\langle \xi, Av\rangle}_{=\langle A^* \xi, v\rangle} + \sup_{w} \langle \xi, w\rangle - \iota_{\mathcal{C}_\gamma}(w)\\
	&= \left\{ \begin{array}{rl}
		\frac{1}{N_1N_2N_3}
	\sum_{i,j,k} \gamma{\scriptstyle[i,j,k]}\,\|\xi{\scriptstyle[i,j,k]}\|, & \xi \in \mathcal{K}_{\bar{\xi}},\\
	+ \infty, & \text{otherwise},
	\end{array}
	\right.
	\end{split}
\]
in terms the set of compatible normal fields
\[
	\mathcal{K}_{\bar{\xi}} = \left\{ 
		\xi:\Y_N \rightarrow \R^6
	\,\middle|\,	
	\text{there is some} \quad \phi:\Y_N \rightarrow \R\text{, s.t.}\quad A^*\xi = \bar{\xi} + \nabla^+ \phi
	\right\}.
\]
The final form
\begin{equation}\label{eq:CCMF_FFT_problem_primal_final}
	\frac{1}{N_1N_2N_3}
	\sum_{i,j,k} \gamma{\scriptstyle[i,j,k]}\,\|\xi{\scriptstyle[i,j,k]}\| \longrightarrow \min_{\xi \in \mathcal{K}_{\bar{\xi}}}
\end{equation}
of the dual to the \CCMF{} problem \eqref{eq:CCMF_FFT_problem} is remarkably close to the original minimum cut formulation \eqref{eq:theory_cell_formulae_cell_formula}, cf. the more involved formulas in Couprie et al.~\cite[Sec. 2.3]{CCMF}.

\subsection{An FFT-based ADMM solver}
\label{sec:numerics_polarization}

To proceed, we rewrite the optimization problem \eqref{eq:CCMF_FFT_problem_primal_final} as an equivalent convex program that is amenable to operator-splitting approaches
\begin{equation}\label{eq:CCMF_FFT_cvx_problem}
	f(\xi) + g(\xi) \longrightarrow \min_{\xi}
\end{equation}
in terms of the convex functions
\[
	f(\xi) = \iota_{\mathcal{K}_{\bar{\xi}}}(\xi) \quad \text{and} \quad g(\xi) = \frac{1}{N_1N_2N_3}	\sum_{i,j,k} \gamma{\scriptstyle[i,j,k]}\,\|\xi{\scriptstyle[i,j,k]}\|.
\]
The starting point of operator-splitting approaches is the rewriting of the unconstrained problem \eqref{eq:CCMF_FFT_cvx_problem} in constrained form
\begin{equation}\label{eq:CCMF_FFT_cvx_problem2}
	f(\xi) + g(e) \longrightarrow \min_{\xi = e}.
\end{equation}
For solving the problem \eqref{eq:CCMF_FFT_cvx_problem2}, we utilize the alternating direction method of multipliers (ADMM)~\cite{ADMM1,ADMM2}, which was pioneered in the context of FFT-based methods by Michel et al.~\cite{MichelMoulinecSuquet2000,MichelMoulinecSuquet2001}, and applied to non-smooth optimization by Willot~\cite{Willot2020StronglyNonlinearMedia}. For this purpose, we investigate the augmented Lagrangian function
\begin{equation}\label{eq:CCMF_FFT_augmented_Lagrangian}
	L_\rho(\xi, e, v) = f(\xi) + g(e) + \langle v, \xi - e  \rangle + \frac{\rho}{2} \left\| \xi - e \right\|^2,
\end{equation}
involving a penalization factor $\rho>0$ and the Lagrange multiplier $v:\Y_N \rightarrow \R^6$. The ADMM is based on the three-term recursion
\begin{equation}\label{eq:CCMF_FFT_ADMM_abstract}
	\begin{split}
		\xi^{k+1} &= \argmin_{\xi} L_\rho(\xi,e^k,v^k),\\
		e^{k+1} &= \argmin_{e} L_\rho(\xi^{k+1},e,v^k),\\
		v^{k+1} &= v^k + \rho\,(\xi^{k+1} - e^{k+1}).
	\end{split}
\end{equation}
Let us investigate the first line more explicitly,
\[
	\begin{split}
		\xi^{k+1} & = \argmin_{\xi} L_\rho(\xi,e^k,v^k)\\
		& = \argmin_{\xi} f(\xi) + \langle v^k, \xi  \rangle + \frac{\rho}{2} \left\| \xi - e^k \right\|^2\\
		& = \argmin_{\xi \in \mathcal{K}_{\bar{\xi}}} \left\| \xi - e^k + \frac{1}{\rho}\, v^k\right\|^2.
	\end{split}
\]
Thus, $\xi^{k+1}$ arises as the orthogonal projection of the point $e^k - v^k / \rho$ onto the set $\mathcal{K}_{\bar{\xi}}$,
\[
	\xi^{k+1} = \mathcal{P}_{\mathcal{K}_{\bar{\xi}}}\left( e^k - \frac{1}{\rho} \, v^k \right).
\]
Let us write down an explicit expression for the projection operator $\mathcal{P}_{\mathcal{K}_{\bar{\xi}}}$. For given $w:\Y_N \rightarrow \R^6$, we seek $\xi:\Y_N \rightarrow \R^6$, s.t.
\begin{equation}\label{eq:CCMF_FFT_Gamma_0}
	\xi = \mathcal{P}_{\mathcal{K}_{\bar{\xi}}}(w), \quad \text{i.e.,} \quad \xi = \argmin_{\xi \in \mathcal{K}_{\bar{\xi}}} \|\xi - w\|^2 \quad \text{holds}.
\end{equation}
Both, the vectors $\xi$ and $w$ may be decomposed in terms of the orthogonal projector $AA^*$,
\[
	\xi = A A^* \xi + (\Id - A A^*) \xi \quad \text{and} \quad w = A A^* w + (\Id - A A^*) w.
\]
Due to the constraint $\xi \in \mathcal{K}_{\bar{\eps}}$, we may parameterize the vector $\xi$ in the form
\[
	\xi = A(\bar{\xi} + \nabla^+ \phi) + (\Id - A A^*) \xi
\]
for some scalar field $\phi:\Y_N \rightarrow \R$. Splitting the objective function in the optimization problem \eqref{eq:CCMF_FFT_Gamma_0} accordingly
\[
	\begin{split}
		\|\xi - w\|^2 & = \left\| A A^* \xi - A A^* w \right\|^2 + \left\| (\Id - A A^*) \xi - (\Id - A A^* w) \right\|^2\\
		& = \left\| A(\bar{\xi} + \nabla^+ \phi) - A A^* w \right\|^2 + \left\| (\Id - A A^*) \xi - (\Id - A A^* w) \right\|^2\\
		& = \left\| \bar{\xi} + \nabla^+ \phi - A^* w \right\|^2 + \left\| (\Id - A A^*) \xi - (\Id - A A^* w) \right\|^2,
	\end{split}
\]
where we used that the operator $A$ is an isometric embedding, we find that the identity 
\begin{equation}\label{eq:CCMF_FFT_Gamma_1}
	\xi = A(\bar{\xi} + \nabla^+ \phi) + (\Id - A A^*) w
\end{equation}
holds, where $\phi:\Y_N \rightarrow \R$ solves
\[
	\left\| \bar{\xi} + \nabla^+ \phi - A^* w \right\|^2 \rightarrow \min,
\]
i.e.,
\[
	\divm\left[\bar{\xi} + \nabla^+ \phi - A^* w\right] = 0 \quad \iff \quad \divm \nabla^+ \phi = \divm A^*w.
\]
The latter equation may be solved formally to give
\[
	\phi = (\divm \nabla^+)^\dagger \divm A^*w,
\]
i.e., reinserting into the earlier expression \eqref{eq:CCMF_FFT_Gamma_1}, we find
\[
	\xi = A(\bar{\xi} + \nabla^+ (\divm \nabla^+)^\dagger \divm A^* ) + (\Id - A A^*) w,
\]
which we may also write in the more convenient form
\begin{equation}\label{eq:CCMF_FFT_Gamma_2}
	\mathcal{P}_{\mathcal{K}_{\bar{\xi}}}(w) = A\bar{\xi} + \left(\Id - A A^* + A\Gamma A^*\right)w \quad \text{with} \quad \Gamma = \nabla^+ (\divm \nabla^+)^\dagger \divm.
\end{equation}
The second line \eqref{eq:CCMF_FFT_ADMM_abstract} can be rewritten using Moreau's identity~\cite[Eq. (3.8)]{Chambolle2016Overview} in the form
\[
	e^{k+1} = \left[ v^k + \rho\, \xi^{k+1} - \mathcal{P}_{\mathcal{C}_\gamma}\left( v^k + \rho\, \xi^{k+1} \right)\right] / \rho,
\]
where $\mathcal{P}_{\mathcal{C}_\gamma}$ is the orthogonal projector
\[
	\left(\mathcal{P}_{\mathcal{C}_\gamma}(w)\right){\scriptstyle[i,j,k]} =
	\left\{
		\begin{array}{rl}
			{\gamma{\scriptstyle[i,j,k]}} \, w{\scriptstyle[i,j,k]} / \|w{\scriptstyle[i,j,k]}\|, & \|w{\scriptstyle[i,j,k]}\| > \gamma{\scriptstyle[i,j,k]},\\
			w{\scriptstyle[i,j,k]}, & \text{otherwise},
		\end{array}
	\right.
\]
onto the constraint set $\mathcal{C}_\gamma$ \eqref{eq:CCMF_FFT_Cgamma}.
Thus, we are led to the following scheme
\begin{equation}\label{eq:CCMF_FFT_ADMM}
	\begin{split}
		\xi^{k+1} &= A\bar{\xi} - \frac{1}{\rho}\,\left(\Id - A A^* + A\Gamma A^*\right) \left( v^k - \rho \, e^k\right),\\
		e^{k+1} &= \left[ v^k + \rho\, \xi^{k+1} - \mathcal{P}_{\mathcal{C}_\gamma}\left( v^k + \rho\, \xi^{k+1} \right)\right] / \rho,\\
		v^{k+1} &= v^k + \rho\,(\xi^{k+1} - e^{k+1}).
	\end{split}
\end{equation}
A recent study~\cite{ADMM2021} highlighted the importance of utilizing a damping factor and choosing the penalty factor $\rho$ adaptively. For this purpose, we consider the modified scheme
\begin{equation}\label{eq:CCMF_FFT_ADMM}
	\begin{split}
		\xi^{k+1/2} &= A\bar{\xi} - \frac{1}{\rho^k}\,\left(\Id - A A^* + A\Gamma A^*\right) \left( v^k - \rho^k \, e^k\right),\\
		\xi^{k+1} &= 2(1-\damping)\xi^{k+\frac{1}{2}} - (1-2\damping) e^k,\\
		e^{k+1} &= \left[ v^k + \rho^k\, \xi^{k+1} - \mathcal{P}_{\mathcal{C}_\gamma}\left( v^k + \rho^k\, \xi^{k+1} \right)\right] / \rho^k,\\
		v^{k+1} &= v^k + \rho^k\,(\xi^{k+1} - e^{k+1}).
	\end{split}
\end{equation}
with damping $\damping \in (0,1)$ and adaptive penalty parameter $\rho^k$. In general, the over-relaxation $\damping = 1/4$ is recommended~\cite{MonchietBonnet2012,MoulinecSilva,ADMM2021}. Simple choices for the parameter $\rho^k$ are based on the Lorenz-Tran-Dinh scaling~\cite{Lorenz2019adaptiveDR}
\[
	\rho^k = \frac{\|v^k\|}{\|e^k\|}
\]
or the Barzilai-Borwein scaling~\cite{Xu2017ADMM_BB}
\[
	\rho^k = \frac{ \langle v^k - v^{k-1}, e^k - e^{k-1}\rangle }{\| e^k - e^{k-1}\|^2}
\]
and an additional safeguard~\cite[Sec. 2.5]{ADMM2021}. In our computational experiments, the latter two schemes outperform, both, constant penalty parameter $\rho$ and residual balancing~\cite{He2000adaptive}.\\ Last but not least, let us stress that the operator $\Gamma$ has an explicit form in Fourier space, see Willot et al.~\cite[Eq. 18]{WillotAbdallahPellegrini2014}

\section{Computational experiments}
\label{sec:computations}

\begin{figure}[t]
 	\begin{center}
 		\begin{subfigure}{0.06\textwidth}
 			\includegraphics[width = \textwidth]{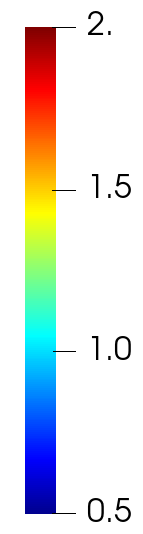}
 			\vspace{0.7cm}
 			\includegraphics[width = \textwidth]{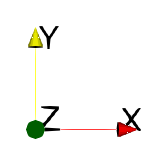} 
 		\end{subfigure}
 		\begin{subfigure}{0.9\textwidth}
			\includegraphics[width=0.3\textwidth, trim = 800 300 800 300, clip]{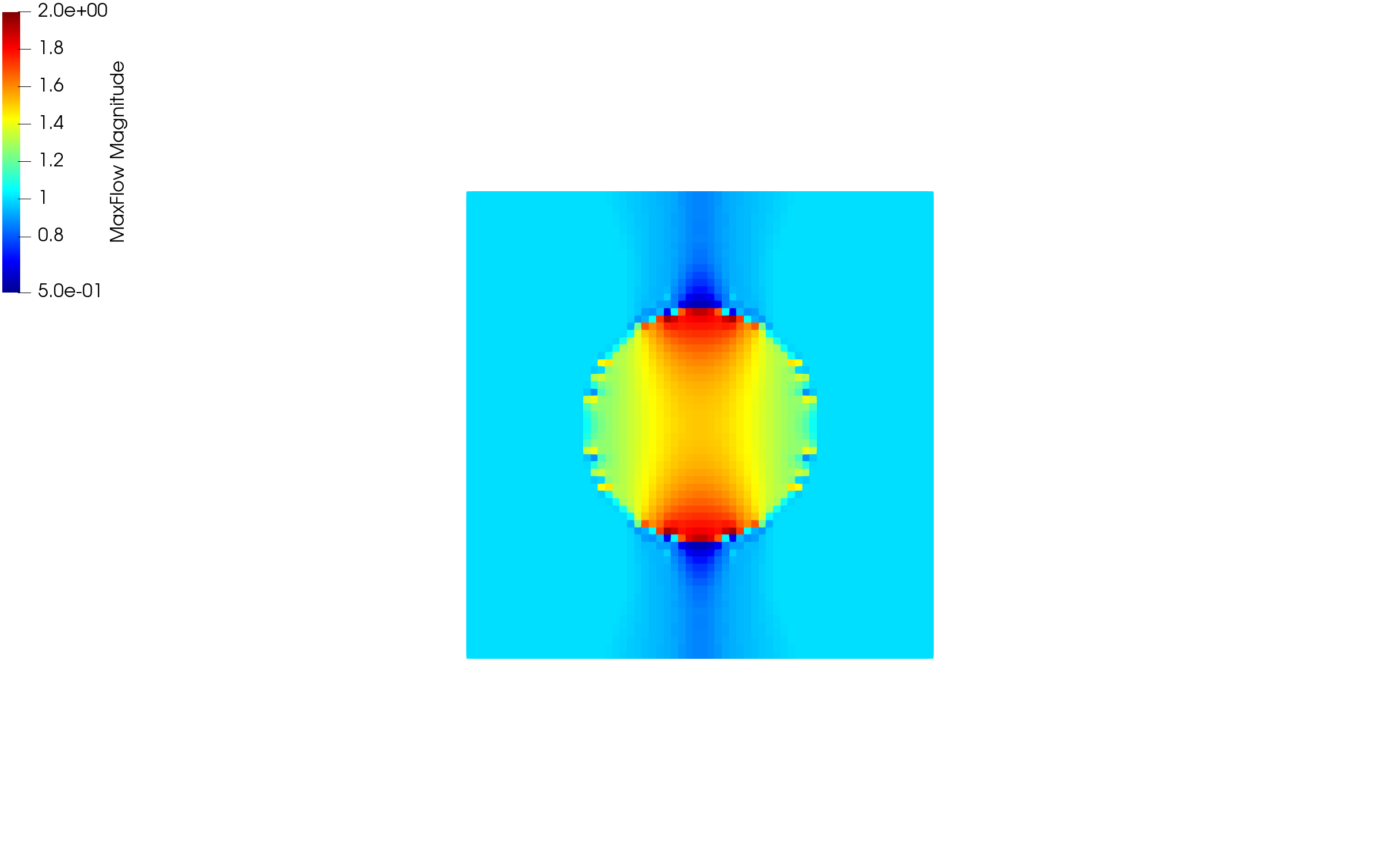}
			\includegraphics[width=0.3\textwidth, trim = 800 300 800 300, clip]{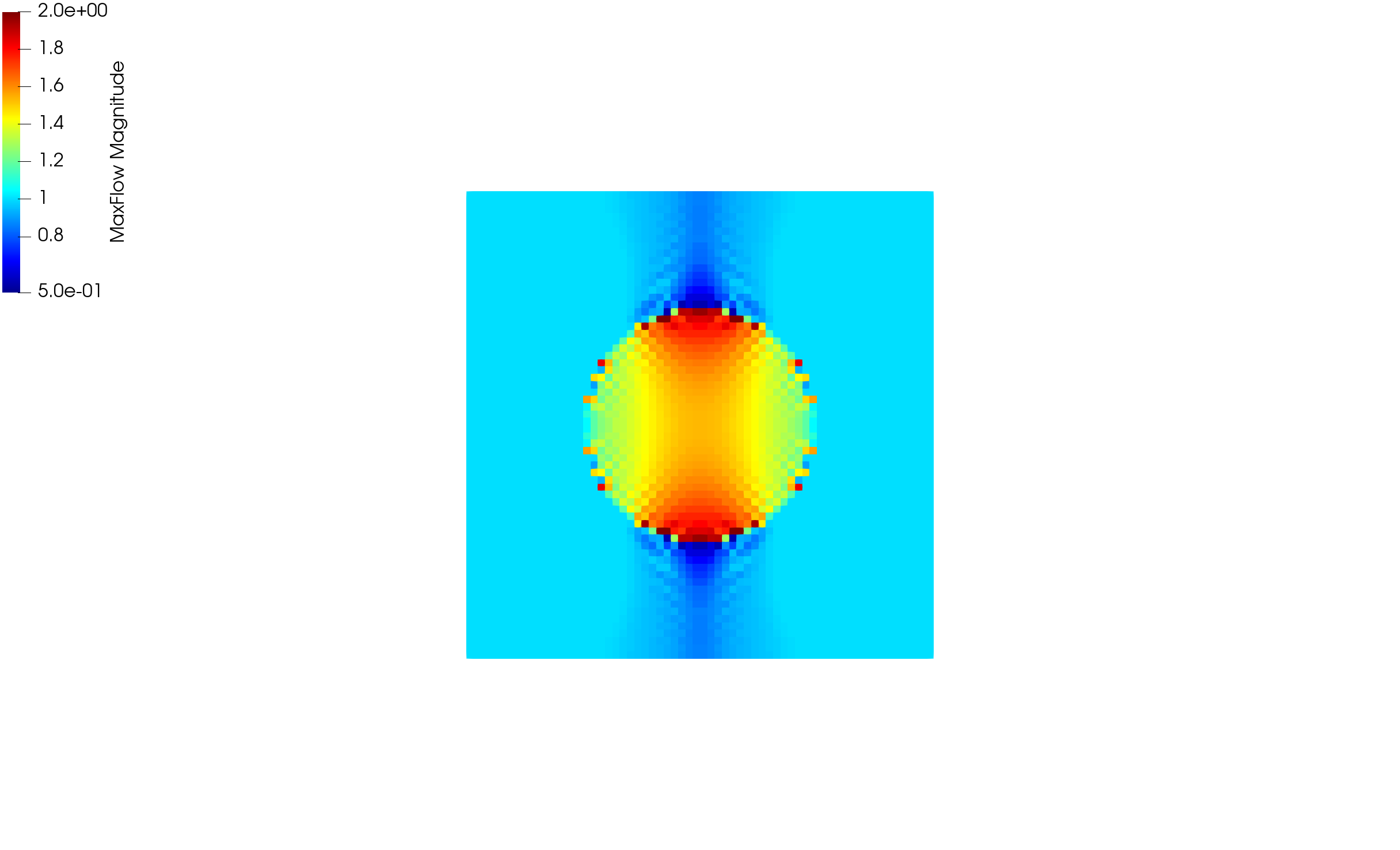}
			\includegraphics[width=0.3\textwidth, trim = 800 300 800 300, clip]{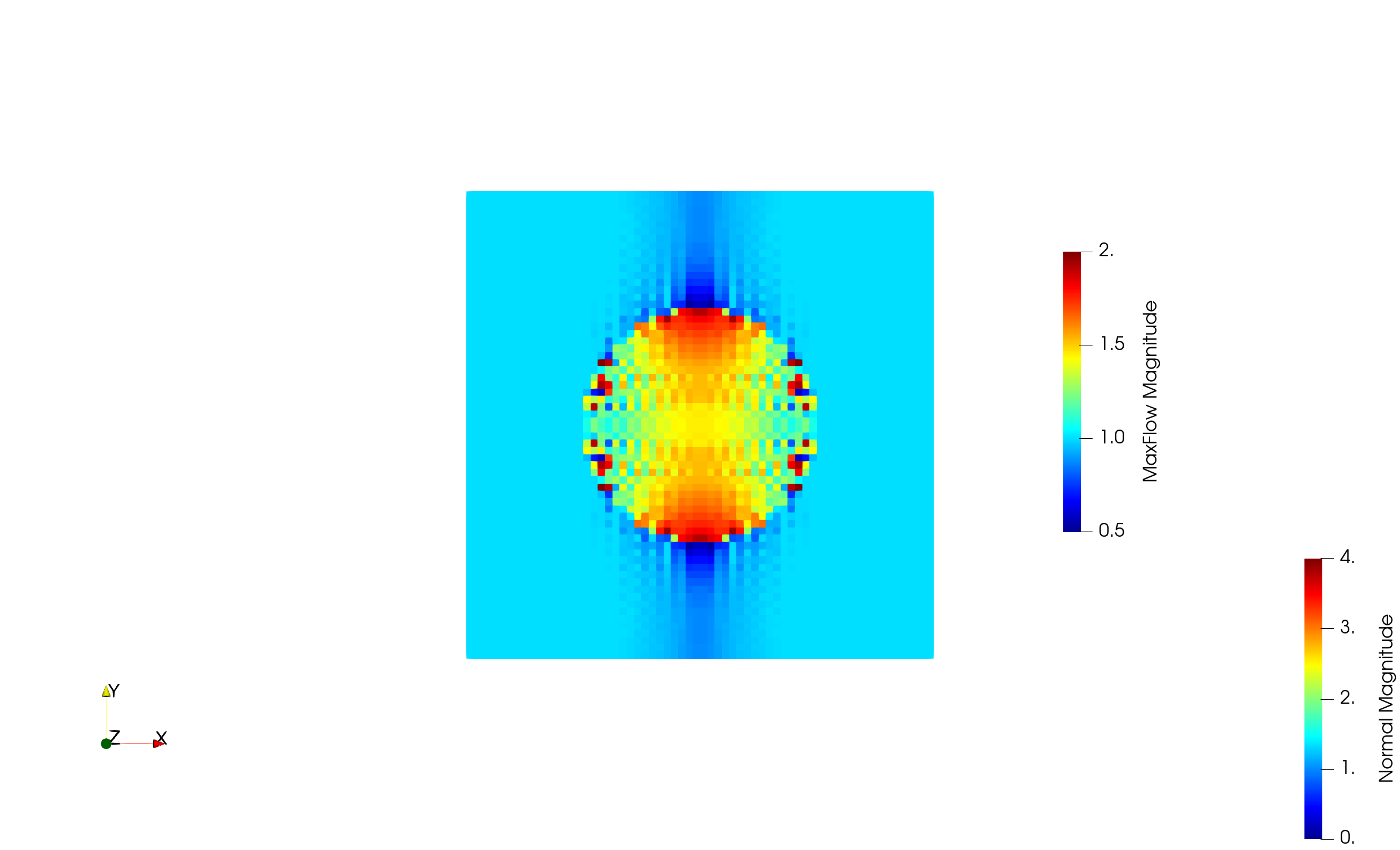}
			\caption{$\|v\|$, \CCMF{} (left), rotated staggered grid (center) and Moulinec-Suquet discretization (right)}
			\label{fig:sphereFlow}
		\end{subfigure}
  		\begin{subfigure}{0.06\textwidth}
 			\includegraphics[width = \textwidth]{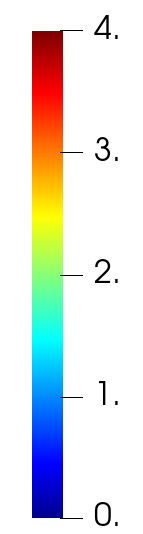}
 			\vspace{0.7cm}
 			\includegraphics[width = \textwidth]{Figures/Figure05_v_KOS} 
 		\end{subfigure}
 		\begin{subfigure}{0.9\textwidth}
			\includegraphics[width=0.3\textwidth,trim = 800 300 800 300, clip]{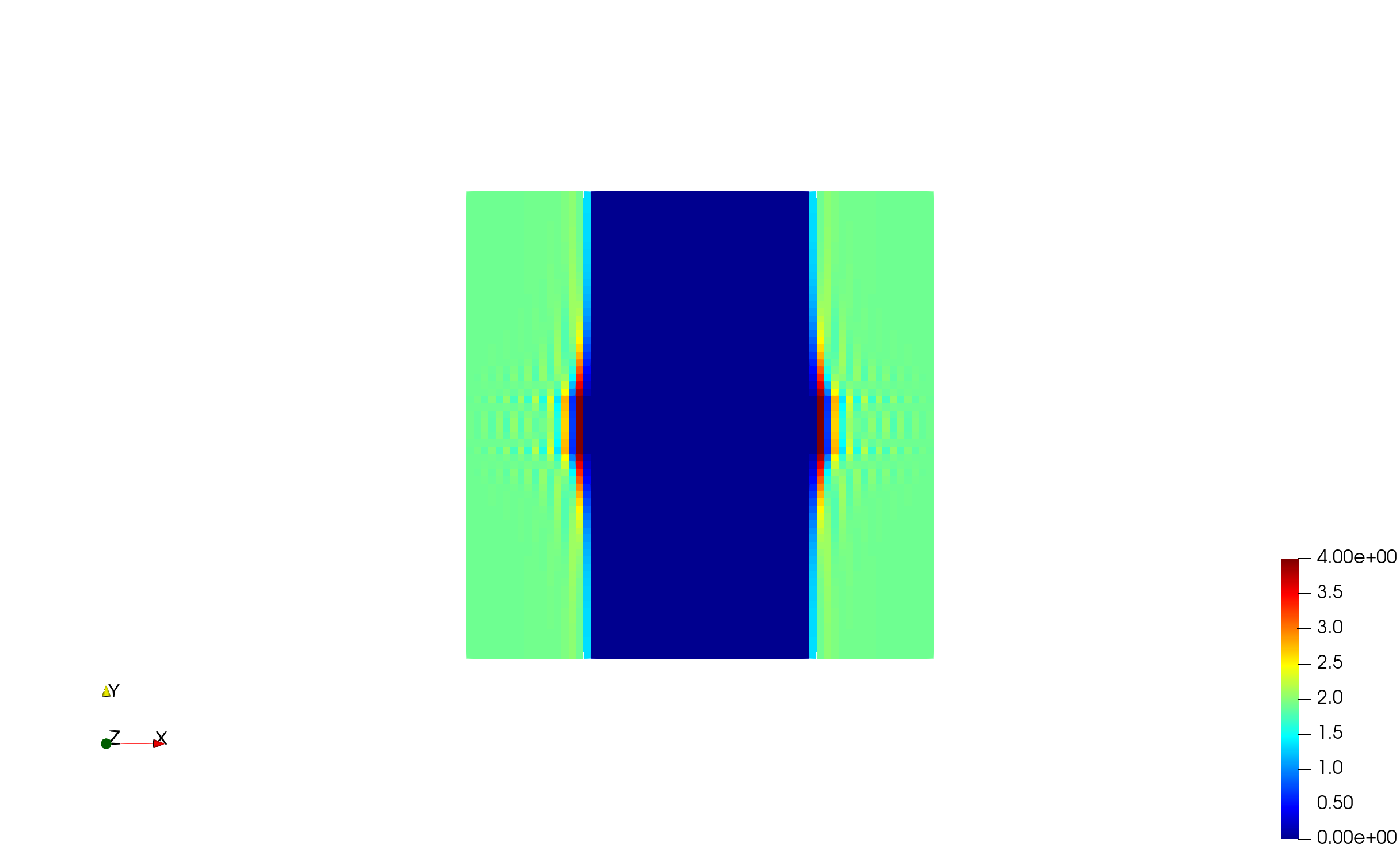}
			\includegraphics[width=0.3\textwidth,trim = 800 300 800 300, clip]{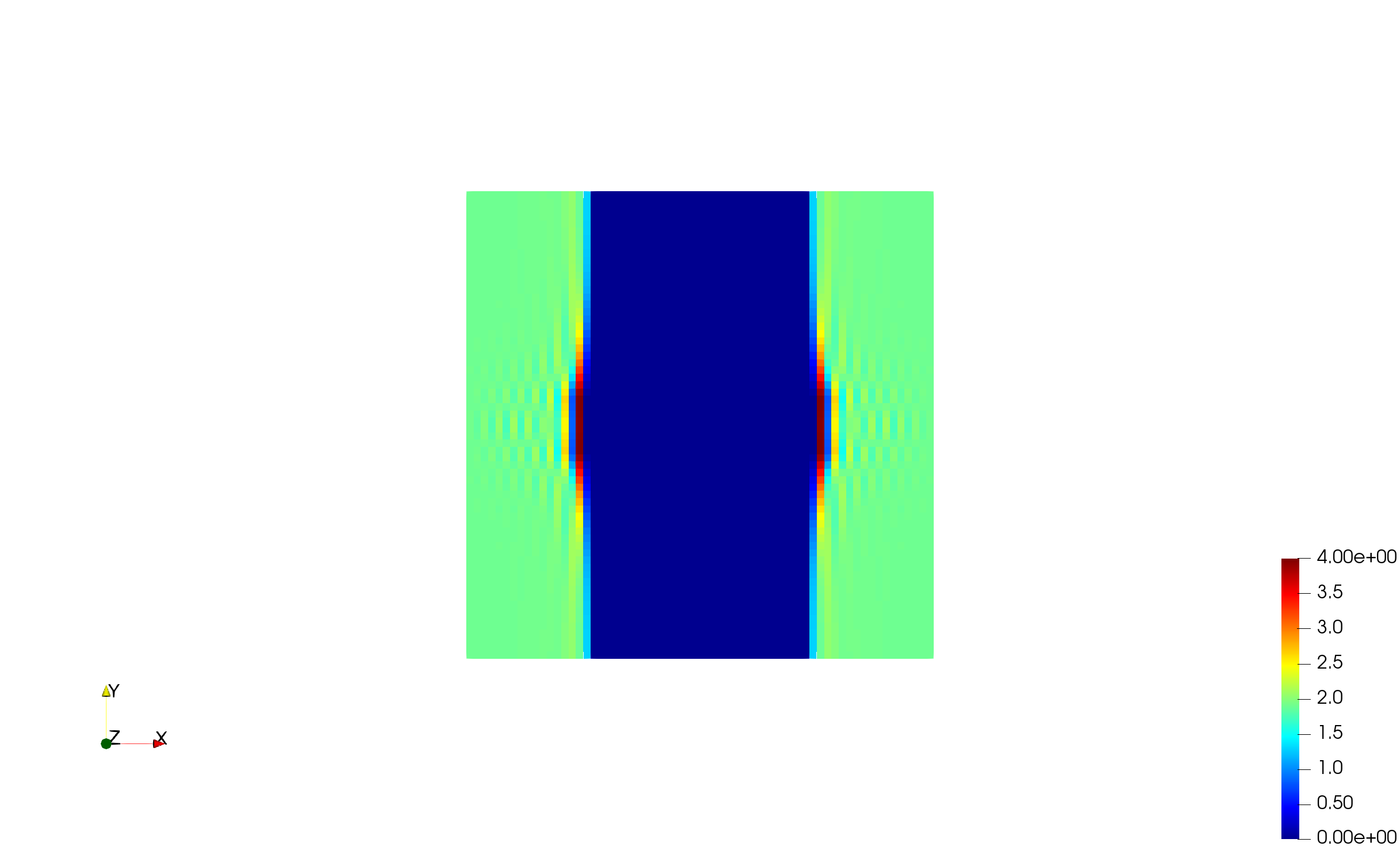}
			\includegraphics[width=0.3\textwidth,trim = 800 300 800 300, clip]{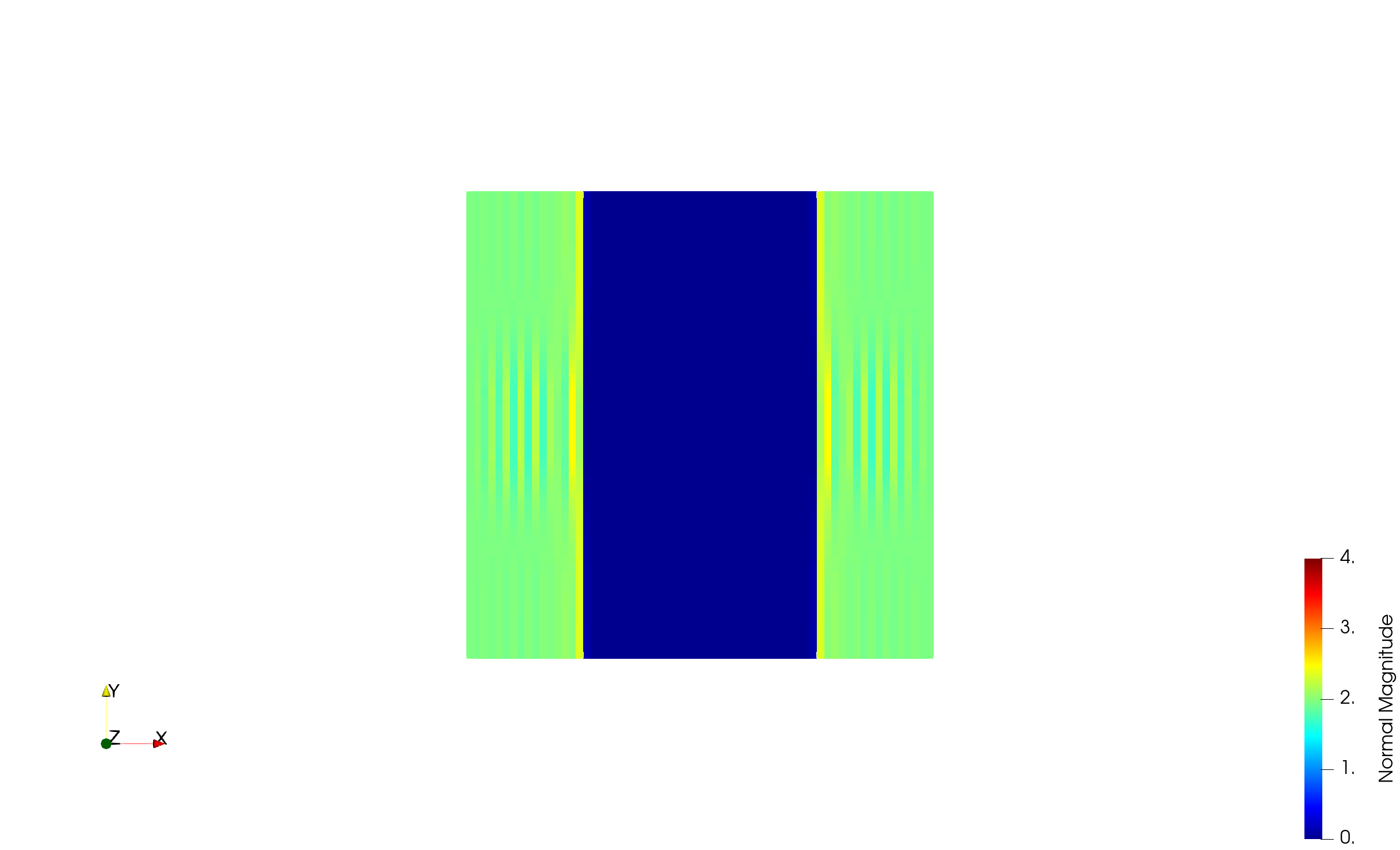}
			\caption{$\|\xi\|$, \CCMF{} (left), rotated staggered grid (center) and Moulinec-Suquet discretization (right)}
			\label{fig:sphereCut}
		\end{subfigure}
 	\end{center}
 	\caption{Flow $v$ and normal $\xi$ fields on a cross section through a $64^3$ single-sphere microstructure for $\bar{\xi} = e_x$, $\gamma_\text{ball} = 10\,\gamma_\text{matrix}$ and different discretizations}
 	\label{fig:sphereFlowCut}
\end{figure}

\subsection{Setup}
\label{sec:computations_setup}

The algorithm \eqref{eq:CCMF_FFT_ADMM} was integrated into an existing FFT-based computational homogenization code for thermal conductivity~\cite{DornSchneider2019}, written in Python with Cython extensions (and OpenMP). In the context of small-strain inelasticity, the implementation of the ADMM \eqref{eq:CCMF_FFT_ADMM} and the memory-efficient computation of the penalty factor is discussed in Schneider~\cite{ADMM2021}. In the same paper, the convergence criterion
\begin{equation}\label{eq:convergence_criterion}
	\left\| e^k - \xi^{k+\frac{1}{2}} \right\|_{L^2} \le \textrm{tol} \, \left\| \mean{v} \right\|,
\end{equation}
for prescribed tolerance $\textrm{tol}$, is identified as suitable. All computational experiments were run on a desktop computer with $32$GB RAM and six $3.7$GHz cores, and on a workstation with $512$ GB RAM and two Intel Xeon(R) Gold $6146$ processors ($12\times 3.20$ GHz), respectively. If not mentioned otherwise, we will use ADMM with damping factor $\damping = 0.25$ and the Barzilai-Borwein adaptive choice for the penalty factor. The default tolerance $\textrm{tol}$ \eqref{eq:convergence_criterion} was set to $\textrm{tol}=10^{-4}$.

\subsection{A single spherical inclusion}
\label{sec:computations_single_sphere}

As a first example, we build upon previous numerical experiments~\cite[Sec. 4.2.2]{HomFrac2019} and compare the \CCMF{}-discretization to previously investigated discretization schemes, namely the rotated staggered grid~\cite{RotatedStaggeredGrid1,RotatedStaggeredGrid2,Willot} and the Moulinec-Suquet discretization~\cite{MoulinecSuquet1994,MoulinecSuquet1998}. We consider a $64^3$ box containing a single spherical inclusion with a diameter of $32$ voxels. The \crackres{} of the inclusion is chosen as $\gamma_\text{sphere}=10\,\gamma_\text{matrix}$. We prescribe a unit vector $\bar{\xi}=e_x$ in $x$-direction as the crack normal. We solved the problem up to a tolerance of $10^{-4}$ using ADMM and chose the penalty factor as lower bound $\rho = \min\{\gamma_\text{sphere},\gamma_\text{matrix}\}$, which was the preferred choice for the primal-dual hybrid gradient method~\cite[Sec. 3]{HomFrac2019}. Solution fields on a central cross section are shown in Fig.~\ref{fig:sphereFlowCut}.\\
The local flow fields $v$ are shown in Fig.~\ref{fig:sphereFlow}. The Moulinec-Suquets discretization shows significant artifacts, which is characteristic for Fourier spectral discretizations. The rotated staggered grid discretization, on the other hand, features checkerboard artifact, although at a lower degree. In contrast, the flow field corresponding to the \CCMF{}-discretization is much smoother, similar to the explicit jump discretization in the context of thermal conductivity\cite{Wiegmann2006EJ,DornSchneider2019}. The differences in the local crack-normal field $\xi$ for the \CCMF{} and the rotated staggered grid discretization are negligible and differ from the Moulinec-Suquet discretization in the local maximum values close to the central inclusion, see Fig.~\ref{fig:sphereCut}.\\
All discretization methods give rise to the same \effCrack{}, i.e., $\geff = \gamma_\text{matrix}$, as the crack bypasses the inclusion in a plane. This is independent of the material contrast, as long as the crack resistance of the matrix exceeds the crack resistance of the single sphere~\cite[Sec. 4.2.2]{HomFrac2019}.\\
As the Moulinec-Suquet discretization shows the strongest artifacts, we focus on the remaining two discretization methods for the remaining investigations.

\subsection{A continuously fiber-reinforced composite}
\label{sec:computations_UD}

\begin{figure}
 	\begin{center}
 		\begin{subfigure}{0.03\textwidth}
 			\includegraphics[width = \textwidth]{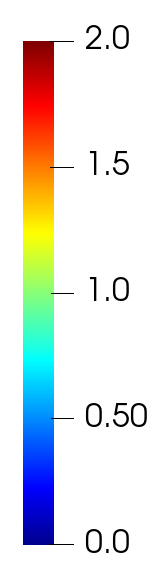}
 			\vspace{0.7cm}
 			\includegraphics[width = \textwidth]{Figures/Figure05_v_KOS}
 		\end{subfigure}
 		\begin{subfigure}{0.45\textwidth}
			\includegraphics[width=.45\textwidth,trim = 650 200 650 200,clip]{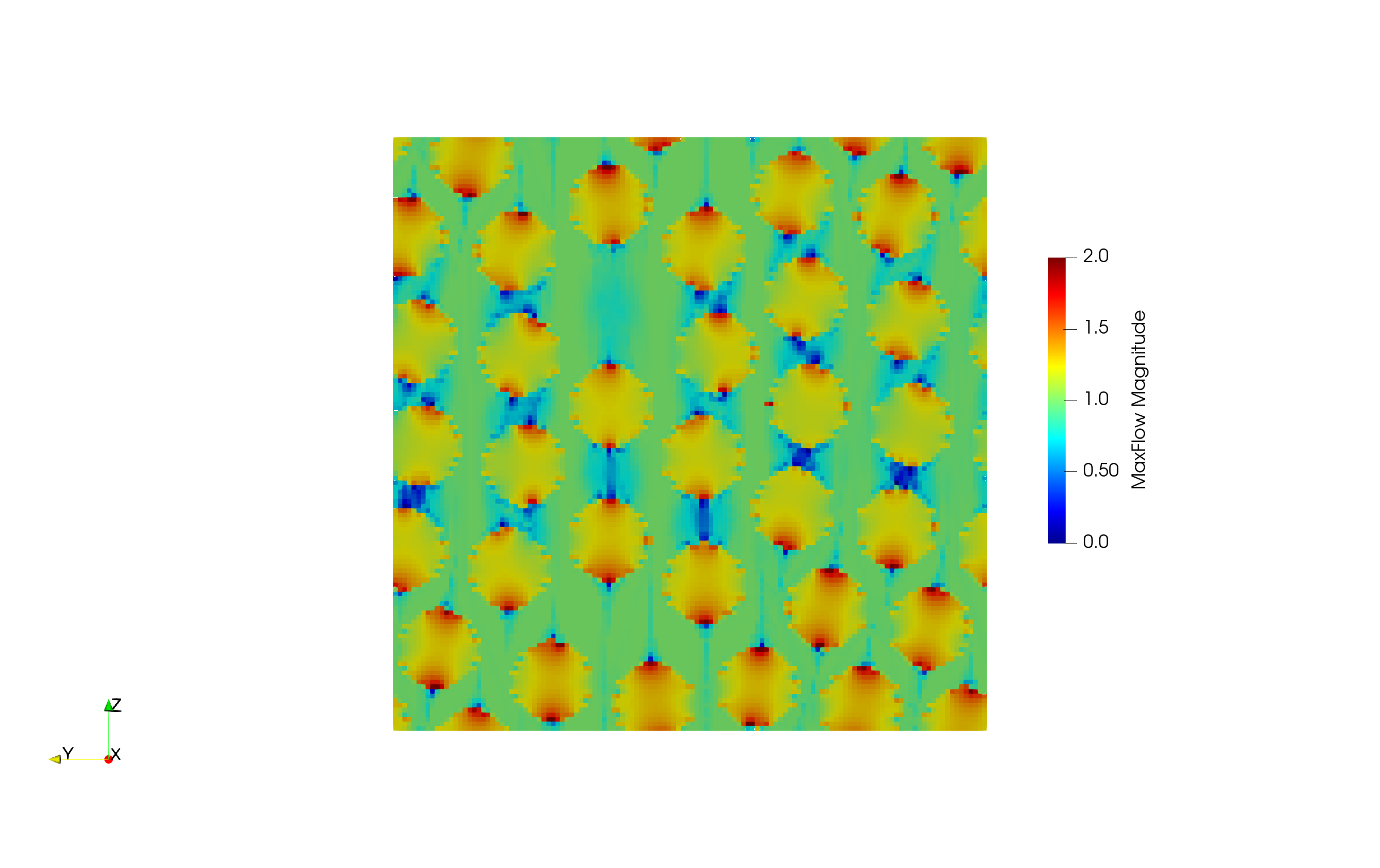}\quad
			\includegraphics[width=.45\textwidth,trim = 650 200 650 200,clip]{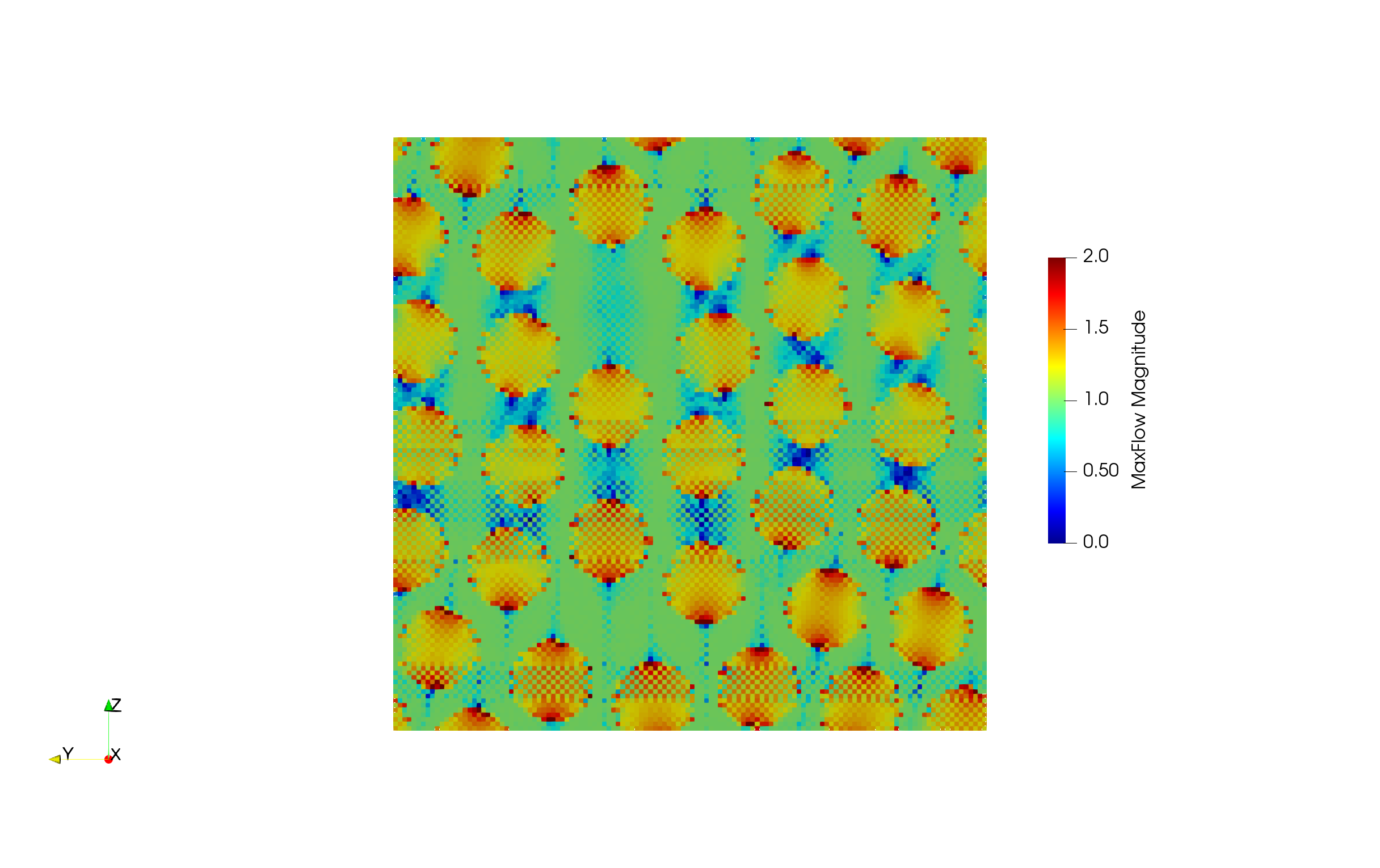}
     		\vspace{-0.1cm}
    		\caption{$\|v\|$, \CCMF{} (left) and rotated staggered grid discretization (right)}\label{fig:UD_Flow}
    	\end{subfigure}
 		\begin{subfigure}{0.03\textwidth}
 			\includegraphics[width = \textwidth]{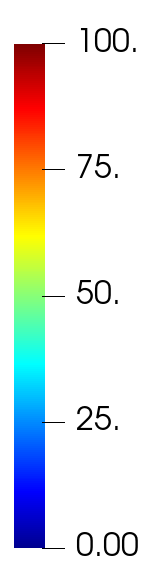}
 			\vspace{0.7cm}
 			\includegraphics[width = \textwidth]{Figures/Figure05_v_KOS}
 		\end{subfigure}
     	\begin{subfigure}{0.45\textwidth}
			\includegraphics[width=.45\textwidth,trim = 650 200 650 200,clip]{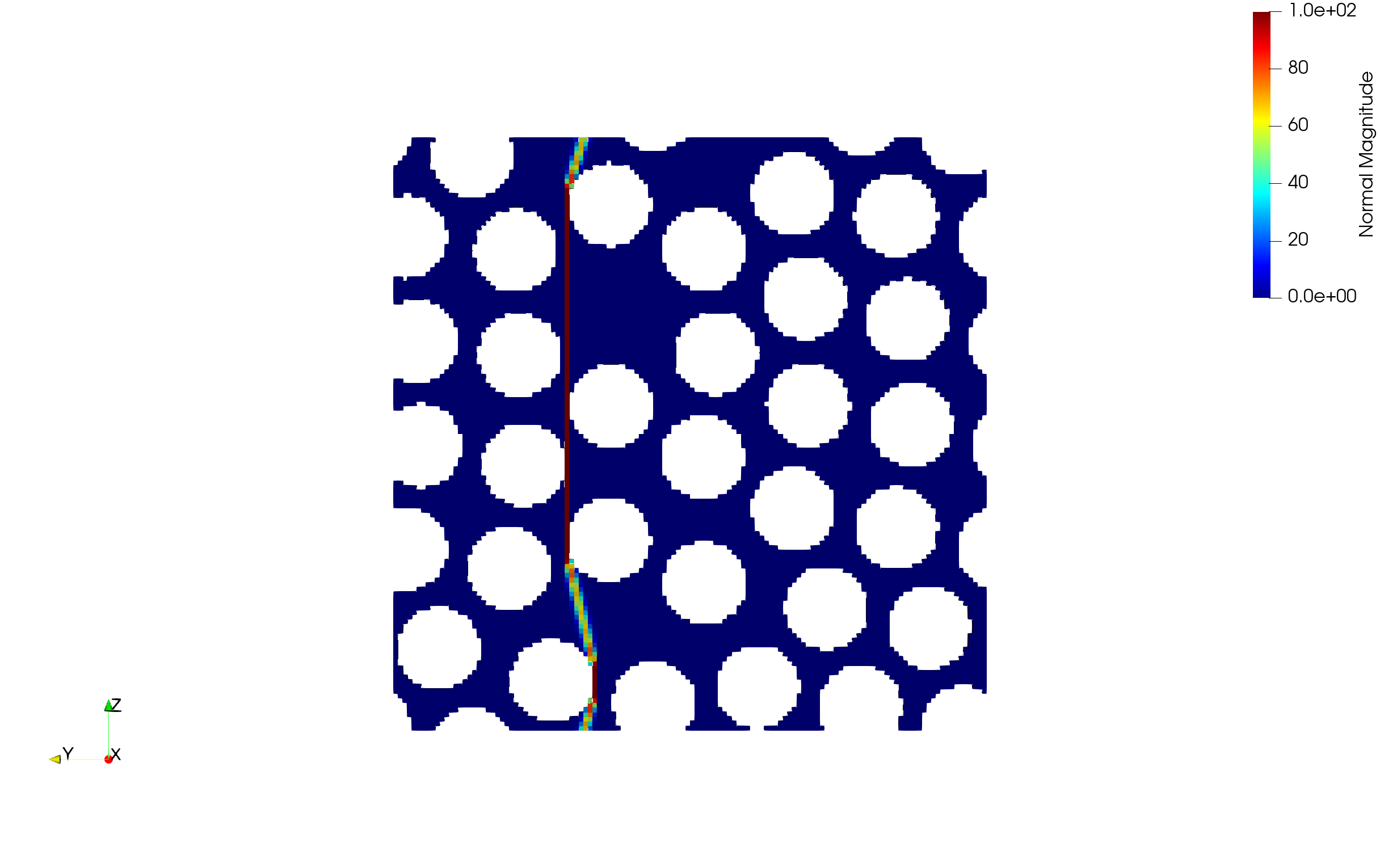}\quad
			\includegraphics[width=.45\textwidth,trim = 650 200 650 200,clip]{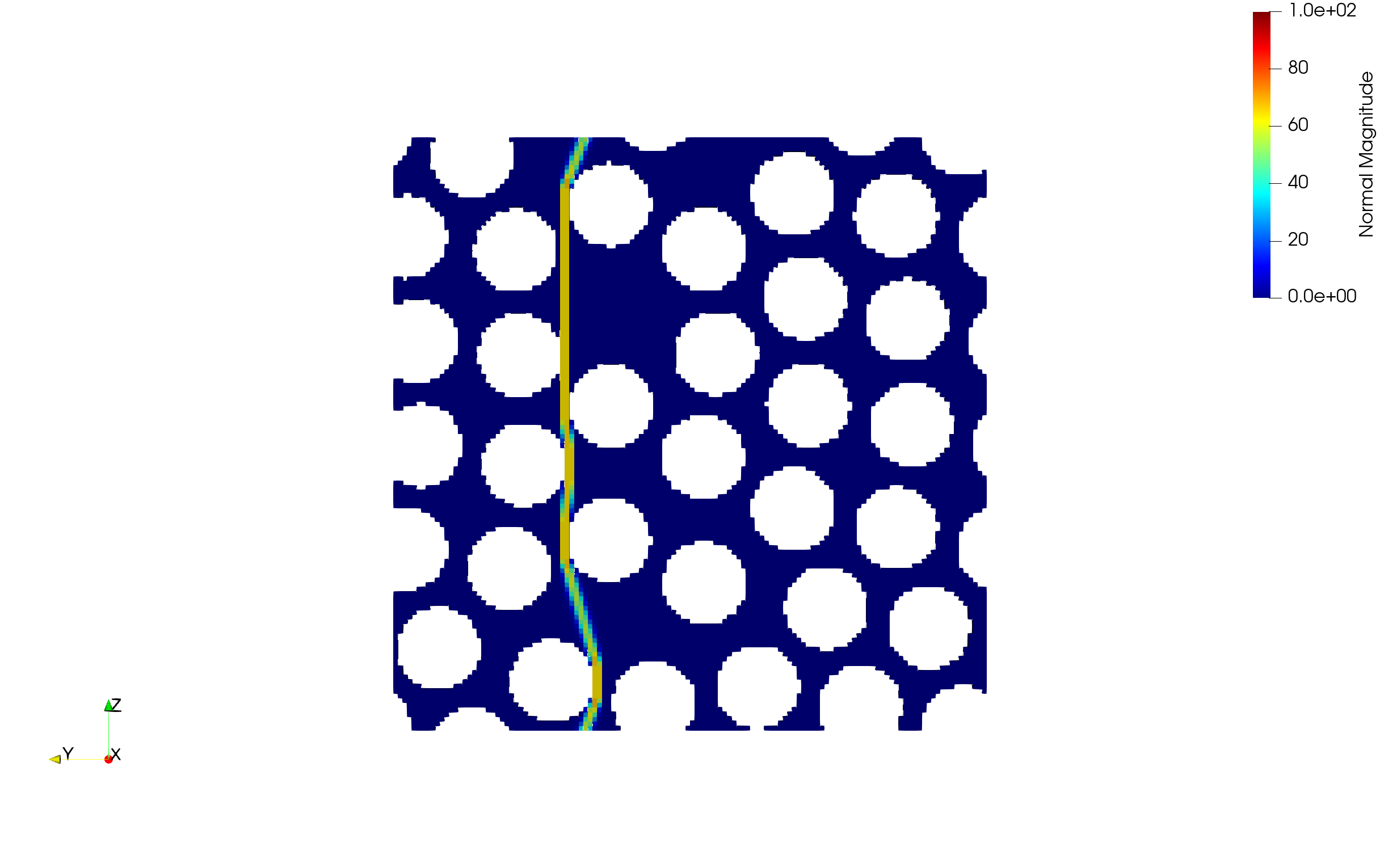}
     		\vspace{-0.1cm}
    		\caption{$\|\xi\|$, \CCMF{} (left) and rotated staggered grid discretization (right)}\label{fig:UD_Cut}
    	\end{subfigure}
 	\end{center}
 	\caption{Cross section through the solution fields $v$ and $\xi$ for a $128^2$ microstructure, containing $32$ circular inclusions for \CCMF{} and rotated staggered grid discretization for $\bar{\xi} = e_x$ and $\gamma_\text{fiber} =10 \gamma_\text{matrix}$ }
 	\label{fig:UD_FlowCut}
\end{figure}

\begin{figure}
 	\begin{center}
 		\begin{subfigure}{\textwidth}
 			\centering
 			\includegraphics[width = 0.2\textwidth]{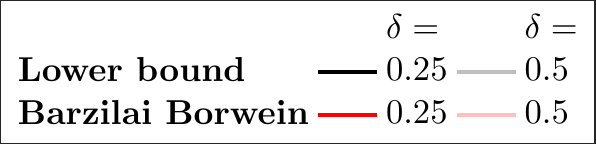}
 		\end{subfigure}
 		\begin{subfigure}{.49\textwidth}
 			\includegraphics[width=.45\textwidth]{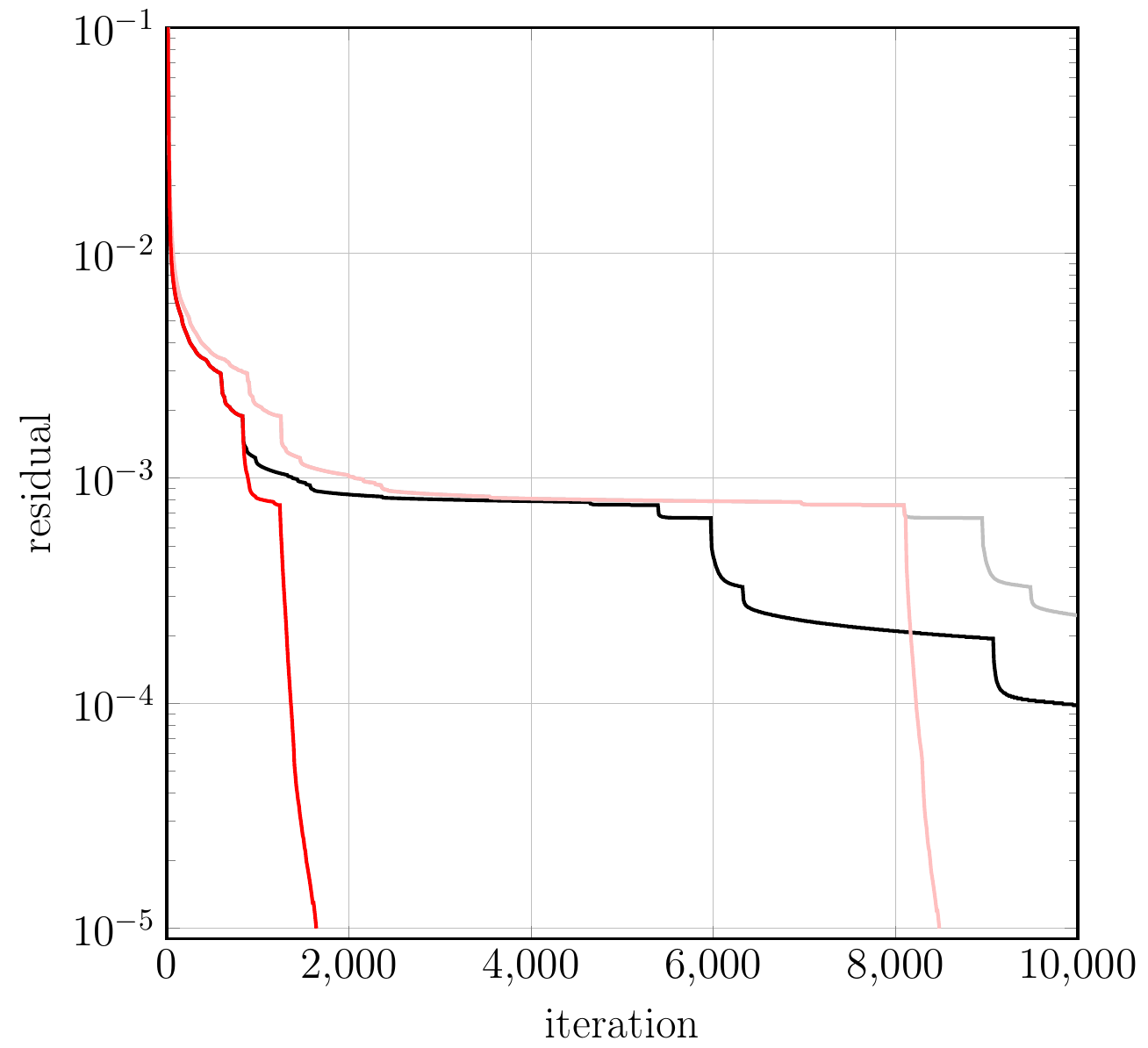} 
 			\includegraphics[width=.45\textwidth]{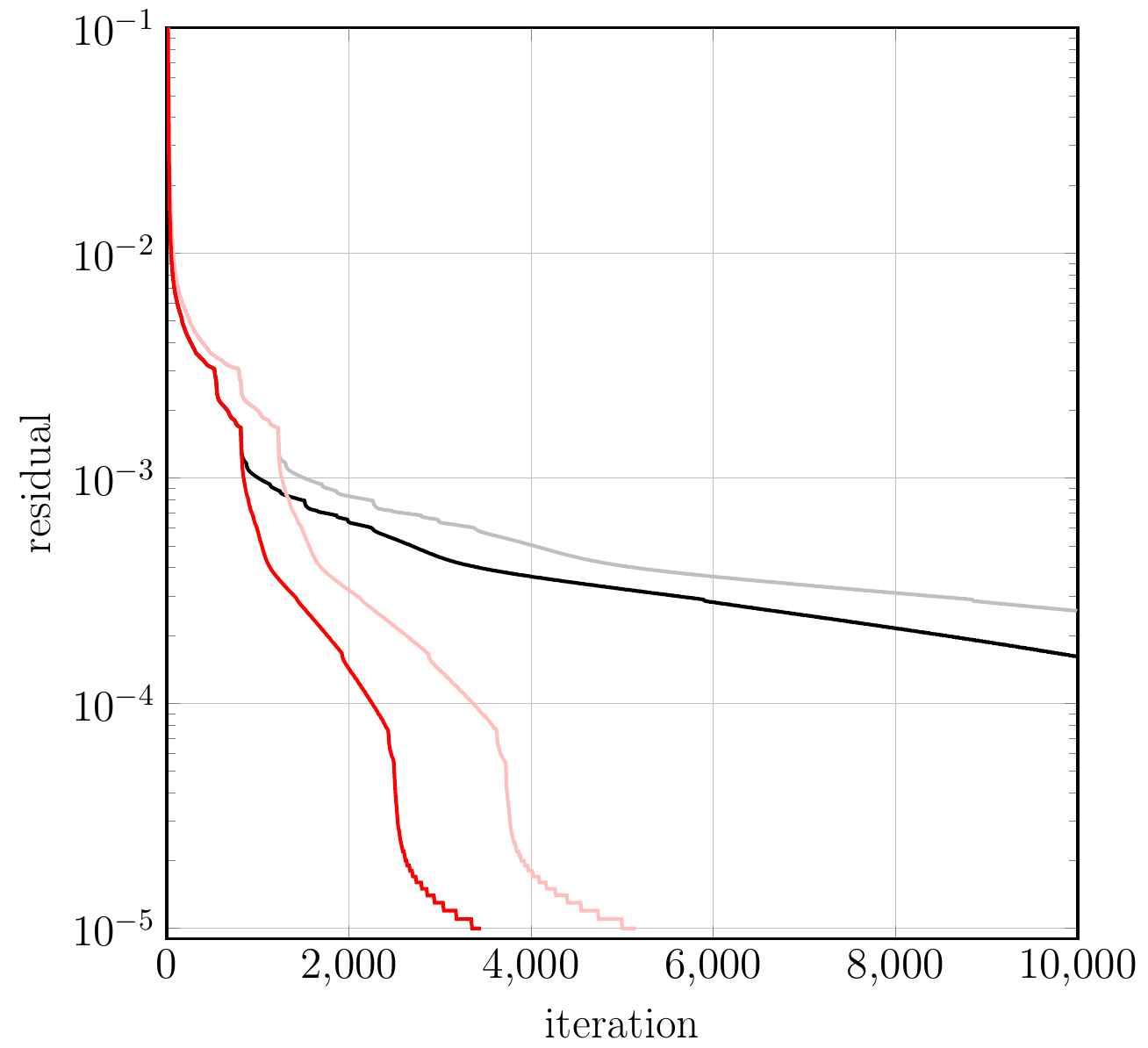}
    		\caption{Residual vs iteration count, \CCMF{} (left) and rotated staggered grid (right)}
    		\label{fig:UD_Res}
    	\end{subfigure}
 		\begin{subfigure}{.49\textwidth}
 			\includegraphics[width=.45\textwidth]{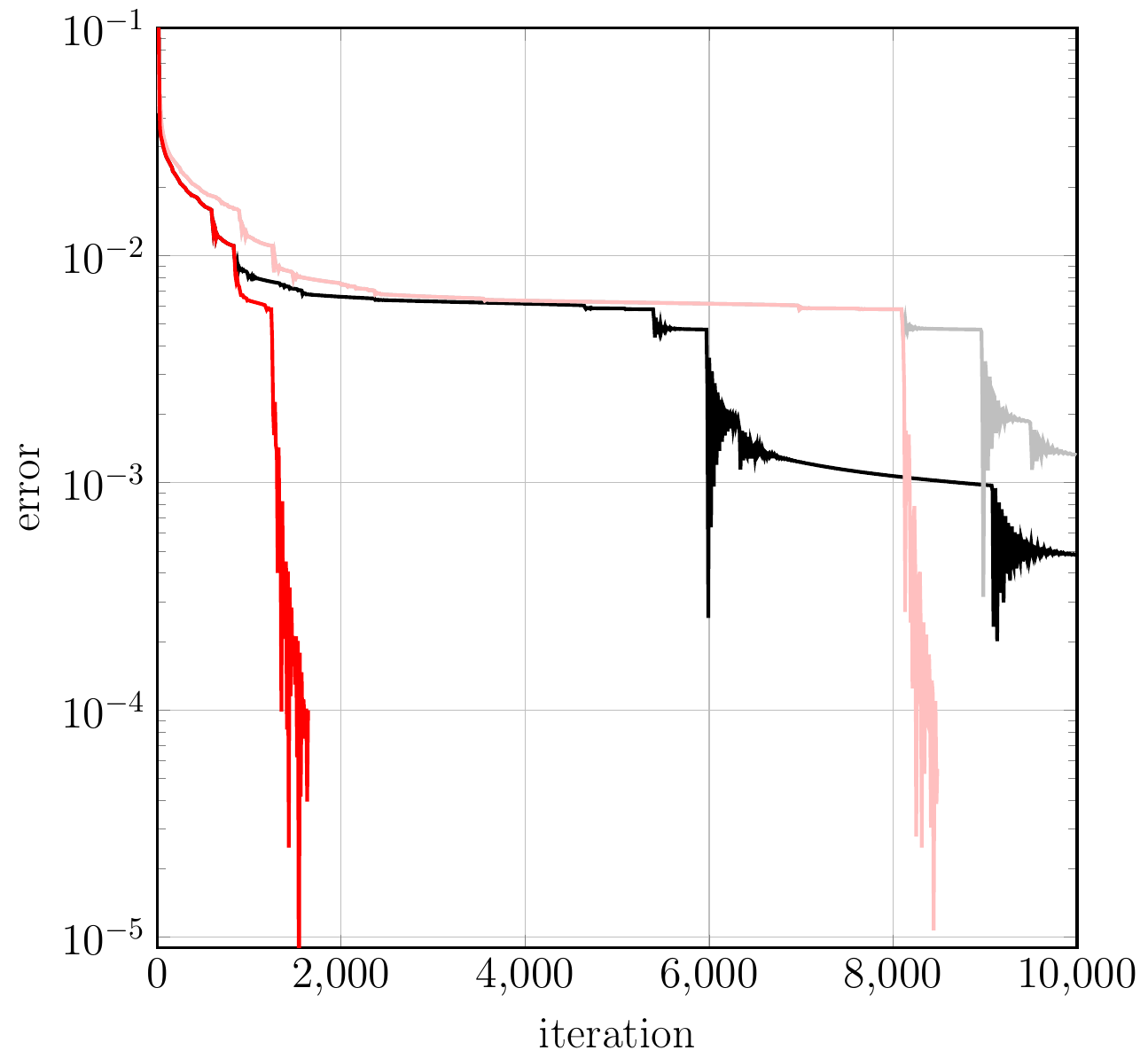}
 			\includegraphics[width=.45\textwidth]{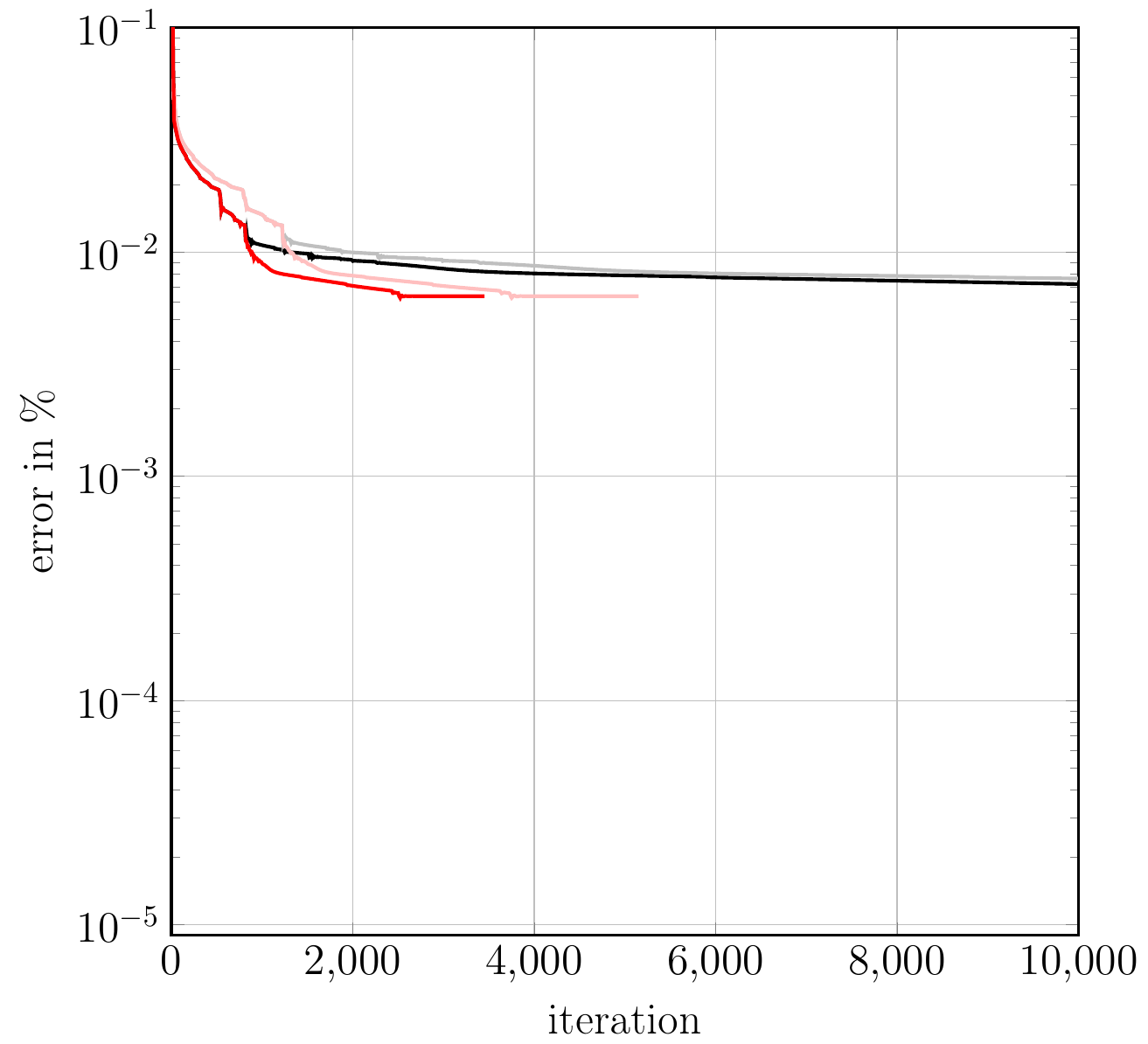}
    		\caption{Error vs iteration count, \CCMF{} (left) and rotated staggered grid (right)}
    		\label{fig:UD_Error}
    	\end{subfigure}
 		\hspace{0.2cm}
 		\caption{Residual and error measure \eqref{eq:UD_relative_error} for \CCMF{} and rotated staggered grid discretizations, comparing different solver parameters}
 		\label{fig:UD_ResError}
 	\end{center}
\end{figure}

In this section, we wish to assess the performance of the ADMM solver introduced in section \ref{sec:numerics}. As a measure of verification, we choose a comparatively simple microstructure which enables us to employ a high-fidelity interior-point solver~\cite{ECOS} for second-order conic programs. The latter produces high-precision solutions, but is limited in terms of problem size.\\
Accounting for this limitation, we consider a continuously fiber-reinforced composite with $50\%$ filler content. The two-dimensional microstructure, containing $32$ circular inclusions, was generated by the mechanical-contraction method~\cite{WilliamsPhilipse} and discretized on a $128^2$ voxel grid. The inclusions were furnished by a \crackres{} of $\gamma_\text{fiber}=10\,\gamma_\text{matrix}$. We investigate the \effCrack{} in direction $\bar{\xi}=e_x$ and compare the CCMF{} discretization and the rotated staggered grid discretization, as well as different ADMM damping parameters $\damping$, namely $\damping = 0.25$ and $\damping=0.5$. Furthermore, we investigate different selection strategies for ADMM penalty-factor, the lower bound $\rho = \min \{\gamma_\text{fiber},\gamma_\text{matrix}\}$, preferred in Schneider~\cite{HomFrac2019} and the Barzilai-Borwein scaling~\cite{Xu2017ADMM_BB}, as well as the Lorenz-Tran Dinh scaling~\cite{Lorenz2019adaptiveDR} and residual balancing~\cite{He2000adaptive}. As announced earlier, we compare the \effCrack{} to solutions obtained by the high-fidelity solver ECOS~\cite{ECOS} applied to conic reformulations of the minimum-cut problem \eqref{eq:CCMF_FFT_problem_primal_final} for the \CCMF{} scheme and the discretization on a rotated staggered grid. We assess the solver quality in terms of the relative error
\begin{equation}\label{eq:UD_relative_error}
	\text{error}  = \frac{|\geff-\geff^\text{accurate}|}{\geff^\text{accurate}}
\end{equation}
in the \effCrack{}.\\
Fig.~\ref{fig:UD_Flow} shows the local flow field for, both, the \CCMF{} discretization and the rotated staggered grid discretization. For the rotated staggered grid, the flow field exhibits significant checkerboard artifacts in the inclusions as well as the matrix. The \CCMF{} solution, on the other hand, is devoid of such artifacts. The corresponding crack paths are shown in Fig.~\ref{fig:UD_Cut}. The cracks bypass the inclusions and look qualitatively similar for both discretizations. However, the rotated staggered grid discretization shows a wider crack path, whereas the \CCMF{} crack path is sharper. This allows the \CCMF{} crack path to avoid several inclusions in a straight line, whereas the rotated staggered grid crack path has to avoid them, resulting in a less straight crack path. This observation is also reflected in the resulting \effCrack{}, i.e. $\geff =  1.021\,\gamma_\text{matrix}$ for the rotated staggered grid and $\geff = 1.014\,\gamma_\text{matrix}$ for the \CCMF{} discretization.\\
Fig.~\ref{fig:UD_Res} shows the residual of the solver vs the iteration count for the two strategies for selecting the penalty factor, 
two damping factors and the two discretizations under consideration. During the first $1000$ iterations, all solvers behave similarly, with a slight advantage for the choice $\damping=0.25$. After $2000$ iterations, all solvers result in a residual below $10^{-3}$. For the \CCMF{} discretization, the ADMM solver with $\damping = 0.25$ and Barzilai-Borwein penalty-choice speeds up at $1600$ iterations and reaches the required tolerance of $10^{-5}$ shortly thereafter. For $\damping = 0.5$ a similar acceleration occurs after slightly more than $8000$ iterations. Selecting the lower bound for the penalty factor $\rho$ does not reach the required tolerance within $10000$ iterations.\\
For the rotated staggered grid discretization, the Barzilai-Borwein penalty-factor outperforms the constant choice, as well. For this discretization, the difference between the two damping factor choices is much smaller than for \CCMF{}.\\
The investigations are supplemented by Fig.~\ref{fig:UD_Error}, which records the associated relative error \eqref{eq:UD_relative_error}. 

\begin{figure}
	\begin{subfigure}{.48\textwidth}
		\includegraphics[width=.15\textwidth]{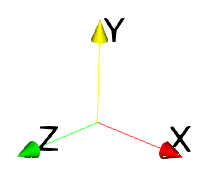}
 		\hspace{-.15\textwidth}
		\includegraphics[width=\textwidth, trim = 500 20 500 100, clip]{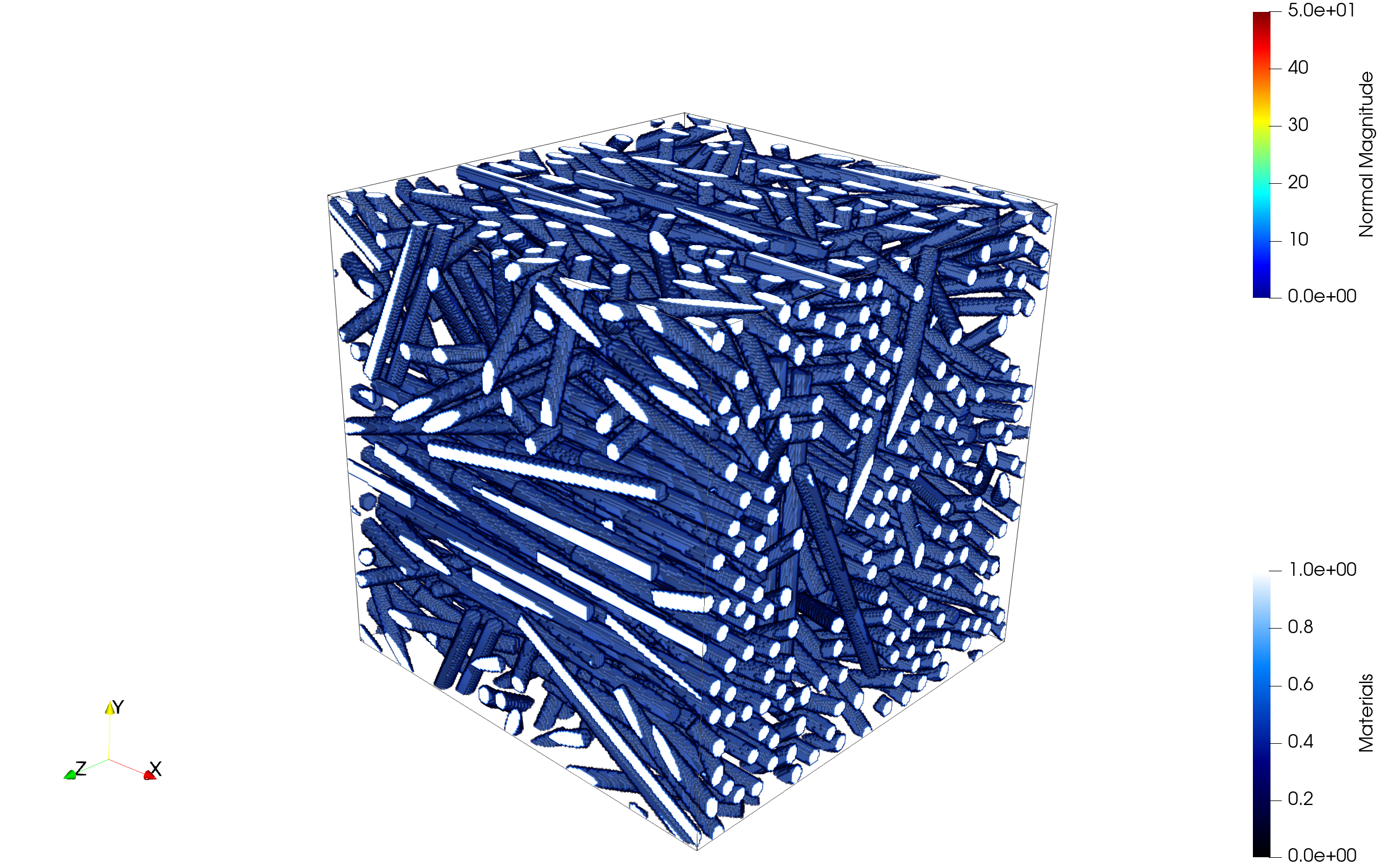}
		\caption{Considered fiber-reinforced microstructure}
    	\label{fig:fiber_structure}
	\end{subfigure}
	\begin{subfigure}{.4925\textwidth}
		\includegraphics[width = \textwidth]{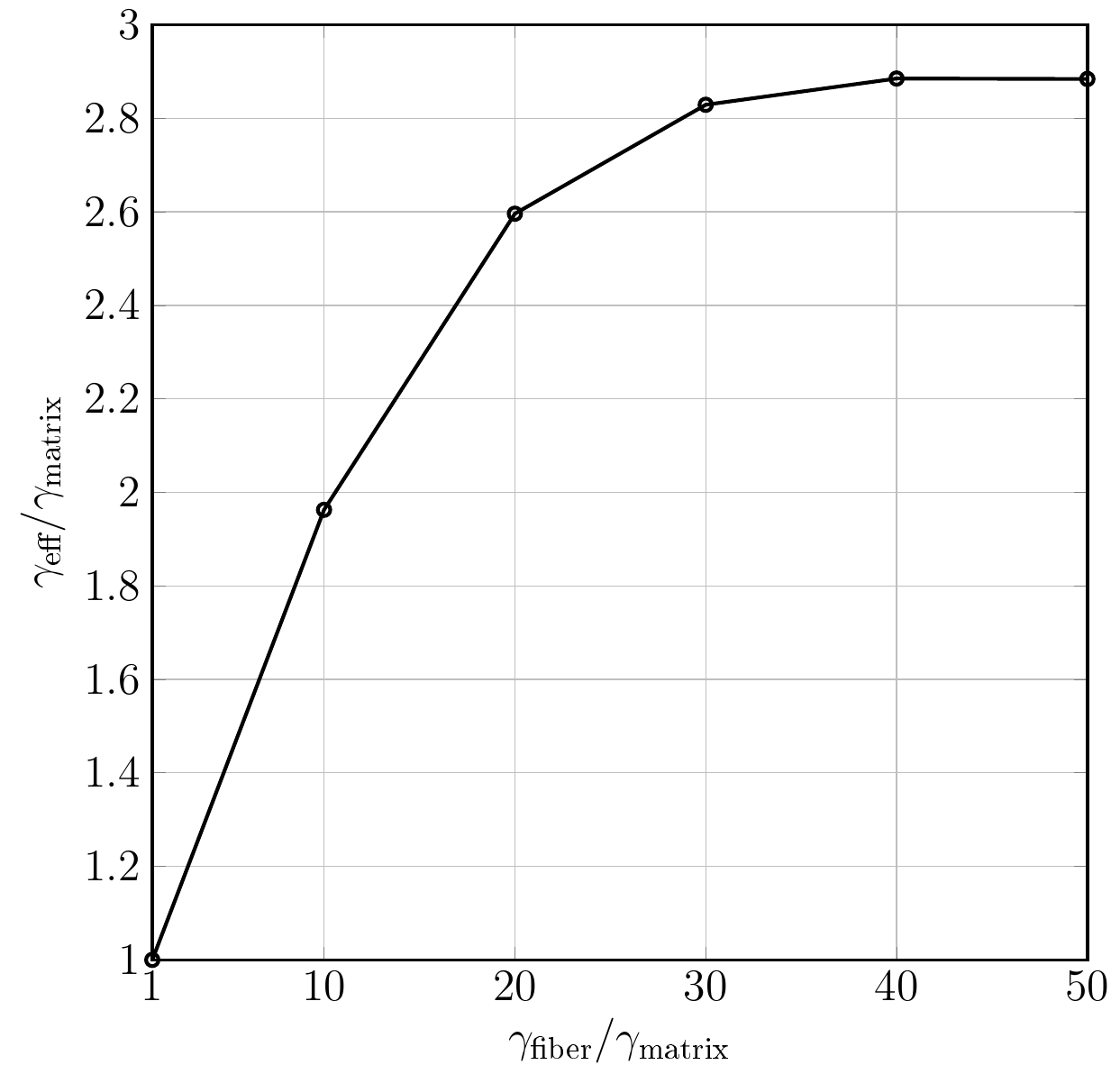}
		\caption{\EffCrack{} in $e_x$ vs. material contrast }					
		\label{fig:fiberContrast}
	\end{subfigure}
	\caption{Microstructure and \effCrack{} for the fiber-reinforced composite}					
	\label{fig:fiber_structure_Contrast}
\end{figure}

Indeed, the relation between the residual \eqref{eq:convergence_criterion} and the error in the quantity of interest \eqref{eq:UD_relative_error} is not directly apparent. Indeed, we know that convergence of effective properties is implied by convergence of the fields. However, the quantitative relation between these may only be determined by comparison to a ground truth. For the \CCMF{} discretizations, the relative error \eqref{eq:UD_relative_error} correlates with the residual rather well, reaching an accuracy below $10^{-4}$ at convergence. In contrast, the solution for the rotated staggered grid  leads to an error of only $0.5\%$, i.e., hits a "stall".\\
In addition to the mentioned penalty-factor choices, we studied two further (less competitive) approaches, namely residual balancing~\cite{He2000adaptive}, which is often recommended in the literature, as well as an the approach suggested by Lorenz-Tran-Dinh~\cite{Lorenz2019adaptiveDR}, which proved to be promising in small-strain micromechanics~\cite{ADMM2021}. To increase readability, the residual and the error \eqref{eq:UD_relative_error} were moved to Fig.~\ref{fig:UD_Error_Appendix} of Appendix \ref{sec:Appendix:A}.\\
With this validation at hand, we restrict to the \CCMF{} discretization in combination with ADMM, damping factor $\damping = 0.25$ and the Barzilai-Borwein penalty-factor for the remainder of this work.

\subsection{A fiber-reinforced composite}
\label{sec:computations_fiber}

\begin{figure}[t]
 \begin{center}
 \begin{subfigure}{0.06\textwidth}
 \includegraphics[width=\textwidth]{Figures/Figure06_xiLegend}
 \vspace{0.7cm}
 \includegraphics[width=\textwidth]{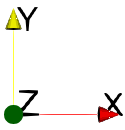}
 \end{subfigure}
 \begin{subfigure}{0.3\textwidth}
\includegraphics[width=\textwidth,trim = 700 250 700 270,clip]{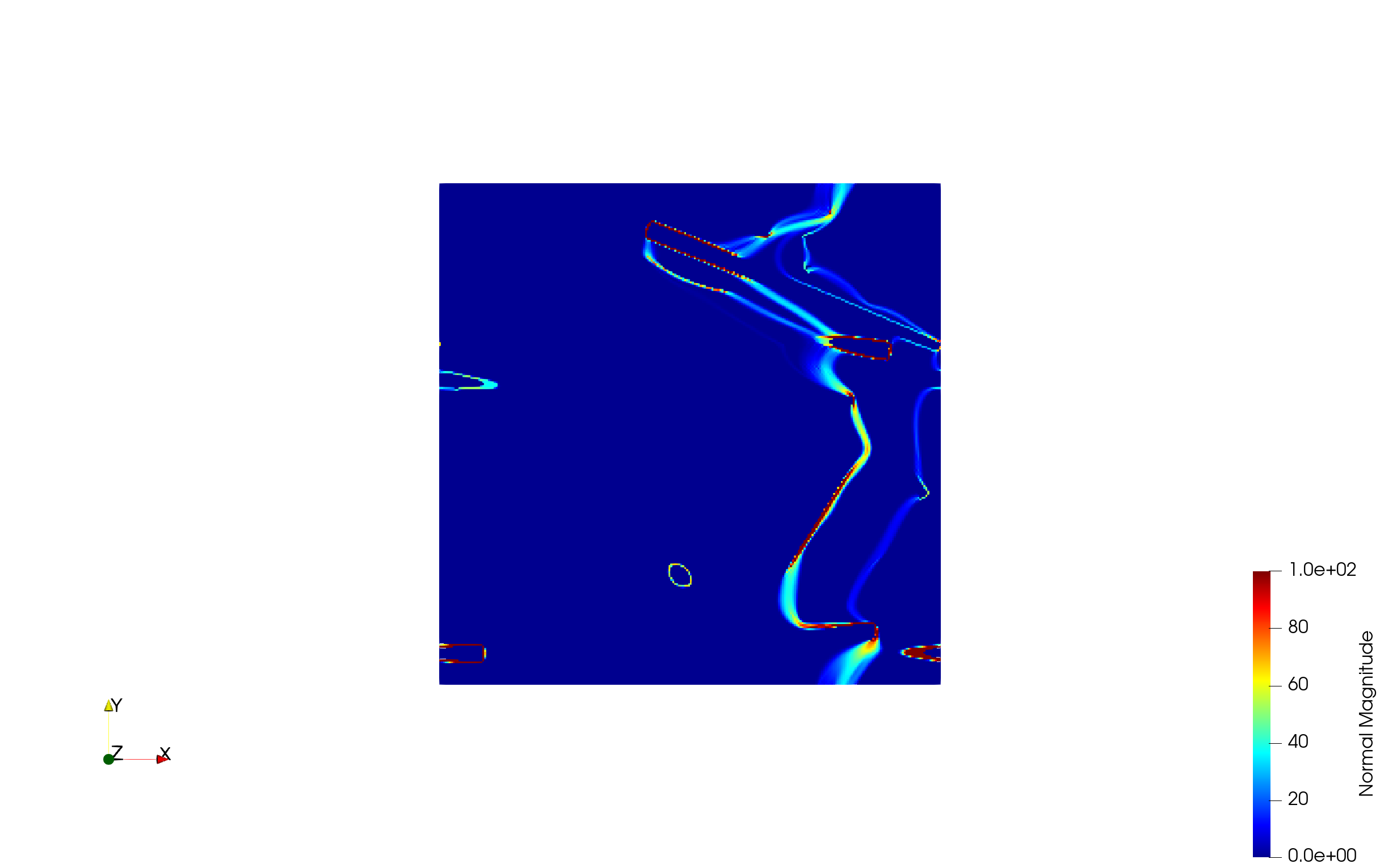}
     \vspace{-0.5cm}
    \caption{$1000$ iterations}
    \label{fig:fiber1000}
   \end{subfigure}
 \begin{subfigure}{0.3\textwidth}
\includegraphics[width=\textwidth,trim = 700 250 700 270,clip]{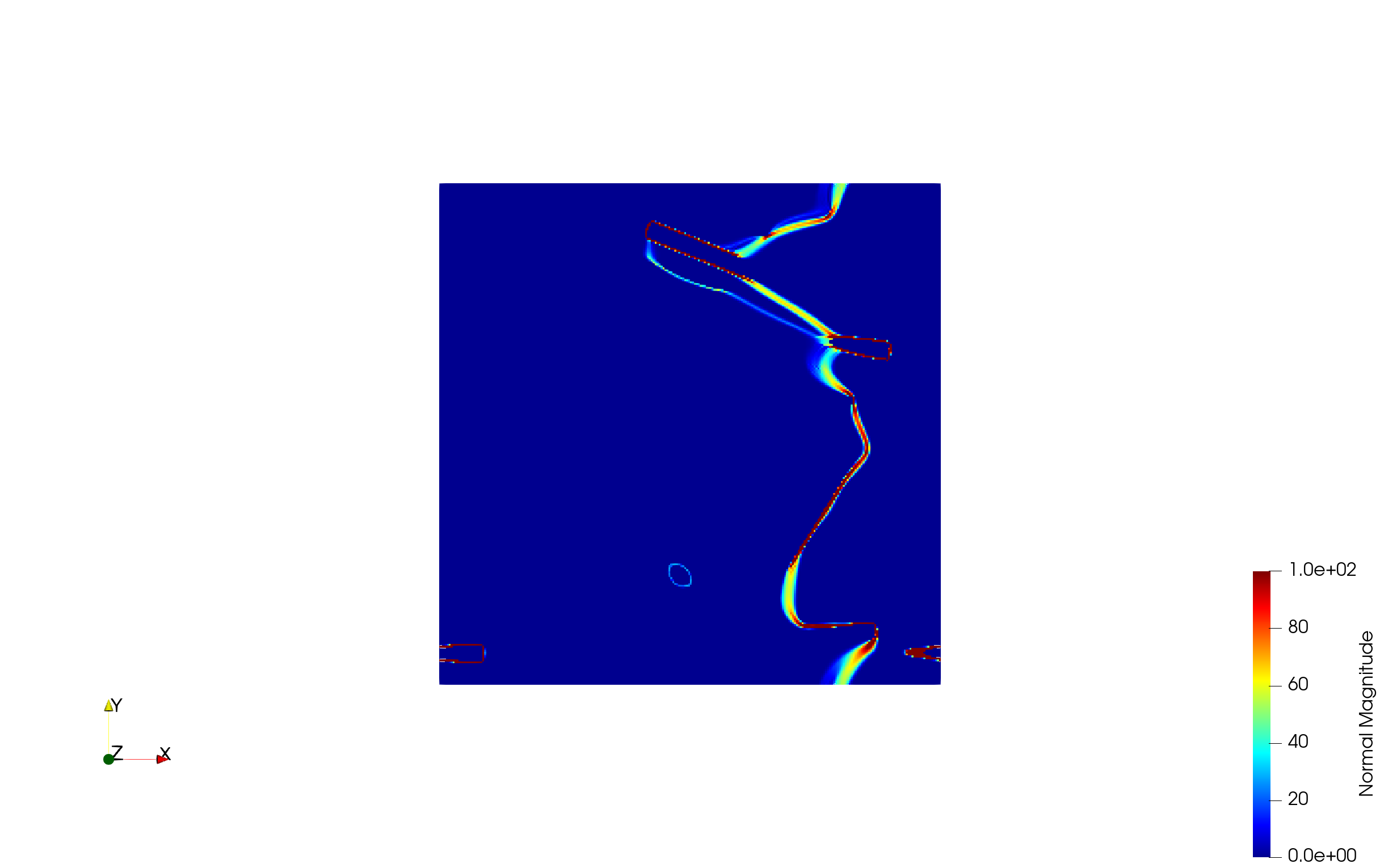}
     \vspace{-0.5cm}
    \caption{$2500$ iterations}
    \label{fig:fiber2500}
   \end{subfigure}
    \begin{subfigure}{0.3\textwidth}
\includegraphics[width=\textwidth,trim = 700 250 700 270,clip]{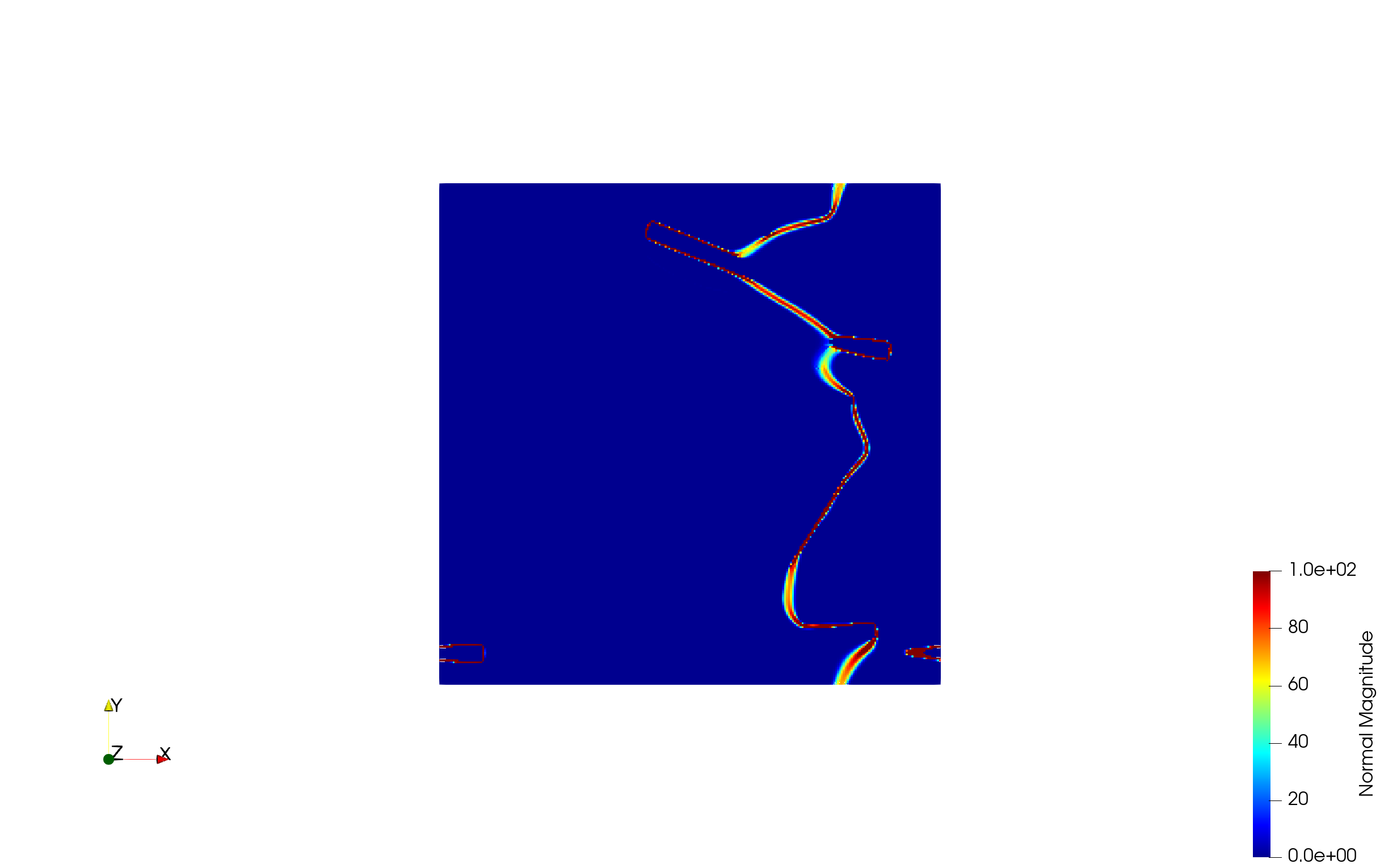}
     \vspace{-0.5cm}
    \caption{$5000$ iterations}
    \label{fig:fiber5000}
   \end{subfigure}
 \end{center}\caption{Cross section through crack surface for $\gamma_\text{fiber} = 50\,\gamma_\text{matrix}$ at different ADMM iterations}
 \label{fig:fiberNumber}
\end{figure}

\begin{figure}
 	\begin{subfigure}{0.06\textwidth}
 		\vspace{2.5cm}
 		\includegraphics[width=\textwidth]{Figures/Figure08_KOS_3D}
 	\end{subfigure}
 	\begin{subfigure}{0.3\textwidth}
		\includegraphics[width=\textwidth, trim = 570 20 560 190, clip]{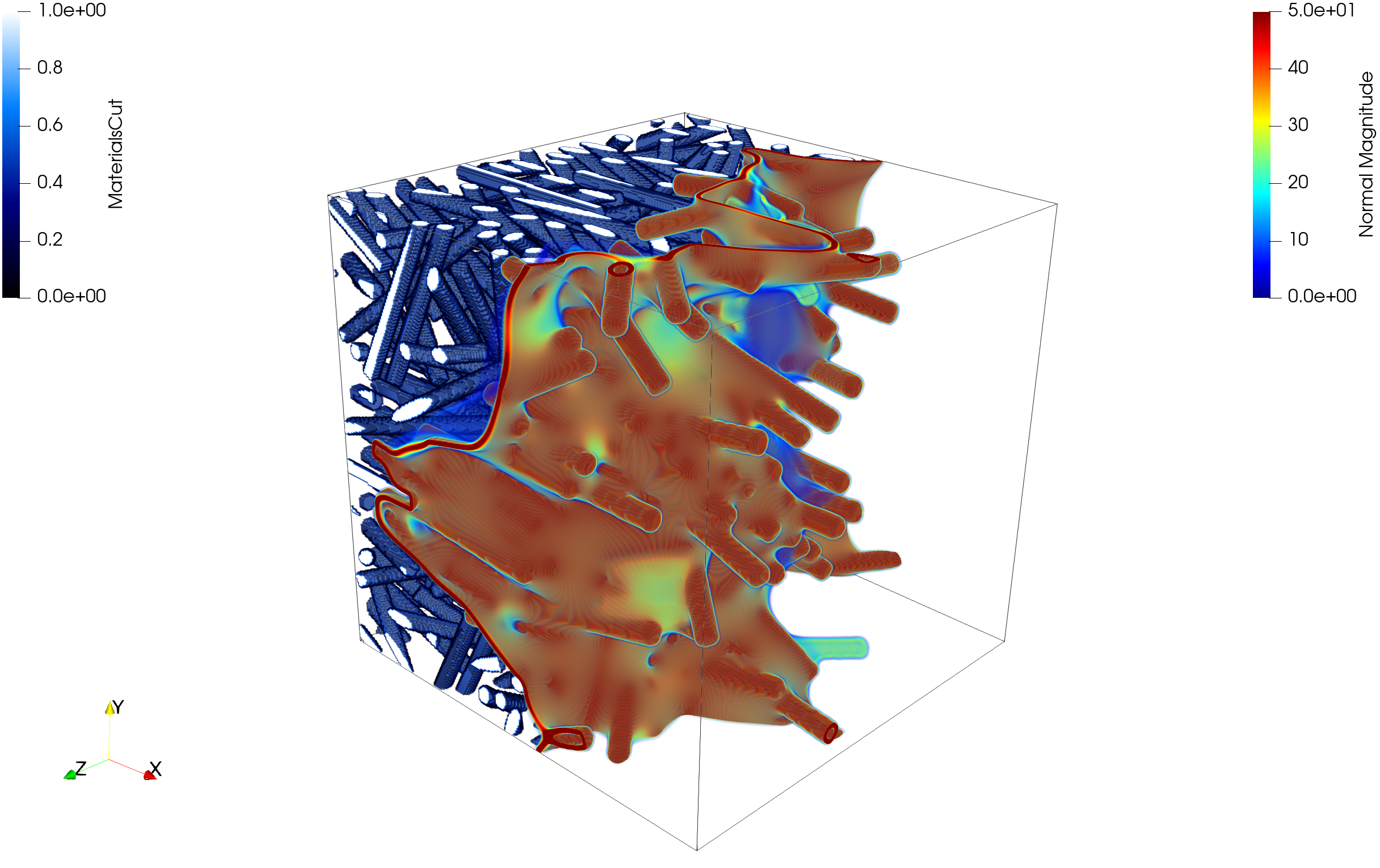}
    	\caption{$\bar{\xi} = e_x, \geff = 2.88\,\gamma_\text{matrix}$}
    	\label{fig:fiberX}
	\end{subfigure}
    \begin{subfigure}{0.3\textwidth}
		\includegraphics[width=\textwidth, trim = 570 20 560 190, clip]{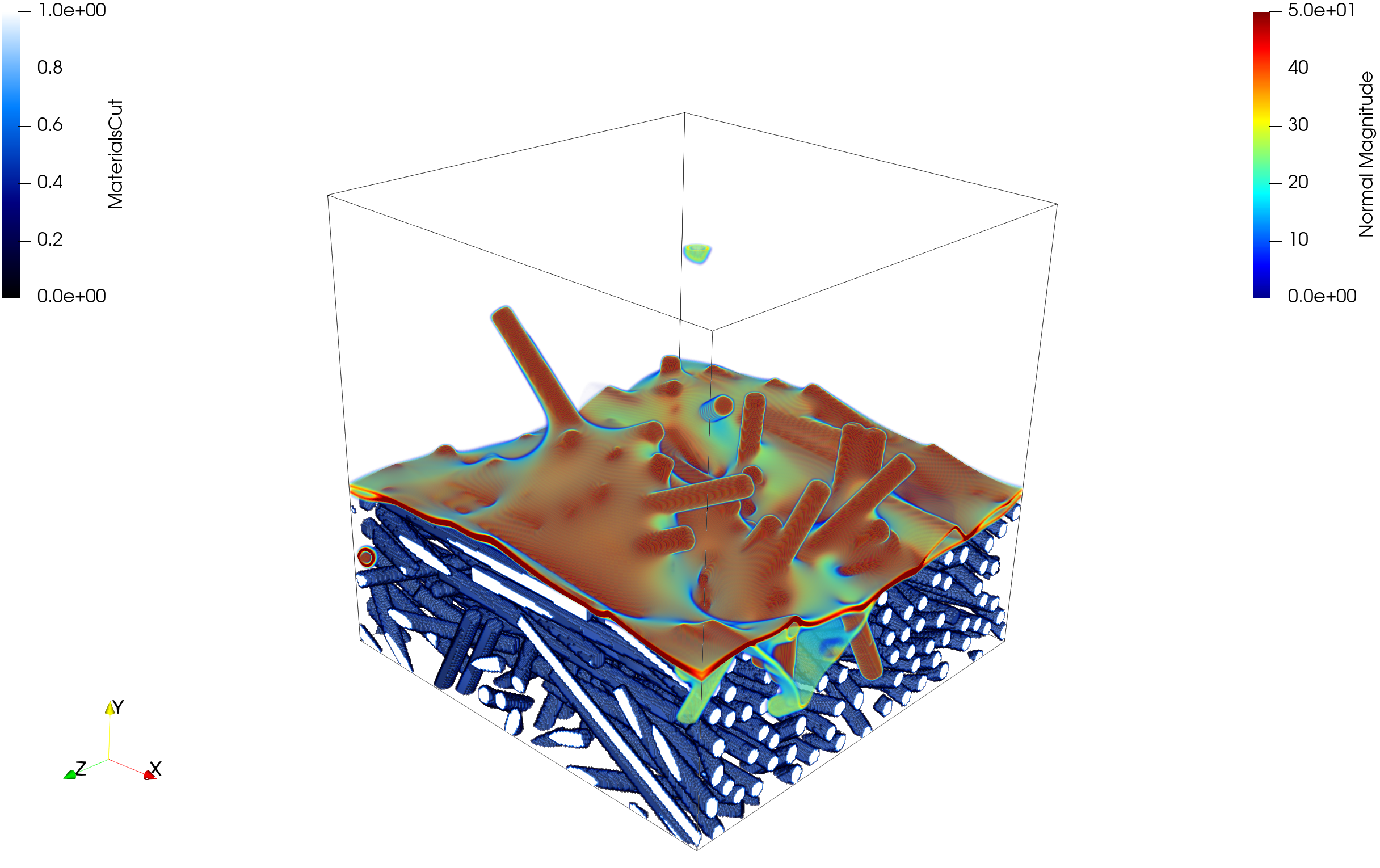}
    	\caption{$\bar{\xi} = e_y, \geff = 1.67\,\gamma_\text{matrix}$}
    	\label{fig:fiberY}
   	\end{subfigure}
 	\begin{subfigure}{0.3\textwidth}
		\includegraphics[width=\textwidth, trim = 570 20 560 190, clip]{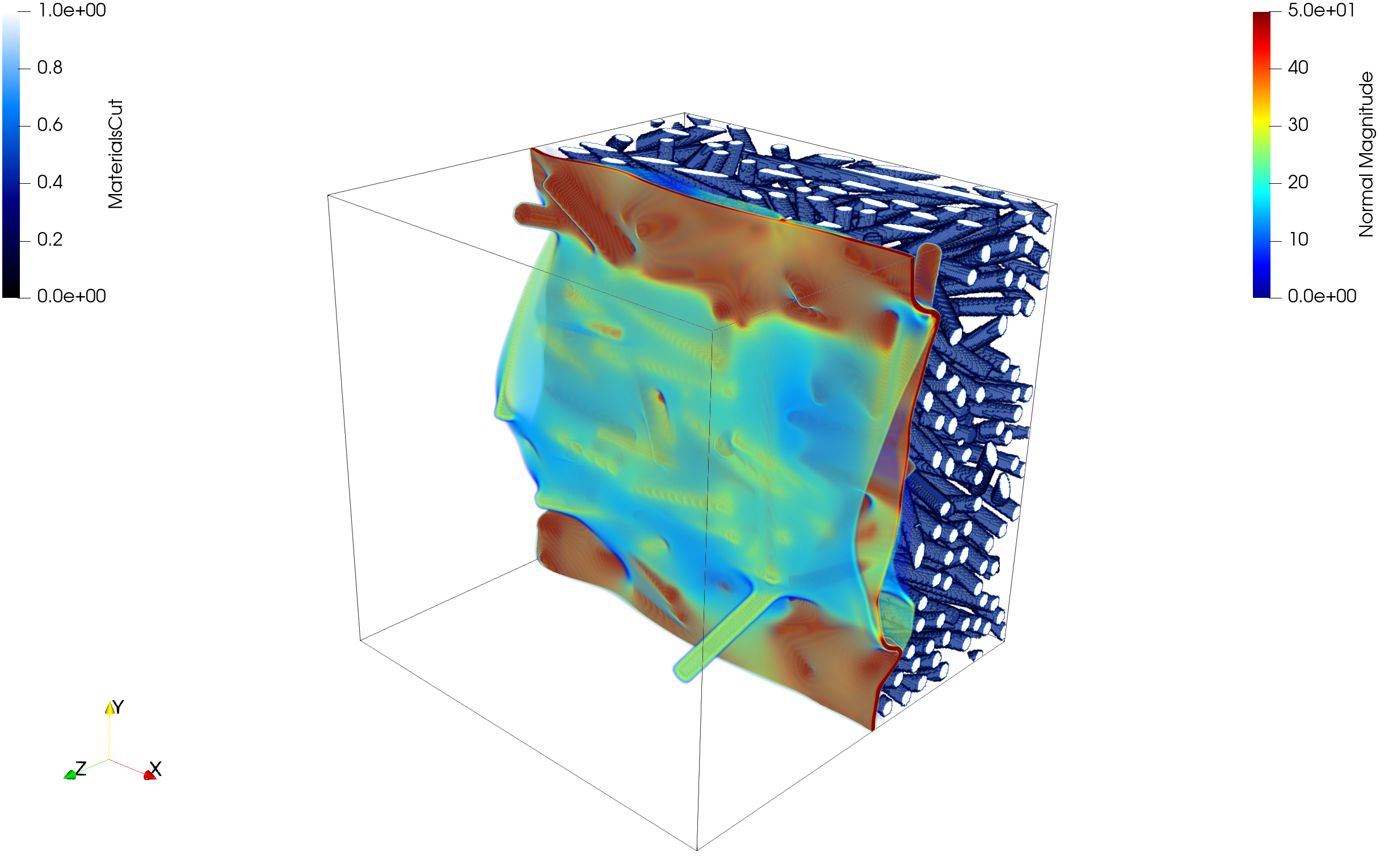}
    	\caption{$\bar{\xi} = e_z, \geff = 1.20\,\gamma_\text{matrix}$}
    	\label{fig:fiberZ}
   	\end{subfigure}
	\caption{Crack surfaces for the Cartesian normals, material contrast $\gamma_\text{fiber}/\gamma_\text{matrix} = 50$ and the fiber-reinforced composite, see Fig.~\ref{fig:fiber_structure}}
	\label{fig:fiberXYZ}
\end{figure}

After the necessary verification steps, we turn our attention to problems with a higher degree of complexity. We consider a short-fiber reinforced composite with $18\%$ filler content. The synthetic structure contains $376$ fibers with an aspect ratio (length/diameter) of $20$, and was generated by the SAM algorithm~\cite{SAM}. The prescribed fiber-orientation tensor of second order~\cite{Kanatani1984,AdvaniTucker} was $\text{diag}(0.75,0.19,0.06)$, i.e., the fibers lie almost exclusively in the $x$-$y$-plane with a strong preference in $x$-direction. The fibers are discretized with eight voxels per diameter, resulting in a volume image with $256^3$ voxels, see Fig.~\ref{fig:fiber_structure}. Since the computations on such large structures are costly, we first investigate the influence of the tolerance entering the stopping criterion ~\eqref{eq:convergence_criterion}. For a configuration $\gamma_\text{fiber}=50\,\gamma_\text{matrix}$, we computed the \effCrack{} in direction $\bar{\xi}=e_x$. After $1000, 2500, 5000, 7500,$ and $10000$ iterations, we take a look at the corresponding residual and the computed \effCrack{}, see Tab.~\ref{tab:fiber_IterInfluence}. 
\begin{table}
\begin{center}
\begin{tabular}{|l | c|c|}
\hline
 $\#$ iteration &residual& $\geff/\gamma_\text{matrix}$\\
 \hline
 \hline
$ 1000 $ & $ 1.7\cdot 10^{-3} $ & $ 2.94 $ \\
$ 2500 $ & $ 6.1\cdot 10^{-4} $ & $ 2.89 $ \\
$ 5000 $ & $ 1.9\cdot 10^{-4} $ & $ 2.87 $ \\
$ 7500 $ & $ 1.0\cdot 10^{-4} $ & $ 2.87 $ \\
$ 10000 $ & $ 5.5\cdot 10^{-5} $ & $ 2.87 $ \\
  \hline
\end{tabular}
\end{center}
  \caption{Residual and computed \effCrack{} with normal $\bar{\xi} = e_x$ depending on the number of ADMM iterations for a fiber-reinforced composite, see Fig.~\ref{fig:fiber_structure}, with material parameters $\gamma_\text{fiber}=50\,\gamma_\text{matrix}$}
  \label{tab:fiber_IterInfluence}
\end{table}
We observe that, after $1000$ iterations we reach a residual of almost $10^{-3}$ with a relative deviation in \effCrack{} about $2\%$ compared to the prediction after $10000$ iterations. After $2500$ iterations, the relative error is below $1\%$ with a residual at about $6\cdot 10^{-4}$. For more than $5000$ iterations, the \effCrack{} does not change in the third significant digit, whereas the residual decreases only slowly.\\
To complement these numbers, we take a look at a cross section through the computed crack surface at different iteration counts, see Fig.~\ref{fig:fiberNumber}. We observe an influence of the solver accuracy on the solution field $\xi$. Indeed, after $1000$ iterations, several distinct crack paths are present in the vicinity of the solution. These different cracks, however, come with different "intensities", as well. This ambiguity is reduced after $2500$ iterations. Only after $5000$ iterations, the solver finds a unique crack surface.\\
Please note that, in general, we do not expect the minimum-cut problem \eqref{eq:theory_cell_formulae_cell_formula} to have a unique solution. Rather, for the problem at hand, a unique crack is formed and, at low levels of the residual, additional cracks appear, compare also Schneider~\cite[Sec. 4.1.2]{HomFrac2019}. These vanish, however, at high accuracy.\\
To Balance accuracy and ensuing computational costs, we fix the tolerance to $5\cdot 10^{-4}$.\\
Next, we investigate the resulting crack surfaces and {effective crack energies} corresponding to different crack normals, see Fig.~\ref{fig:fiberXYZ}. In $e_x$-direction, the \effCrack{} is highest. This is caused by preferred fiber direction in this direction, forcing the crack surface to bypass the numerous inclusions. In $e_y$-direction, see Fig.~\ref{fig:fiberY}, the crack surface looks roughly similar. However, the crack needs to avoid fewer fibers, resulting in a lower \effCrack{}. In $e_z$-direction, the crack surface is almost straight, see Fig.~\ref{fig:fiberY}, resulting in the lowest \effCrack{}.\\
Last but not least, we investigate the influence of the material contrast on the computed \effCrack{} in $\bar{\xi}=e_x$-direction, see Fig.~\ref{fig:fiberContrast}. This contrast is responsible for the allowed crack-inclusion interaction-mechanisms. Indeed, for high contrast, the inclusions can only be avoided, i.e., inclusion bypass is the only viable option. In general, the particles' anisotropy (encoded by the aspect ratio and the fiber orientation for the example at hand) and the filler content determine the threshold in contrast where only inclusion bypass is permitted. For the example at hand, Fig.~\ref{fig:fiberContrast} reveals that this threshold is roughly at a material contrast of $40$.\\
For lower contrast, it may be energetically more favorable to cross some of the inclusion. For decreasing material contrast, this inclusion-crossing mechanism occurs more frequently.

\subsection{Microstructures with a monodisperse pore distribution}
\label{sec:computations_porous}

\begin{figure}
	\begin{subfigure}{0.33\textwidth}
		\begin{tikzpicture}[      
        	every node/.style={anchor=south west,inner sep=0pt},
        	x=1mm, y=1mm,
      	]   
     	\node (fig1) at (0,0) {\includegraphics[width=\textwidth, trim = 545 45 620 210, clip]{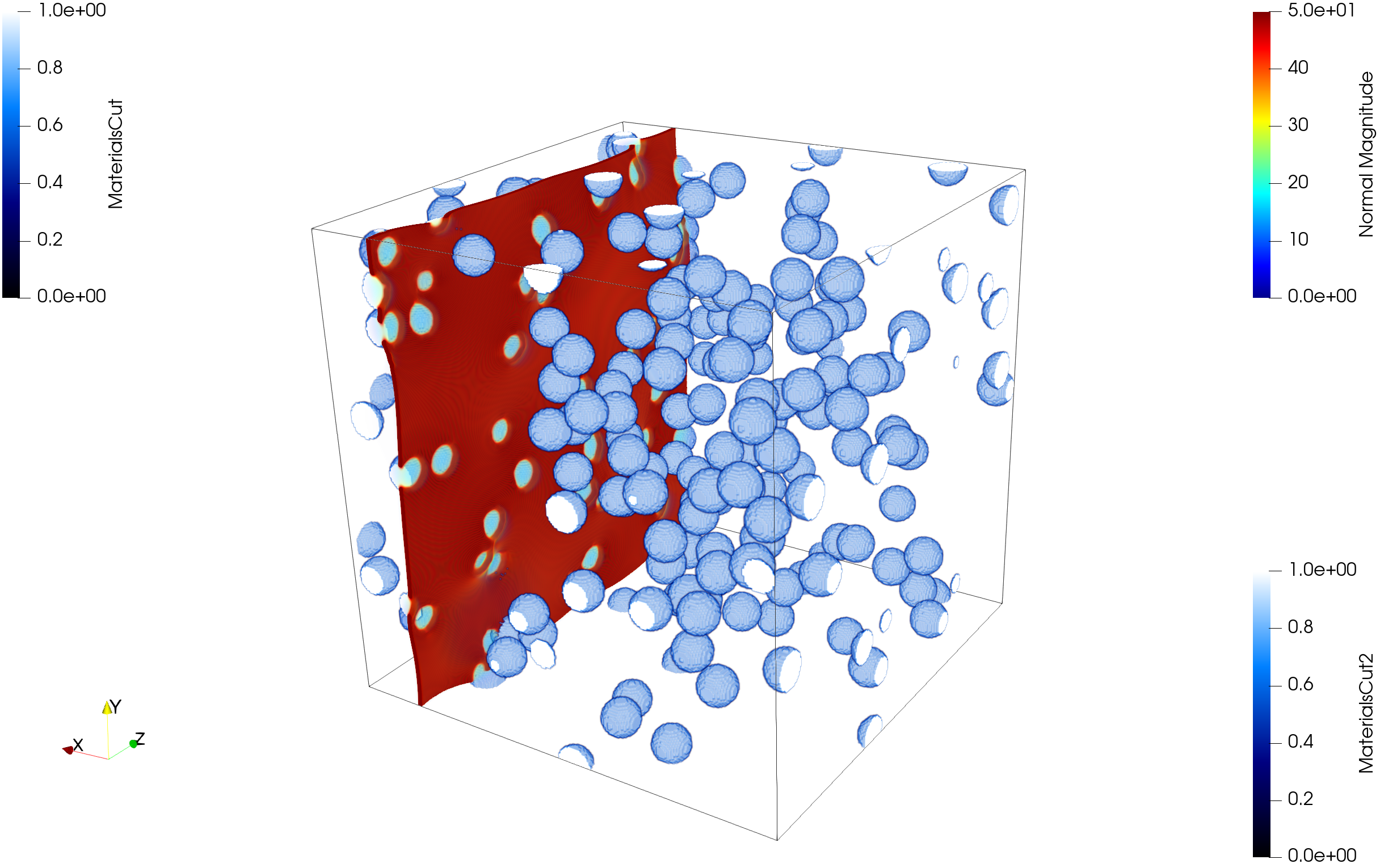}};
     	\node (fig2) at (0,0) {\includegraphics[width=.15\textwidth]{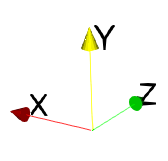}};  
		\end{tikzpicture}
		\caption{$5\%$ porosity}
    	\label{fig:Porous5}
	\end{subfigure}
	\begin{subfigure}{0.33\textwidth}
		\includegraphics[width=\textwidth, trim = 545 45 620 210, clip]{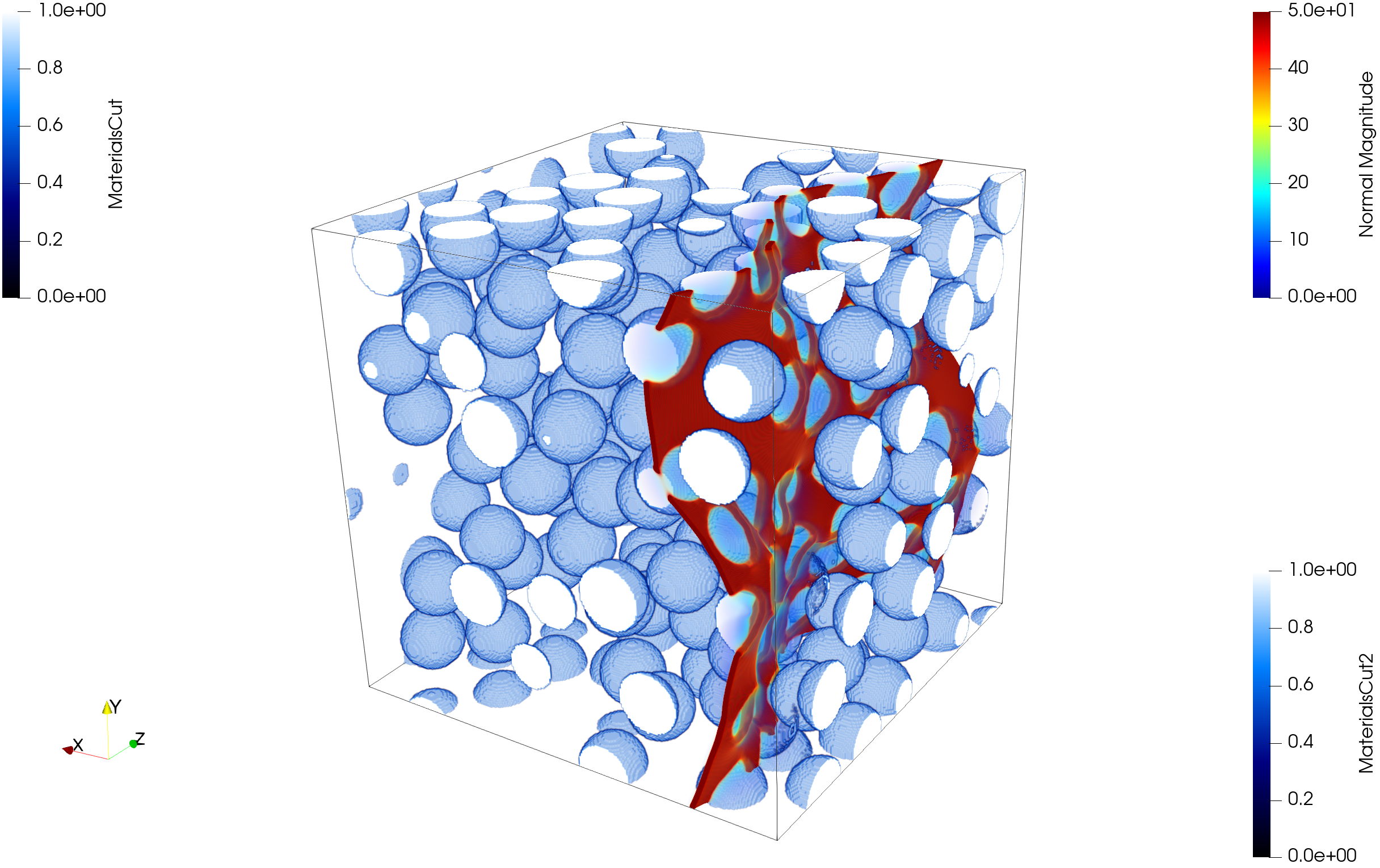}
    	\caption{$25\%$ porosity}
    	\label{fig:Porous25}
   	\end{subfigure}
 	\begin{subfigure}{0.33\textwidth}
		\includegraphics[width=\textwidth, trim = 545 45 620 210, clip]{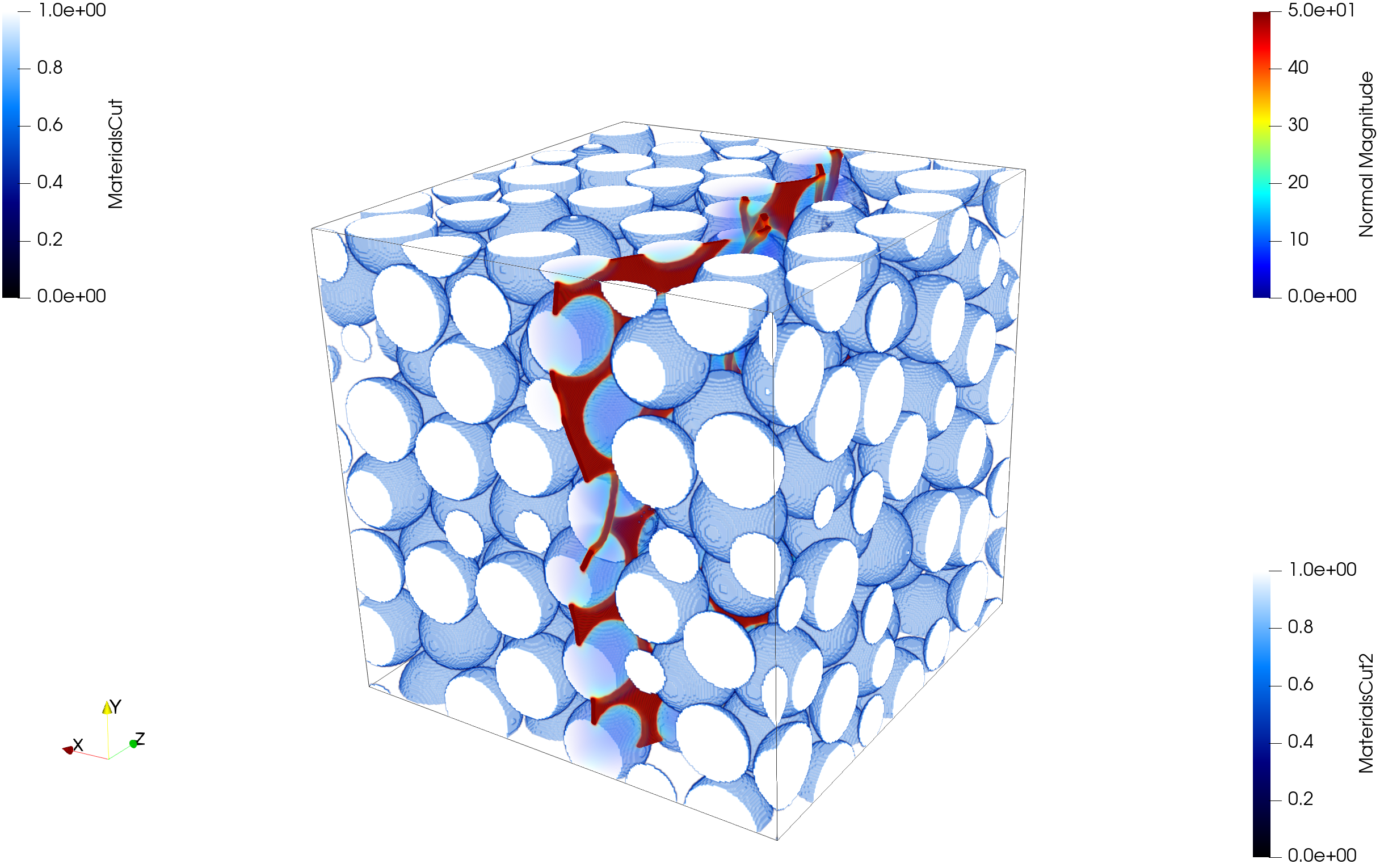}
    	\caption{$50\%$ porosity}
    	\label{fig:Porous50}
   \end{subfigure}
	\caption{Crack surface through microstructures with varying porosity}
	\label{fig:Porous}
\end{figure}

As our next example, we consider microstructures with monodisperse, spherical pores and varying degree of porosity. For a porosity between $5$ and $50\%$, we generated microstructures with $200$ spheres by the mechanical contraction method~\cite{WilliamsPhilipse}, see Fig.~\ref{fig:Porous}. All structures were discretized on a $256^3$ voxel grid. The solid material has \crackres{} $\gamma$, and the spherical pores are furnished with a vanishing \crackres{}, resulting in an infinite material contrast. Please note that in the previous study~\cite[Sec. 4.1.1]{HomFrac2019}, the pores were furnished with a non-vanishing (yet small) \crackres{} to ensure robust convergence of the utilized solution scheme. Such a restriction appears unnecessary for the improved solution method presented in this article.\\
The \effCrack{} in direction $\bar{\xi}=e_x$ and the required ADMM iterations are listed in Tab.~\ref{tab:PorousIter}. 
\begin{table}
\begin{center}
\begin{tabular}{|c | c|c|}
\hline
 porosity in $\%$ &$\geff/\gamma$& iterations	\\
 \hline
 \hline
$ 5 $ & $ 0.871 $ & $ 4986 $ \\
$ 25 $ & $ 0.535 $ & $ 3300 $ \\
$ 40 $ & $ 0.412 $ & $ 3327 $ \\
$ 50 $ & $ 0.305 $ & $ 2358 $ \\
  \hline
\end{tabular}
\end{center}
  \caption{Influence of the porosity on the \effCrack{} and solver performance for varying porosity}
  \label{tab:PorousIter}
\end{table}
Following physical intuition, the \effCrack{} decreases for increasing porosity. The iteration count appears to be uncorrelated with the porosity. As a remark, we found the iteration count to be strongly dependent on the specific realization of the microstructure, in general. Thus, we expect that no such correlation may be inferred from a single sample, but would require a more elaborate study.\\
Fig.~\ref{fig:Porous} presents the computed crack surfaces. The crack surface is not straightly planar, but adjusts to cross a larger number of pores in order to reduce its $\gamma$-weighted surface area.

\subsection{Sand-binder composite}
\label{sec:computations_grains}

\begin{figure}
	\begin{subfigure}{0.33\textwidth}
		\begin{tikzpicture}[      
        	every node/.style={anchor=south west,inner sep=0pt},
        	x=1mm, y=1mm,
      	]   
     	\node (fig1) at (0,0) {\includegraphics[width=\textwidth, trim = 550 60 650 220, clip]{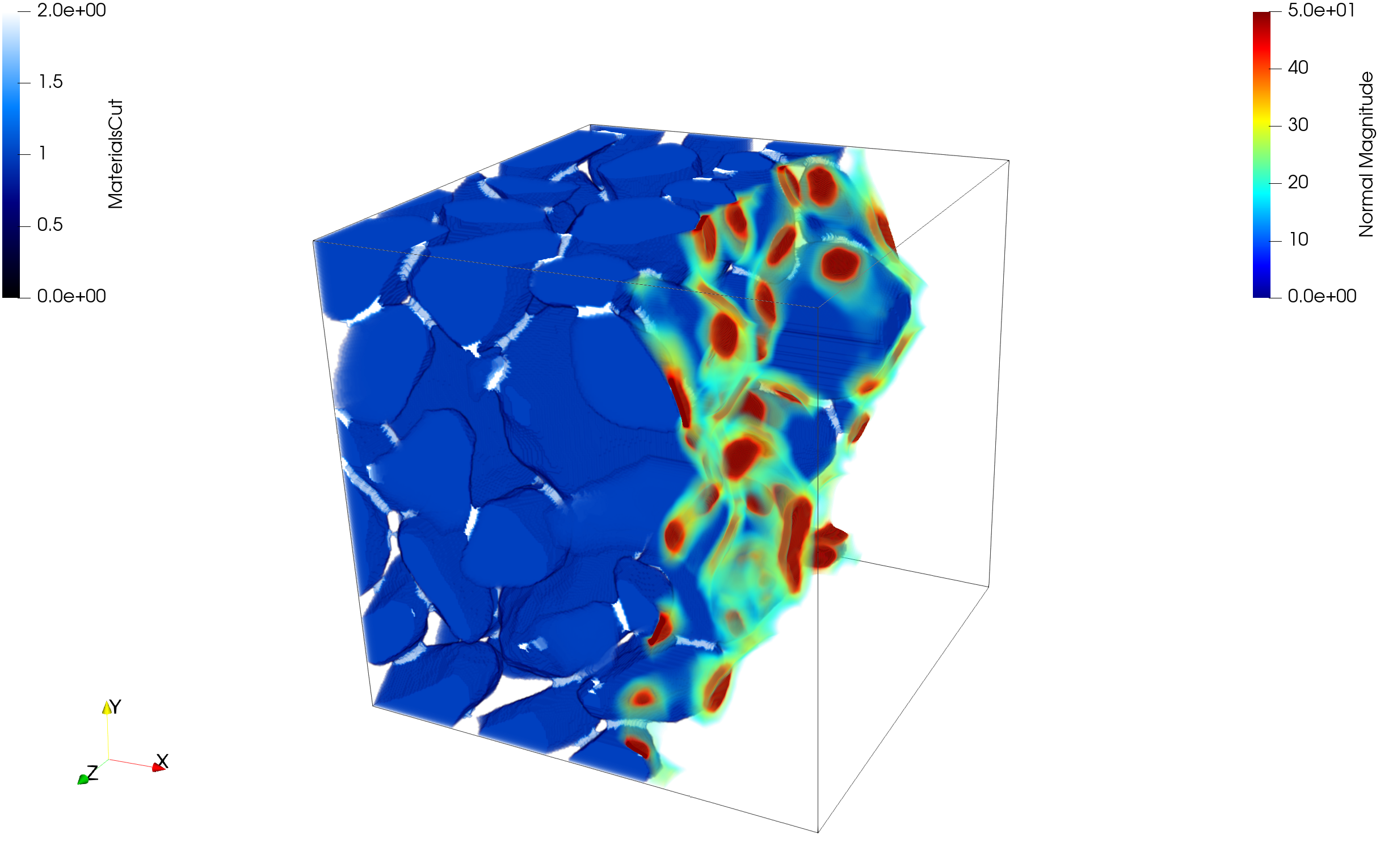}};
     	\node (fig2) at (0,0) {\includegraphics[width=0.15\textwidth]{Figures/Figure01_KOS}};  
		\end{tikzpicture}
		\caption{Case $\#1$ - porous composite}
    	\label{fig:Sand1}
	\end{subfigure}
	\begin{subfigure}{0.33\textwidth}
		\includegraphics[width=\textwidth, trim = 550 60 650 220, clip]{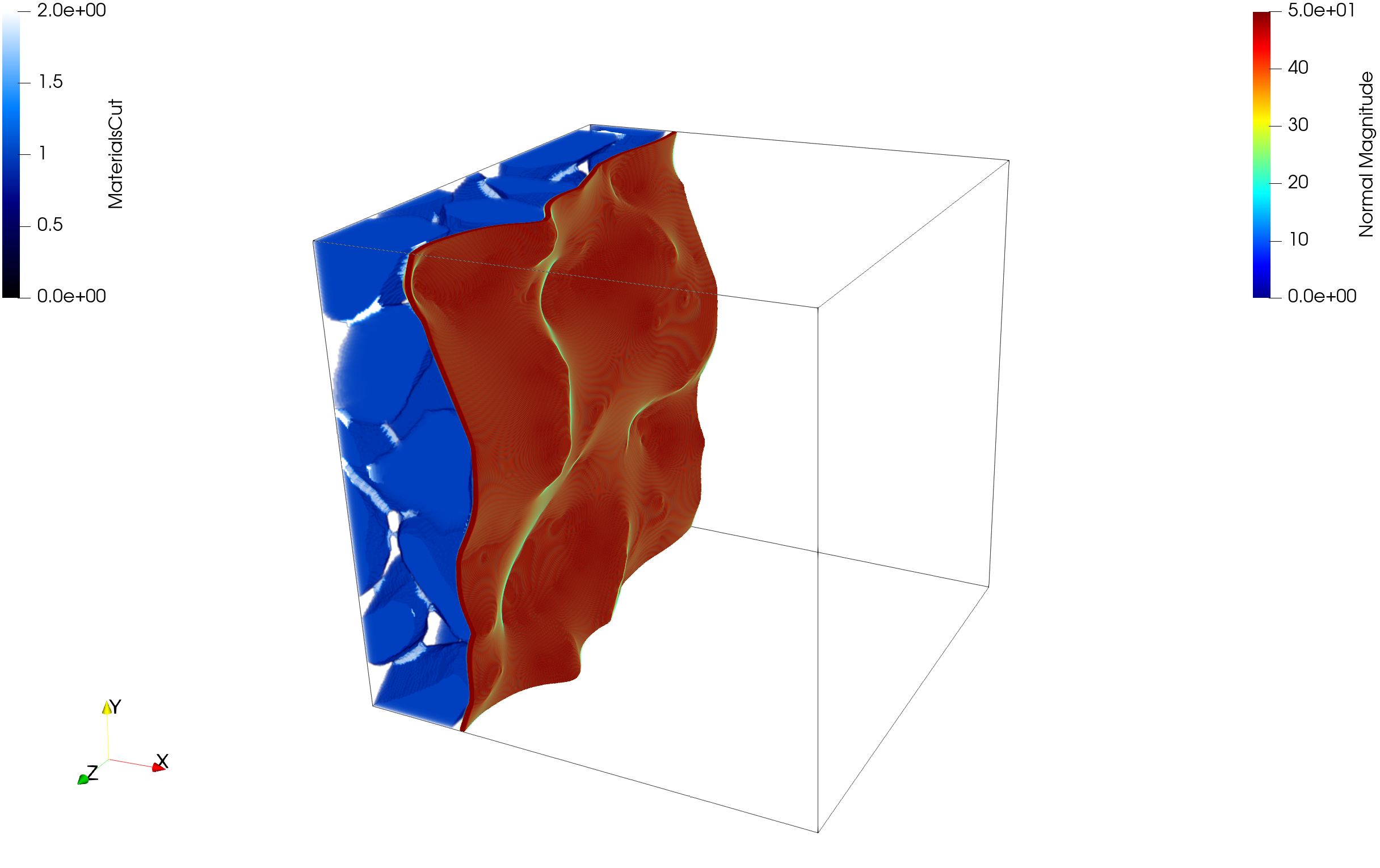}
    	\caption{Case $\# 2$ - grain-matrix composite}
    	\label{fig:Sand2}
   	\end{subfigure}
   	\begin{subfigure}{0.33\textwidth}
		\includegraphics[width=\textwidth, trim = 550 60 650 220, clip]{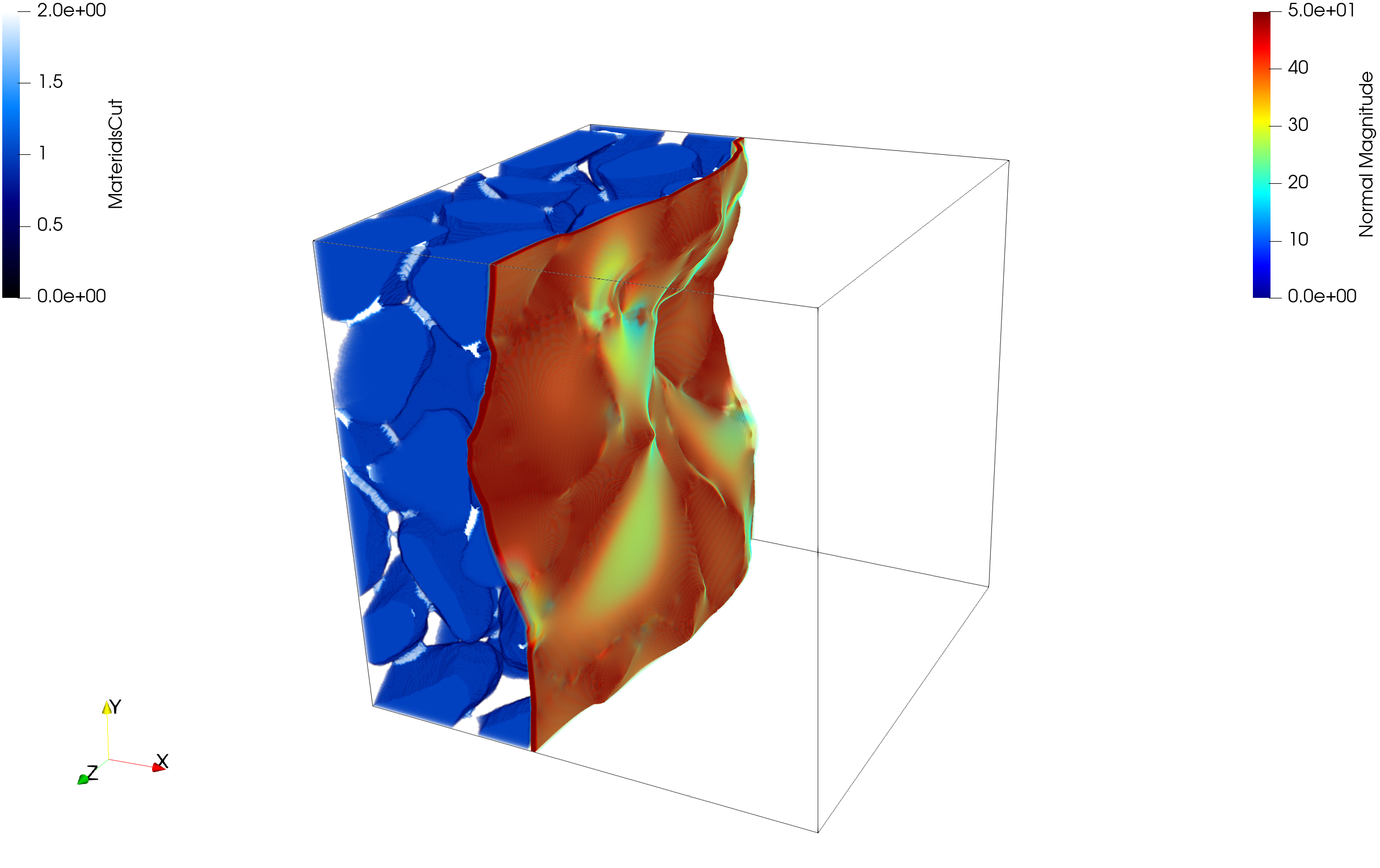}
    	\caption{Case $\# 3$ - porous inclusions}
    	\label{fig:Sand3}
   	\end{subfigure}
	\caption{Crack surfaces through the bound sand-grain microstructure for three different combinations of crack resistances for matrix, inclusion and binder}
	\label{fig:Sand}
\end{figure}

In our final example, we examine the microstructure of a sand-binder aggregate which is characteristic for inorganically bound sand cores used in casting applications. The synthetic structure was generated by a mechanical-contraction type method~\cite{Sand,SandMultiPhysics}, and is shown in Fig.~\ref{fig:SandStructure}. The microstructure consists of three phases: The sand grains ($58.6\%$), connected by a binder phase ($1.3\%$), and a third phase ($40.1\%$). In the physical applications, the latter phase represents the pore space. We wish to utilize the microstructure to get insights for a number of physical scenarios, and we will refer to the third phase more generally as the "matrix" for reasons that will become clear shortly.\\
 The \crackres{}s associated to the phases are denoted by $\gamma_\text{grain}$, $\gamma_\text{binder}$ and $\gamma_\text{matrix}$, respectively. To investigate the \effCrack{} and possible crack surfaces through the microstructure, we consider three different parameter scenarios, where the single phases model different physical scenarios. The governing parameters and their resulting \effCrack{} are listed in Tab.~\ref{tab:Sand}, 
 \begin{table}
\begin{center}
\begin{tabular}{|l | c|c|c|c|c|}
\hline
 &$\gamma_\text{matrix}$ in \unit{MPa$\cdot\mu$m}&$\gamma_\text{grain}$ in \unit{MPa$\cdot\mu$m}&$\gamma_\text{binder}$ in \unit{MPa$\cdot\mu$m}&$\gamma_\text{eff}$ in \unit{MPa$\cdot\mu$m}& iterations	\\
 \hline
 \hline
 $ \# 1 $ & $ 0 $ & $ 1 $ & $ 1 $ & $ 0.074 $& $ 3204 $\\
\hline
$ \# 2 $ & $ 1 $ & $ 10$ & $ 1 $ & $ 1.133 $& $ 1711 $\\
\hline
$ \# 3 $ & $ 10 $ & $ 1 $ & $ 10$ & $ 3.246 $& $ 3971 $\\
  \hline
\end{tabular}
\end{center}
  \caption{Material parameters, {effective crack energies} as well as iteration count for the three cases under consideration}
  \label{tab:Sand}
\end{table}
together with the required iteration count. In parameter case $\#1$, the \crackres{} of the grains and the binder are equal, and the matrix material corresponds to a pore space. The resulting crack surface is shown in Fig.~\ref{fig:Sand1}. We notice that the crack is fully contained in the binder phase. The \effCrack{} is reduced to $7.5\%$ of the crack resistance which grain and binder share. The second parameter case models the structure as a matrix material with tougher sand-grain inclusions. The binder phase is treated as additional matrix material. Fig.~\ref{fig:Sand2} shows the crack surface avoiding the sand-grain shaped inclusions. The resulting \effCrack{} of the composite is $1.133\,\gamma_\text{matrix}$. The third case deals with the same contrast, i.e., the binder phase is once again treated as additional matrix. This time, however, the sand-grain shaped inclusions are weaker than the surrounding material. The \effCrack{} is $32\%$ of the matrix \crackres{}. Fig.~\ref{fig:Sand3} shows the crack surface crossing several grains in order to avoid the matrix phase as much as possible.

\section{Conclusions}
\label{sec:conclusion}

In this work, we presented a powerful FFT-based solution method for computing the \effCrack{} of industrial-scale composite microstructures. Based on a homogenization result for the Francfort-Marigo model~\cite{FrancfortMarigo} in an anti-plane shear setting, see Braides et al.~\cite{Braides1996}, a cell formula for computing the \effCrack{} was investigated. This cell formula may be interpreted as a minimum cut / maximum flow problem~\cite{Strang1983}, which finds various applications, for instance in graph networks and image segmentation. Following Couprie et al.~\cite{CCMF}, we considered the \CCMF{} discretization on regular voxel data and integrated it into an FFT-based computational homogenization framework. In comparison to traditional spectral and finite-difference discretizations, we found the \CCMF{} discretization to significantly reduce artifacts in the local fields.\\
For solving the discretized equations, we investigated the alternating direction method of multipliers (ADMM) with various adaptive strategies, and found a damping parameter $\delta = 0.25$ combined with the Barzilai-Borwein penalty-factor choice to be the most effective. We demonstrated the applicability of our approach to various large-scale problems, considering complex microstructures, as well as large or even infinite contrast in the local crack resistance. The presented framework was implemented into an existing homogenization code for thermal conductivity, and, although we ran some computations on a workstation, all presented computations could be done on a conventional desktop computer.\\
Future work on this topic may focus on accounting for anisotropic \crackres{}. This may be of interest in a two-step homogenization framework for braided or SMC composites~\cite{Gorthofer2020}, and for brittle fracture in polycrystalline materials. Furthermore, accounting for weak interfaces may extend the applicability of the presented framework.

\appendix
\section{Performance of additional penalty factor choices}
\label{sec:Appendix:A}

\begin{figure}[t]
 	\begin{center}
 		\begin{subfigure}{\textwidth}
 			\centering
 			\includegraphics[width = 0.2\textwidth]{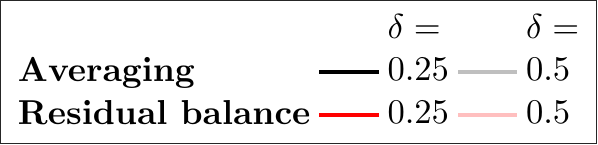}
 		\end{subfigure}
 		\begin{subfigure}{.49\textwidth}
 			\includegraphics[width=.45\textwidth]{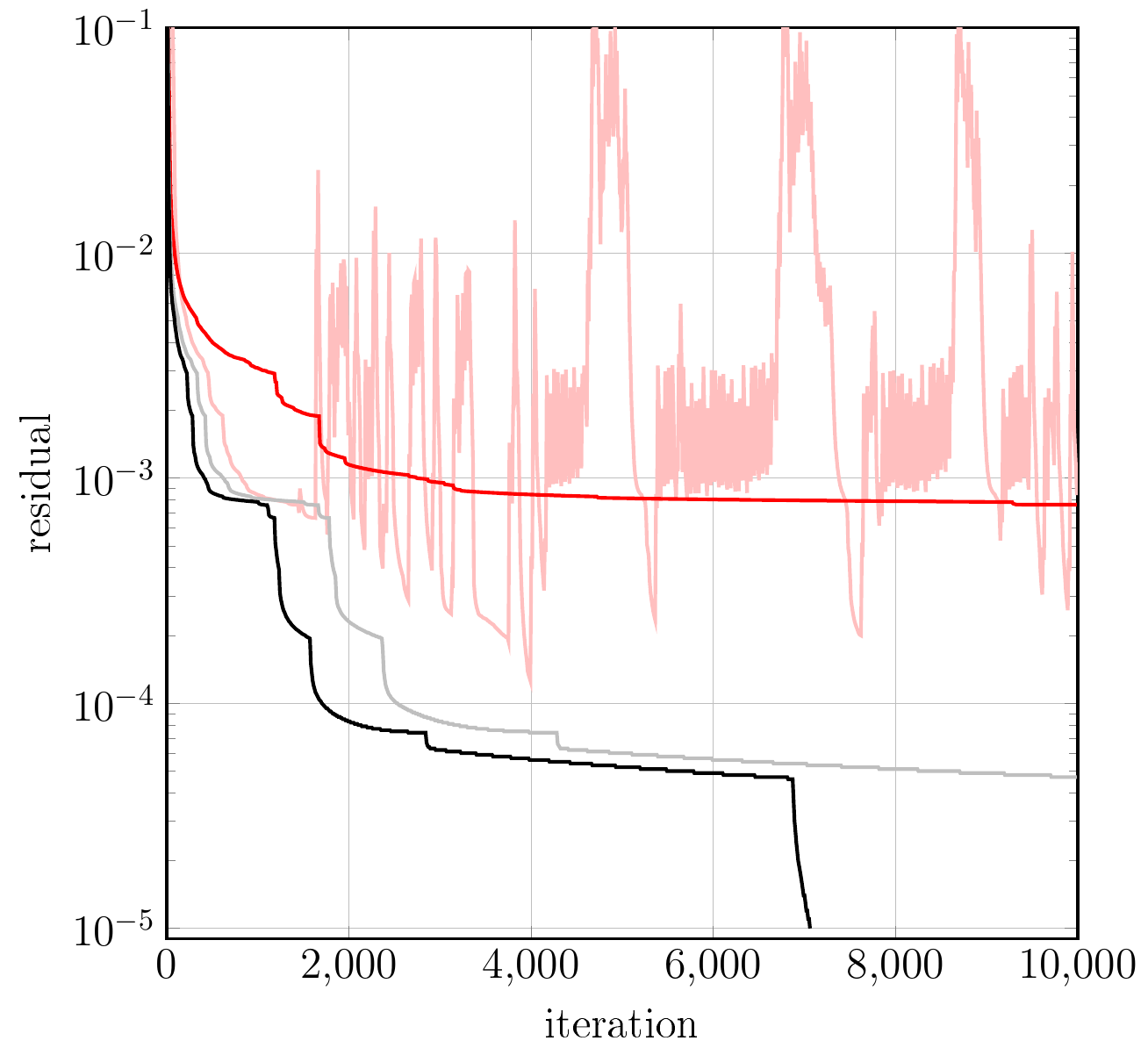} 
 			\includegraphics[width=.45\textwidth]{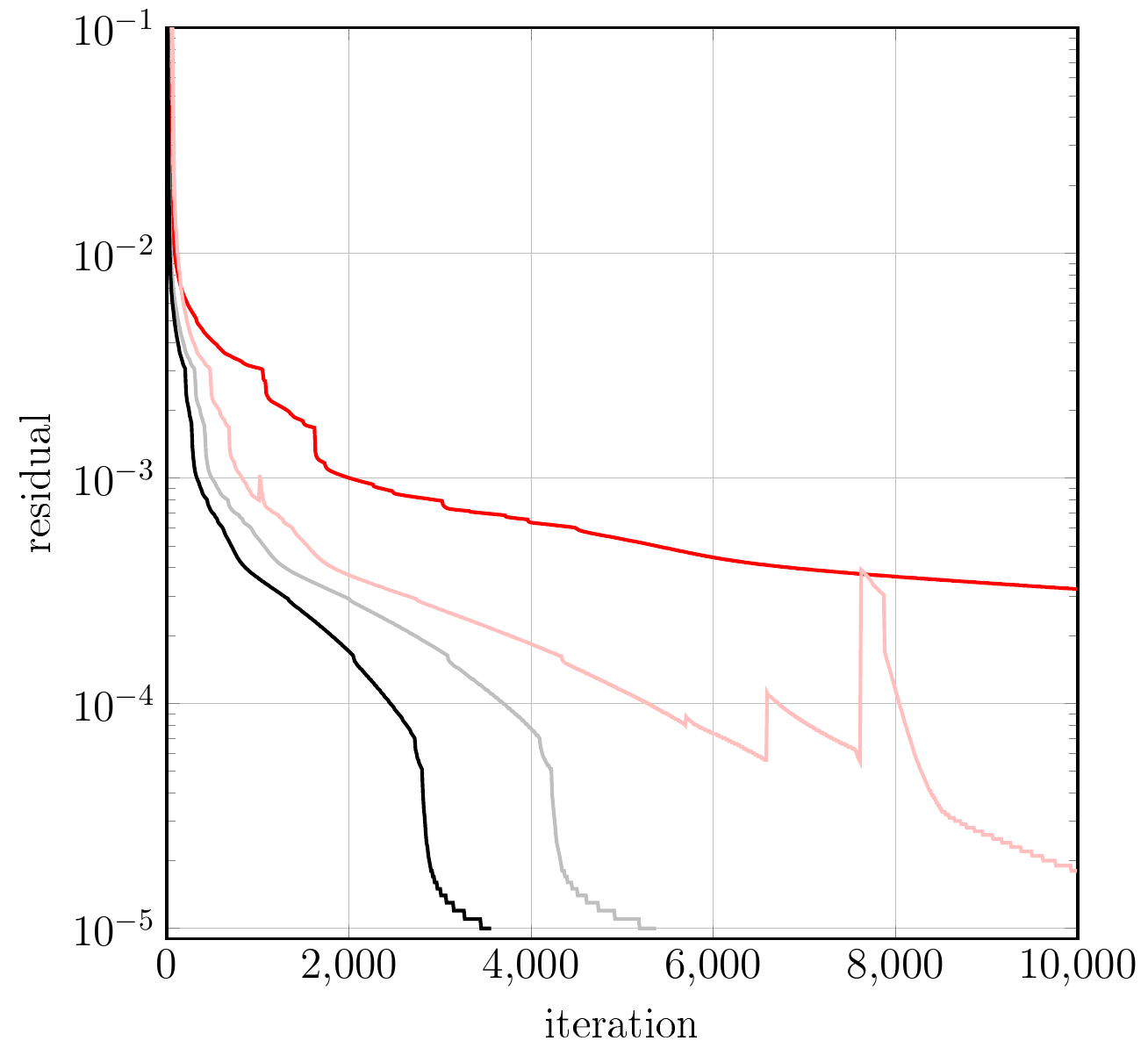}
    		\caption{Residual vs iteration count, \CCMF{} (left) and rotated staggered grid (right)}
    		\label{fig:UD_Res_Appendix}
    	\end{subfigure}
 		\begin{subfigure}{.49\textwidth}
 			\includegraphics[width=.45\textwidth]{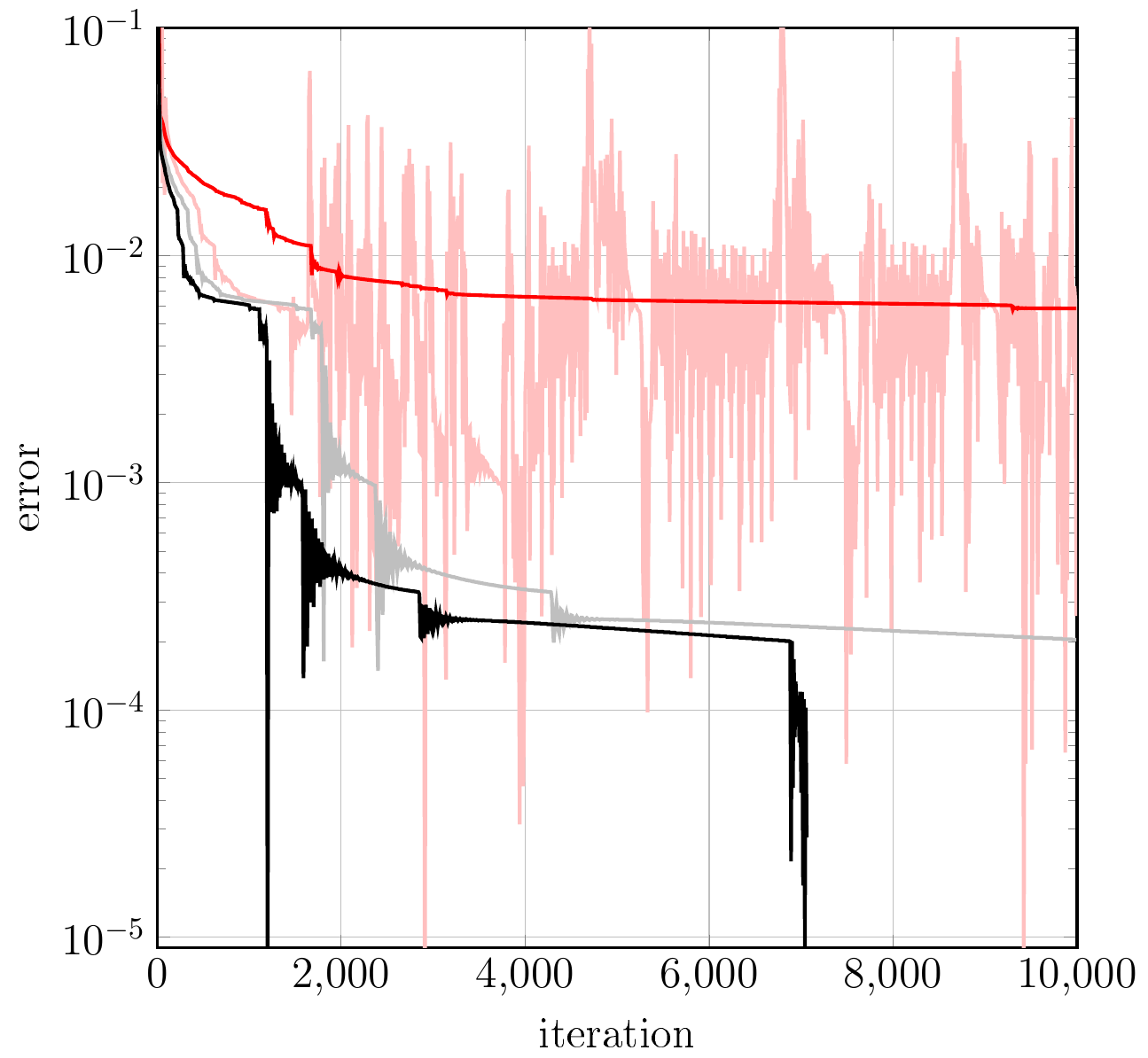}
 			\includegraphics[width=.45\textwidth]{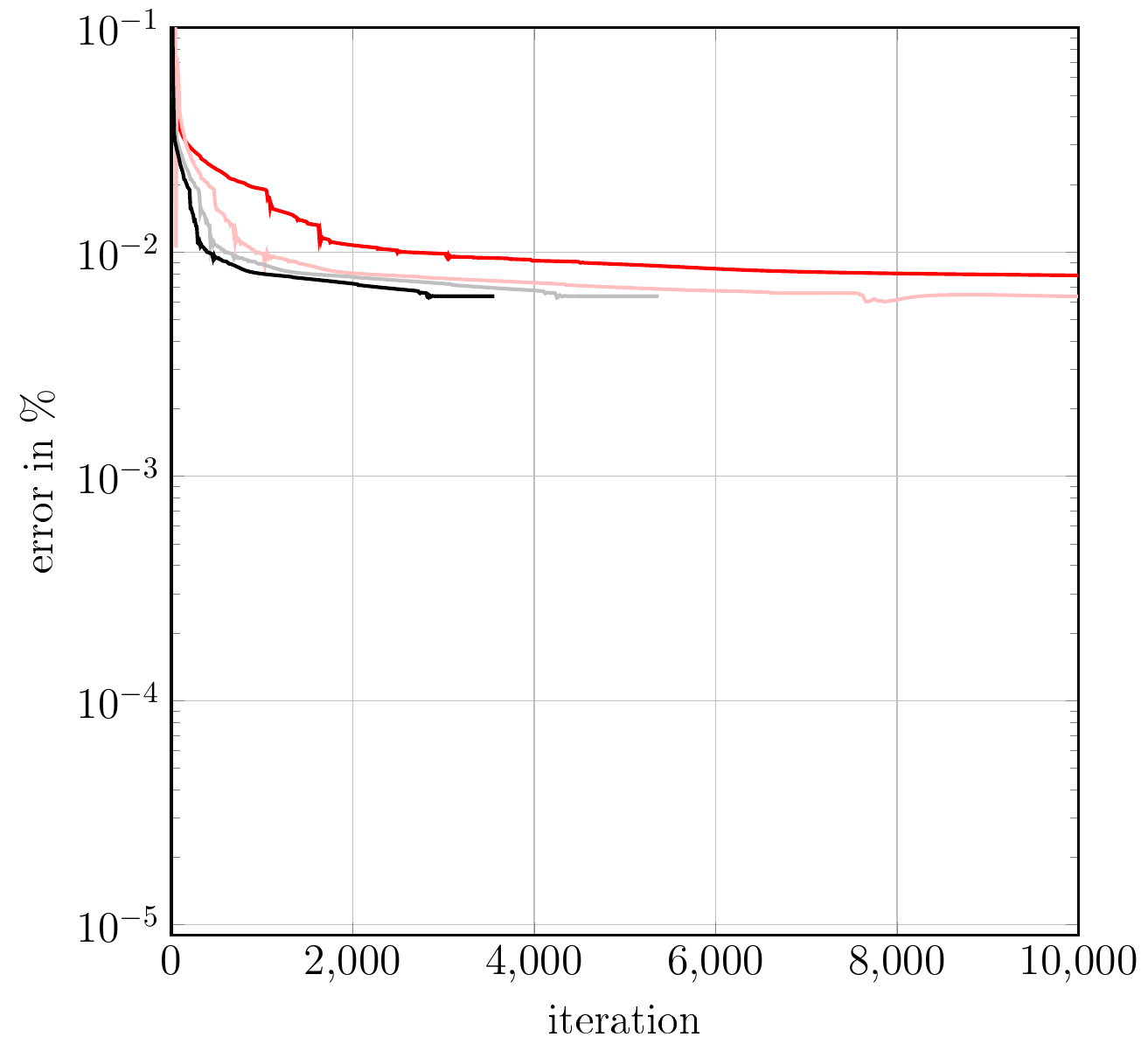}
    		\caption{Error vs iteration count, \CCMF{}(left) and rotated staggered grid (right)}
    		\label{fig:UD_Error_Appendix}
    	\end{subfigure}
 		\hspace{0.2cm}
 		\caption{Residual and error measure for \CCMF{} and rotated staggered grid discretizations, comparing different solver parameters}
 		\label{fig:UD_ResError_Appendix}
 	\end{center}
\end{figure}

In addition to the lower bound and the Barzilai-Borwein strategy for choosing the penalty factor $\rho$ in combination with different damping parameters $\delta$, see Section \ref{sec:computations_UD}, we investigated two additional choices which are popular in the literature. More precisely, we consider residual balancing~\cite{He2000adaptive} and the Lorenz-Tran-Dinh strategy~\cite{Lorenz2019adaptiveDR}, which perform admirably for linear elastic and inelastic homogenization problems~\cite{ADMM2021}. The resulting residual and error plots are shown in Fig.~\ref{fig:UD_ResError_Appendix}. For the \CCMF{}-discretization and the damping parameter $\delta=0.5$, the residual balancing strategy led to an unstable behavior. The choice $\delta=0.25$ resolves this instability. However, this approach does not lead to a high accurate solution. The Lorenz-Tran-Dinh strategy~\cite{Lorenz2019adaptiveDR} shows more promising results, reaching a tolerance of $10^{-4}$ in fewer than $2000$ iterations and $\delta=0.25$. However, this parameter choice turns out to be inferior to the Barzilai-Borwein approach. The relative error \eqref{eq:UD_relative_error}, shown in Fig.~\ref{fig:UD_Error_Appendix} correlates with the residual in a similar way as for the choices considered in Section \ref{sec:computations_UD}. For the rotated staggered grid discretization, the Lorenz-Tran-Dinh scaling with $\delta=0.25$ shows the best performance. However, only low accuracy in terms of the relative error \eqref{eq:UD_relative_error} may be reached.
 
\section*{Acknowledgements}

Support by the German Research Foundation (DFG) within the International Research Training Group "Integrated engineering of continuous-discontinuous long fiber reinforced polymer structures" (GRK 2078) and in terms of the project SCHN 1595/2-1 is gratefully acknowledged. The authors thank S. Gajek (KIT), M. Lebihain (EPFL Lausanne) and D. Kondo (Sorbonne) for fruitful discussions.

\bibliographystyle{ieeetr}
{\footnotesize\bibliography{literature}}

\end{document}